\documentclass[11pt]{article}
\usepackage{indentfirst}
\usepackage{float}
\usepackage{needspace}

\usepackage{tikz}
\usetikzlibrary{shapes, fit, backgrounds, calc, positioning}
\usetikzlibrary{arrows.meta}
\usetikzlibrary{decorations.pathmorphing}

\usepackage[utf8]{inputenc}
\usepackage{amsmath}
\usepackage{amssymb}
\usepackage{amsthm}
\usepackage{cite}
\usepackage{braket}
\usepackage{booktabs}
\usepackage{arydshln}
\usepackage{xcolor}

\newtheorem{theorem}{Theorem}
\newtheorem{corollary}{Corollary}

\newtheorem{lemma}{Lemma}
\newtheorem{example}{Example}

\newtheorem{remark}{Remark}

\newcommand\blfootnote[1]{%
  \begingroup
  \renewcommand\thefootnote{}\footnote{#1}%
  \addtocounter{footnote}{-1}%
  \endgroup
}

\newcommand{\Fq}{\mathbb{F}_q}
\newcommand{\Fp}{\mathbb{F}_p}

\newcommand{\ba}{\boldsymbol a}
\newcommand{\bb}{\boldsymbol b}
\newcommand{\bw}{\boldsymbol w}
\newcommand{\by}{\boldsymbol y}
\newcommand{\bz}{\boldsymbol z}
\newcommand{\bwr}{\widetilde{\boldsymbol w}}

\DeclareMathOperator{\nullspace}{nullspace}

\usepackage[hidelinks]{hyperref}
\usepackage{pgfplots}
\usetikzlibrary{patterns}
\pgfplotsset{compat=1.18}
\definecolor{capblue}{RGB}{28,74,140}
\definecolor{alignorange}{RGB}{204,108,24}
\pgfplotsset{
 capplot/.style={axis lines=left,
  axis line style={-{Stealth[length=2mm]},black!70,line width=.6pt},
  tick style={black!70},tick align=outside,tick label style={font=\small},
  ymajorgrids,grid style={black!12},
  xlabel style={at={(axis description cs:1.01,0)},anchor=west,font=\small},
  ylabel style={rotate=-90,at={(axis description cs:0,1.03)},anchor=south,font=\small},
  legend style={font=\small,draw=black!20,fill=white,inner sep=2pt,at={(0.03,0.94)},anchor=north west},
  legend cell align=left,clip=false},
 capline/.style={capblue,line width=1.3pt,no marks},
 alignline/.style={alignorange,densely dashed,line width=1pt,no marks}}
\usepgfplotslibrary{groupplots}
\DeclareMathOperator{\col}{col}

\title{Capacity of Generic Symmetric Distributed\\Linear Computation: The Geometry of Relations}
\author{Hua Sun}
\date{}
\begin{document}
\maketitle

\blfootnote{
Hua Sun (email: hua.sun@unt.edu) is with the Department of Electrical Engineering at the University of North Texas.}

\begin{abstract}
In distributed linear computation, $K$ users observe linear functions of common finite-field data and send separate messages to a server that wants another linear function. We characterize the generic capacity $C$, the minimum total number of field symbols sent per computed instance, when all observations and the demand have dimension $\nu$, the data have dimension $\mu$, and the coefficients of these linear functions are drawn independently and uniformly from the field and stay fixed during coding, with the field size growing through arbitrary prime powers. For fixed $K$, $C/\nu$ depends only on $\tau=\mu/\nu$: the curve has finitely many linear pieces for at most three users, infinitely many pieces approaching one value of $\tau$ for four users, and a rational part between two sequences of linear pieces for five or more users. The geometry of the relation space, formed by the linear relations among the observations and the demand, determines three things: the shape of the curve, which relations the coding scheme aligns, and the entropy inequalities used in the converse. Generically, the user subspaces of this space are either balanced, where the capacity lies on a single rational curve, or split into two balanced components, a sparse and a dense one, which produces the linear pieces. Achievability uses asymptotic set alignment, in which monomial precoders confine the combined observation differences to a small set. The converse uses the same geometry to lower-bound the sum of the users' remaining observation entropies in terms of the remaining data entropy.
\end{abstract}
\allowdisplaybreaks
\emergencystretch=3em

\section{Introduction}\label{sec:intro}
Distributed linear computation asks a basic question: how much must users communicate to compute a linear function of data that they observe only through other linear functions? Write $q=p^n$, where $p$ is prime and $n\ge1$, so $\Fq$ has characteristic $p$ and contains the prime field $\Fp$. The four-line problem in~\cite{FourLine} takes a microscopic view of this question. Three users observe $w_1$, $w_2$, and $w_1+w_2$, and a server wants $y=w_1+\lambda w_2$, where $\lambda\in\Fq\setminus\{0,1\}$. Although every choice of four distinct lines (one-dimensional subspaces) in a plane (a two-dimensional subspace) has the same dimensions of sums and intersections, its capacity can depend on $\lambda$. This parameter records the relative positions of the four lines and is called their cross-ratio. Its relation to $\Fp$ also matters. Subspace dimensions alone therefore do not determine the capacity of a fixed instance.

\begin{figure}[H]
\centering
\begin{minipage}[b]{.43\textwidth}
\centering
\begin{tikzpicture}[
  user/.style={draw=capblue!70, fill=capblue!7, rounded corners=3pt, line width=.7pt, minimum width=2.0cm, minimum height=0.72cm, font=\small},
  server/.style={draw=black!60, fill=black!4, rounded corners=3pt, line width=.7pt, minimum width=1.4cm, minimum height=3.6cm, font=\small},
  msg/.style={-{Stealth[length=2mm]}, line width=.8pt, black!75},
  lab/.style={font=\small, fill=white, inner sep=1pt}]
\begin{scope}
\node[user] (u1) at (0,1.45) {$\mathbf H_1\bw$};
\node[user] (u2) at (0,0.4) {$\mathbf H_2\bw$};
\node at (0,-0.52) {$\vdots$};
\node[user] (uK) at (0,-1.45) {$\mathbf H_K\bw$};
\node[server] (s) at (3.1,0) {server};
\draw[msg] (u1.east) -- node[lab, above] {$X_1$} (u1.east -| s.west);
\draw[msg] (u2.east) -- node[lab, above] {$X_2$} (u2.east -| s.west);
\draw[msg] (uK.east) -- node[lab, above] {$X_K$} (uK.east -| s.west);
\draw[msg] (s.east) -- ++(0.75,0) node[right, font=\small] {$\mathbf D\bw$};

\end{scope}
\end{tikzpicture}
\par\vspace{14pt}
{\small (a) The general problem}
\end{minipage}\hspace{.012\textwidth}
\begin{minipage}[b]{.43\textwidth}
\centering
\begin{tikzpicture}
\begin{axis}[width=.94\linewidth,height=4.8cm,capplot,xlabel={$\mu/\nu$},ylabel={$C/\nu$},
 xmin=1,xmax=2,ymin=.75,ymax=2.15,xtick={1,1.5,2},xticklabels={$1$,$3/2$,$2$},ytick={1,2}]
\addplot[capline] coordinates {(1,1) (1.5,1) (2,2)};
\node[capblue,font=\small,anchor=south] at (axis cs:1.25,1.03) {$C=\nu$};
\node[capblue,font=\small,anchor=north west] at (axis cs:1.62,1.25) {$C=2(\mu-\nu)$};
\end{axis}
\end{tikzpicture}
\par\smallskip
{\small (b) Capacity}
\end{minipage}
\caption{(a) Each observation and the demand have dimension $\nu$, and the data have dimension $\mu$. User $k$ sends message $X_k$ to the server. (b) Capacity curve of~\eqref{eq:twouserintro}.}
\label{fig:problem}
\end{figure}
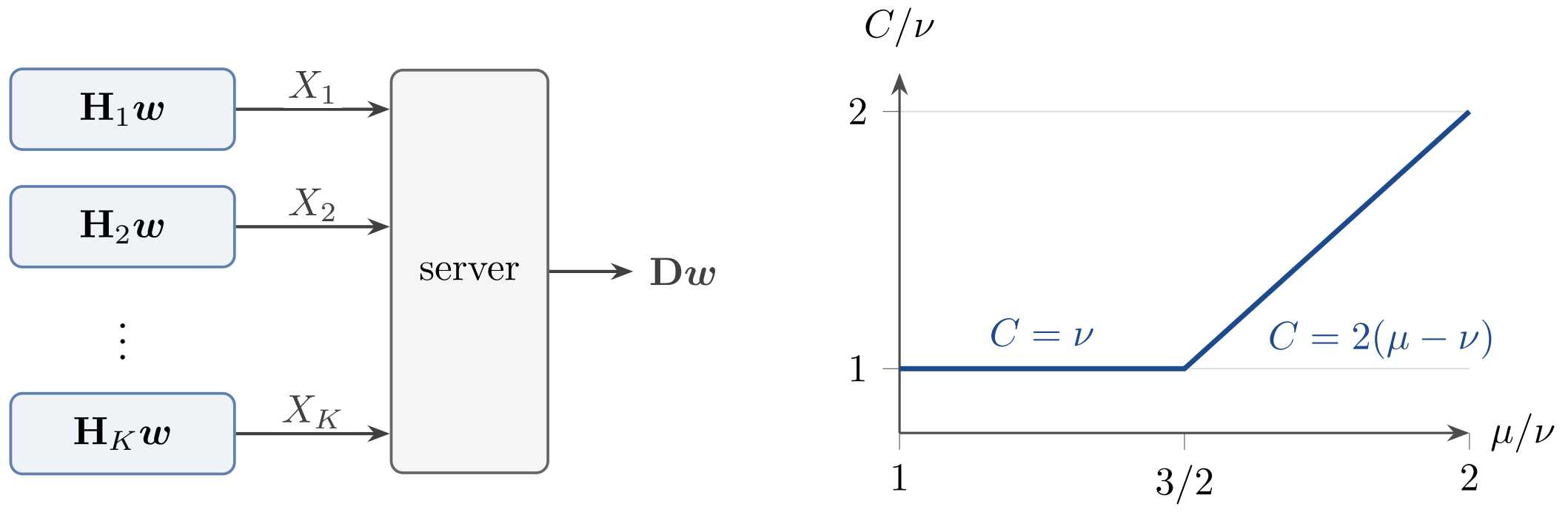

Here we take a macroscopic view. We determine the generic capacity when the coefficients specifying the linear observations and demand are drawn independently and uniformly from a large field. We allow any number of users and focus on the symmetric setting in which all observations and the demand have the same dimension. Specifically, for data $\bw\in\Fq^\mu$, matrices $\mathbf H_k,\mathbf D\in\Fq^{\nu\times\mu}$ specify user $k$'s observation $\mathbf H_k\bw$ and the server's demand $\mathbf D\bw$; see Figure~\ref{fig:problem}(a). The \emph{capacity} $C$ is the minimum total number of $q$-ary symbols the users must send per computed instance with vanishing error for uniform data. We let $q\to\infty$, whether $p$, $n$, or both grow; for every feasible $(K,\mu,\nu)$, the generic capacity is the same along every such sequence. It turns out that, for fixed $K$, $C/\nu$ depends only on the data-to-demand ratio $\tau=\mu/\nu$.

\emph{Two users as a warm-up.} The two-user problem is fully characterized in~\cite{FourLine}, for arbitrary fixed observations and demand. Its generic symmetric specialization is
\begin{equation}\label{eq:twouserintro}
C=\max\{\nu,\;2(\mu-\nu)\},\qquad \nu\le\mu\le2\nu.
\end{equation}
The first term is the demand dimension: at least $\nu$ symbols must be sent. For the second term, even if the server is given one user's observation, the generic demand still contains $\mu-\nu$ unknown symbols. The other user must send at least that much, giving a total of $2(\mu-\nu)$. For achievability, change bases so that each demand coordinate is either observed by a user or is a sum of two separately observed coordinates. Send the former once and both summands for the latter. Generically, there are $\max\{0,2\mu-3\nu\}$ coordinates of the latter kind, giving total communication $\nu+\max\{0,2\mu-3\nu\}$, as in~\eqref{eq:twouserintro}. Thus capacity stays at $\nu$ while $\mu/\nu\le3/2$, and then increases linearly. Figure~\ref{fig:problem}(b) shows this capacity curve as a first illustration. Throughout the figures, both axes are measured relative to the demand dimension: $\mu/\nu$ horizontally and $C/\nu$ vertically.

The observations $\bz_k=\mathbf H_k\bw$ and demand $\by=\mathbf D\bw$ contain $(K+1)\nu$ coordinates generated by $\mu$ data symbols. Their total redundancy is $\kappa=(K+1)\nu-\mu$, the number of independent linear equations connecting them. We collect these equations using matrices $\mathbf A_k,\mathbf B\in\Fq^{\kappa\times\nu}$:
\[
\mathbf A_1\bz_1+\cdots+\mathbf A_K\bz_K=\mathbf B\by.
\]
Both sides are vectors in $\Fq^\kappa$. We call this the \emph{relation space}: it is the space in which the equations connecting observations and demand are written. The directions contributed by user $k$ form the $\nu$-dimensional subspace $\mathcal A_k=\col(\mathbf A_k)$. The demand contributes $\mathcal B=\col(\mathbf B)$. Here $\col$ denotes column space. Generically, the entries of all matrices $\mathbf A_1,\ldots,\mathbf A_K,\mathbf B$ can be modeled as independent, uniformly distributed random variables (Appendix~\ref{app:reduction}). For coding, user $k$ partitions its possible observation values into bins and sends the bin index. If $\bz_k$ and $\bz_k'$ belong to the same bin, the local observation difference $\bz_k-\bz_k'$ leaves its index unchanged. Its signal in the relation equations is $\mathbf A_k(\bz_k-\bz_k')$. The overlap to exploit is between these signals in $\mathcal A_k$, rather than between the originally observed functions.

For two data inputs producing the same bin indices, subtracting their relation equations gives
\begin{equation}\label{eq:subtract}
\sum_k\mathbf A_k(\bz_k-\bz_k')=\mathbf B(\by-\by').
\end{equation}
Our scheme, \emph{asymptotic set alignment}, keeps the bins large while confining the combined observation differences on the left to a small set. It chooses this set to contain no nonzero demand difference $\mathbf B(\by-\by')$, so equal bin indices force $\by=\by'$. Thus large bins save communication while their aligned differences still permit demand recovery. To align, the scheme uses precoding matrices whose entries are monomials in the entries of $\mathbf A_1,\ldots,\mathbf A_K$: multiplying by an entry of some $\mathbf A_k$ changes one exponent, and restricted coefficient ranges keep the number of combined observation differences small. The same scheme achieves capacity in every case below; the geometry of the relation space decides where it is applied.

This geometry, namely how the user subspaces $\mathcal A_1,\ldots,\mathcal A_K$ sit in the relation space, determines three things: the shape of the capacity curve, which relations the coding scheme aligns, and the entropy bounds used in the converse. On average they occupy $K\nu/\kappa$ dimensions per relation dimension, and the configuration is \emph{balanced} if no subspace contains more than this proportional share. For generic channels, the per-user fraction $\nu/\kappa$ alone decides between two regimes (Lemma~\ref{lem:geometry}): either the configuration is balanced, or it splits into a direct sum of two balanced components, a \emph{sparse} one in which each user occupies a smaller fraction and a \emph{dense} one in which each user occupies a larger fraction. We call $\nu/\kappa$ a \emph{balanced fraction} or a \emph{split fraction} accordingly. In the balanced regime, applying asymptotic set alignment to all relation coordinates together, called \emph{direct alignment}, turns out to be optimal, and the capacity lies on a single rational curve. In the split regime, direct alignment is applied to each component, using the sparse component first: it computes as much of the demand as it can, since sparse relations need less communication per demand dimension, and the dense component computes the rest; here the capacity turns out to be piecewise linear in the data-to-demand ratio.

The capacity curves have three forms, naturally organized by the representation theory of subspaces~\cite{DerksenWeyman}. The geometry of the relation space involves the $K$ user subspaces, with the demand kept separate, and its classical representation types are finite for $K\le3$, tame for $K=4$, and wild for $K\ge5$~\cite{DerksenWeyman}, \cite{Kirillov}. Finite type has only finitely many configurations that cannot be split further; tame type permits one-parameter families at each fixed dimension, as in the four-subspace problem, whose smallest case is the four-line problem with its cross-ratio; wild type contains the classification of arbitrary pairs of square matrices up to simultaneous similarity. These names refer to classifying all configurations; the capacity needs only the two regimes above.

\subsection{Finite type: finitely many linear pieces}
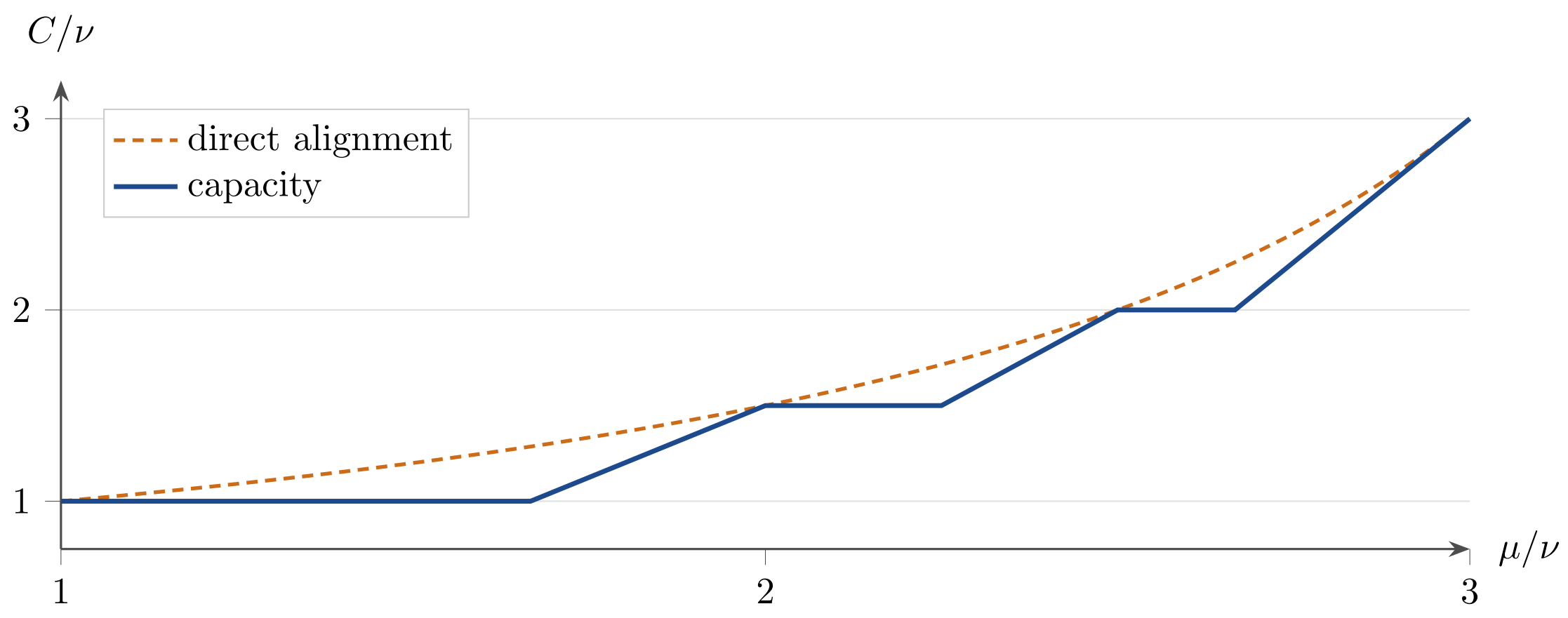
\begin{figure}[H]
\centering
\begin{tikzpicture}
\begin{axis}[width=.9\textwidth,height=6cm,capplot,xlabel={$\mu/\nu$},ylabel={$C/\nu$},
 xmin=1,xmax=3,ymin=.75,ymax=3.2,
 xtick={1,2,3},ytick={1,2,3}]
\addplot[alignline,domain=1:3,samples=250] {3/(4-x)};
\addlegendentry{direct alignment}
\addplot[capline] coordinates {(1,1) (1.002227171,1) (1.004464286,1) (1.005,1) (1.006711409,1) (1.00896861,1) (1.01,1) (1.011235955,1) (1.013513513,1) (1.015,1) (1.015801354,1) (1.018099548,1) (1.02,1) (1.020408163,1) (1.022727273,1) (1.025,1) (1.025056948,1) (1.02739726,1) (1.029748284,1) (1.03,1) (1.032110092,1) (1.034482759,1) (1.035,1) (1.036866359,1) (1.03926097,1) (1.04,1) (1.041666667,1) (1.044083527,1) (1.045,1) (1.046511628,1) (1.048951049,1) (1.05,1) (1.051401869,1) (1.053864169,1) (1.055,1) (1.056338028,1) (1.058823529,1) (1.06,1) (1.061320755,1) (1.063829787,1) (1.065,1) (1.066350711,1) (1.06888361,1) (1.07,1) (1.071428571,1) (1.07398568,1) (1.075,1) (1.076555024,1) (1.079136691,1) (1.08,1) (1.081730769,1) (1.084337349,1) (1.085,1) (1.086956522,1) (1.089588378,1) (1.09,1) (1.09223301,1) (1.094890511,1) (1.095,1) (1.097560976,1) (1.1,1) (1.100244499,1) (1.102941176,1) (1.105,1) (1.105651106,1) (1.108374384,1) (1.11,1) (1.111111111,1) (1.113861386,1) (1.115,1) (1.11662531,1) (1.119402985,1) (1.12,1) (1.122194514,1) (1.125,1) (1.125,1) (1.127819549,1) (1.13,1) (1.130653266,1) (1.133501259,1) (1.135,1) (1.136363636,1) (1.139240506,1) (1.14,1) (1.14213198,1) (1.145,1) (1.145038168,1) (1.147959184,1) (1.15,1) (1.150895141,1) (1.153846154,1) (1.155,1) (1.156812339,1) (1.159793814,1) (1.16,1) (1.162790698,1) (1.165,1) (1.165803109,1) (1.168831169,1) (1.17,1) (1.171875,1) (1.174934726,1) (1.175,1) (1.178010471,1) (1.18,1) (1.181102362,1) (1.184210526,1) (1.185,1) (1.187335092,1) (1.19,1) (1.19047619,1) (1.193633952,1) (1.195,1) (1.196808511,1) (1.2,1) (1.2,1) (1.203208556,1) (1.205,1) (1.206434316,1) (1.209677419,1) (1.21,1) (1.212938005,1) (1.215,1) (1.216216216,1) (1.219512195,1) (1.22,1) (1.222826087,1) (1.225,1) (1.226158038,1) (1.229508197,1) (1.23,1) (1.232876712,1) (1.235,1) (1.236263736,1) (1.239669421,1) (1.24,1) (1.243093923,1) (1.245,1) (1.246537396,1) (1.25,1) (1.253481894,1) (1.255,1) (1.25698324,1) (1.26,1) (1.260504202,1) (1.264044944,1) (1.265,1) (1.267605634,1) (1.27,1) (1.271186441,1) (1.274787535,1) (1.275,1) (1.278409091,1) (1.28,1) (1.282051282,1) (1.285,1) (1.285714286,1) (1.289398281,1) (1.29,1) (1.293103448,1) (1.295,1) (1.296829971,1) (1.3,1) (1.300578035,1) (1.304347826,1) (1.305,1) (1.308139535,1) (1.31,1) (1.311953353,1) (1.315,1) (1.315789474,1) (1.319648094,1) (1.32,1) (1.323529412,1) (1.325,1) (1.327433628,1) (1.33,1) (1.331360947,1) (1.335,1) (1.335311573,1) (1.339285714,1) (1.34,1) (1.343283582,1) (1.345,1) (1.347305389,1) (1.35,1) (1.351351351,1) (1.355,1) (1.355421687,1) (1.359516616,1) (1.36,1) (1.363636364,1) (1.365,1) (1.367781155,1) (1.37,1) (1.371951219,1) (1.375,1) (1.376146789,1) (1.38,1) (1.380368098,1) (1.384615385,1) (1.385,1) (1.388888889,1) (1.39,1) (1.393188854,1) (1.395,1) (1.397515528,1) (1.4,1) (1.401869159,1) (1.405,1) (1.40625,1) (1.41,1) (1.410658307,1) (1.415,1) (1.41509434,1) (1.41955836,1) (1.42,1) (1.424050633,1) (1.425,1) (1.428571429,1) (1.43,1) (1.433121019,1) (1.435,1) (1.43769968,1) (1.44,1) (1.442307692,1) (1.445,1) (1.446945338,1) (1.45,1) (1.451612903,1) (1.455,1) (1.45631068,1) (1.46,1) (1.461038961,1) (1.465,1) (1.465798046,1) (1.47,1) (1.470588235,1) (1.475,1) (1.475409836,1) (1.48,1) (1.480263158,1) (1.485,1) (1.485148515,1) (1.49,1) (1.490066225,1) (1.495,1) (1.495016611,1) (1.5,1) (1.5,1) (1.505,1) (1.505016723,1) (1.51,1) (1.510067114,1) (1.515,1) (1.515151515,1) (1.52,1) (1.52027027,1) (1.525,1) (1.525423729,1) (1.53,1) (1.530612245,1) (1.535,1) (1.535836178,1) (1.54,1) (1.54109589,1) (1.545,1) (1.546391752,1) (1.55,1) (1.551724138,1) (1.555,1) (1.557093426,1) (1.56,1) (1.5625,1) (1.565,1) (1.567944251,1) (1.57,1) (1.573426573,1) (1.575,1) (1.578947369,1) (1.58,1) (1.584507042,1) (1.585,1) (1.59,1) (1.590106007,1) (1.595,1) (1.595744681,1) (1.6,1) (1.601423488,1) (1.605,1) (1.607142857,1) (1.61,1) (1.612903226,1) (1.615,1) (1.618705036,1) (1.62,1) (1.624548736,1) (1.625,1) (1.63,1) (1.630434783,1) (1.635,1) (1.636363636,1) (1.64,1) (1.642335766,1) (1.645,1) (1.648351648,1) (1.65,1) (1.654411765,1) (1.655,1) (1.66,1) (1.660516605,1) (1.665,1) (1.666666667,1) (1.67,1.005) (1.672862453,1.00929368) (1.675,1.0125) (1.679104477,1.018656716) (1.68,1.02) (1.685,1.0275) (1.685393259,1.028089888) (1.69,1.035) (1.691729323,1.037593985) (1.695,1.0425) (1.698113208,1.047169811) (1.7,1.05) (1.704545454,1.056818182) (1.705,1.0575) (1.71,1.065) (1.711026616,1.066539924) (1.715,1.0725) (1.717557252,1.076335878) (1.72,1.08) (1.724137931,1.086206897) (1.725,1.0875) (1.73,1.095) (1.730769231,1.096153846) (1.735,1.1025) (1.737451737,1.106177606) (1.74,1.11) (1.744186047,1.11627907) (1.745,1.1175) (1.75,1.125) (1.750972763,1.126459144) (1.755,1.1325) (1.7578125,1.13671875) (1.76,1.14) (1.764705882,1.147058823) (1.765,1.1475) (1.77,1.155) (1.771653543,1.157480315) (1.775,1.1625) (1.778656127,1.16798419) (1.78,1.17) (1.785,1.1775) (1.785714286,1.178571429) (1.79,1.185) (1.792828685,1.189243028) (1.795,1.1925) (1.8,1.2) (1.8,1.2) (1.805,1.2075) (1.807228916,1.210843374) (1.81,1.215) (1.814516129,1.221774194) (1.815,1.2225) (1.82,1.23) (1.821862348,1.232793522) (1.825,1.2375) (1.829268293,1.243902439) (1.83,1.245) (1.835,1.2525) (1.836734694,1.255102041) (1.84,1.26) (1.844262295,1.266393443) (1.845,1.2675) (1.85,1.275) (1.851851852,1.277777778) (1.855,1.2825) (1.859504132,1.289256198) (1.86,1.29) (1.865,1.2975) (1.867219917,1.300829875) (1.87,1.305) (1.875,1.3125) (1.875,1.3125) (1.88,1.32) (1.882845188,1.324267782) (1.885,1.3275) (1.89,1.335) (1.890756302,1.336134454) (1.895,1.3425) (1.898734177,1.348101266) (1.9,1.35) (1.905,1.3575) (1.906779661,1.360169492) (1.91,1.365) (1.914893617,1.372340426) (1.915,1.3725) (1.92,1.38) (1.923076923,1.384615385) (1.925,1.3875) (1.93,1.395) (1.931330472,1.396995708) (1.935,1.4025) (1.939655172,1.409482758) (1.94,1.41) (1.945,1.4175) (1.948051948,1.422077922) (1.95,1.425) (1.955,1.4325) (1.956521739,1.434782609) (1.96,1.44) (1.965,1.4475) (1.965065502,1.447598253) (1.97,1.455) (1.97368421,1.460526316) (1.975,1.4625) (1.98,1.47) (1.982378855,1.473568282) (1.985,1.4775) (1.99,1.485) (1.991150443,1.486725664) (1.995,1.4925) (2,1.5) (2.005,1.5) (2.008928571,1.5) (2.01,1.5) (2.015,1.5) (2.01793722,1.5) (2.02,1.5) (2.025,1.5) (2.027027027,1.5) (2.03,1.5) (2.035,1.5) (2.036199095,1.5) (2.04,1.5) (2.045,1.5) (2.045454545,1.5) (2.05,1.5) (2.05479452,1.5) (2.055,1.5) (2.06,1.5) (2.064220184,1.5) (2.065,1.5) (2.07,1.5) (2.073732719,1.5) (2.075,1.5) (2.08,1.5) (2.083333333,1.5) (2.085,1.5) (2.09,1.5) (2.093023256,1.5) (2.095,1.5) (2.1,1.5) (2.102803738,1.5) (2.105,1.5) (2.11,1.5) (2.112676056,1.5) (2.115,1.5) (2.12,1.5) (2.122641509,1.5) (2.125,1.5) (2.13,1.5) (2.132701422,1.5) (2.135,1.5) (2.14,1.5) (2.142857143,1.5) (2.145,1.5) (2.15,1.5) (2.153110048,1.5) (2.155,1.5) (2.16,1.5) (2.163461539,1.5) (2.165,1.5) (2.17,1.5) (2.173913043,1.5) (2.175,1.5) (2.18,1.5) (2.184466019,1.5) (2.185,1.5) (2.19,1.5) (2.195,1.5) (2.195121951,1.5) (2.2,1.5) (2.205,1.5) (2.205882353,1.5) (2.21,1.5) (2.215,1.5) (2.216748769,1.5) (2.22,1.5) (2.225,1.5) (2.227722772,1.5) (2.23,1.5) (2.235,1.5) (2.23880597,1.5) (2.24,1.5) (2.245,1.5) (2.25,1.5) (2.25,1.5) (2.255,1.51) (2.26,1.52) (2.261306533,1.522613066) (2.265,1.53) (2.27,1.54) (2.272727273,1.545454545) (2.275,1.55) (2.28,1.56) (2.284263959,1.568527919) (2.285,1.57) (2.29,1.58) (2.295,1.59) (2.295918367,1.591836734) (2.3,1.6) (2.305,1.61) (2.307692308,1.615384616) (2.31,1.62) (2.315,1.63) (2.319587629,1.639175258) (2.32,1.64) (2.325,1.65) (2.33,1.66) (2.331606218,1.663212435) (2.335,1.67) (2.34,1.68) (2.34375,1.6875) (2.345,1.69) (2.35,1.7) (2.355,1.71) (2.356020943,1.712041885) (2.36,1.72) (2.365,1.73) (2.368421053,1.736842106) (2.37,1.74) (2.375,1.75) (2.38,1.76) (2.380952381,1.761904762) (2.385,1.77) (2.39,1.78) (2.393617021,1.787234042) (2.395,1.79) (2.4,1.8) (2.405,1.81) (2.406417112,1.812834224) (2.41,1.82) (2.415,1.83) (2.419354839,1.838709678) (2.42,1.84) (2.425,1.85) (2.43,1.86) (2.432432432,1.864864865) (2.435,1.87) (2.44,1.88) (2.445,1.89) (2.445652174,1.891304348) (2.45,1.9) (2.455,1.91) (2.459016393,1.918032786) (2.46,1.92) (2.465,1.93) (2.47,1.94) (2.472527473,1.945054946) (2.475,1.95) (2.48,1.96) (2.485,1.97) (2.486187845,1.972375691) (2.49,1.98) (2.495,1.99) (2.5,2) (2.505,2) (2.51,2) (2.51396648,2) (2.515,2) (2.52,2) (2.525,2) (2.528089887,2) (2.53,2) (2.535,2) (2.54,2) (2.542372882,2) (2.545,2) (2.55,2) (2.555,2) (2.556818182,2) (2.56,2) (2.565,2) (2.57,2) (2.571428571,2) (2.575,2) (2.58,2) (2.585,2) (2.586206896,2) (2.59,2) (2.595,2) (2.6,2) (2.60115607,2) (2.605,2) (2.61,2) (2.615,2) (2.61627907,2) (2.62,2) (2.625,2) (2.63,2) (2.631578947,2) (2.635,2) (2.64,2) (2.645,2) (2.647058823,2) (2.65,2) (2.655,2) (2.66,2) (2.662721893,2) (2.665,2) (2.666666667,2) (2.67,2.01) (2.675,2.025) (2.678571429,2.035714286) (2.68,2.04) (2.685,2.055) (2.69,2.07) (2.694610779,2.083832336) (2.695,2.085) (2.7,2.1) (2.705,2.115) (2.71,2.13) (2.710843373,2.13253012) (2.715,2.145) (2.72,2.16) (2.725,2.175) (2.727272727,2.181818181) (2.73,2.19) (2.735,2.205) (2.74,2.22) (2.743902439,2.231707318) (2.745,2.235) (2.75,2.25) (2.755,2.265) (2.76,2.28) (2.760736196,2.282208589) (2.765,2.295) (2.77,2.31) (2.775,2.325) (2.777777778,2.333333333) (2.78,2.34) (2.785,2.355) (2.79,2.37) (2.795,2.385) (2.795031056,2.385093167) (2.8,2.4) (2.805,2.415) (2.81,2.43) (2.8125,2.437499999) (2.815,2.445) (2.82,2.46) (2.825,2.475) (2.83,2.49) (2.83018868,2.490566039) (2.835,2.505) (2.84,2.52) (2.845,2.535) (2.848101266,2.544303798) (2.85,2.55) (2.855,2.565) (2.86,2.58) (2.865,2.595) (2.866242038,2.598726114) (2.87,2.61) (2.875,2.625) (2.88,2.64) (2.884615384,2.653846153) (2.885,2.655) (2.89,2.67) (2.895,2.685) (2.9,2.7) (2.903225807,2.70967742) (2.905,2.715) (2.91,2.73) (2.915,2.745) (2.92,2.76) (2.922077922,2.766233767) (2.925,2.775) (2.93,2.79) (2.935,2.805) (2.94,2.82) (2.941176471,2.823529412) (2.945,2.835) (2.95,2.85) (2.955,2.865) (2.96,2.88) (2.960526316,2.881578947) (2.965,2.895) (2.97,2.91) (2.975,2.925) (2.98,2.94) (2.98013245,2.94039735) (2.985,2.955) (2.99,2.97) (2.995,2.985) (3,3)};
\addlegendentry{capacity}
\end{axis}
\end{tikzpicture}
\caption{Finite type: three users. The six linear pieces of $C/\nu$ alternate between horizontal and rising line segments. The dashed curve is $R_{\mathrm{align}}/\nu$, achieved by direct alignment.}
\label{fig:finite}
\end{figure}

For $K\le3$, $C/\nu$ has finitely many linear pieces; Figure~\ref{fig:finite} shows the three-user curve. Direct alignment achieves sum-rate
\begin{equation}\label{eq:align-reference}
R_{\mathrm{align}}=\frac{K\nu^2}{\kappa},
\end{equation}
so $C\le R_{\mathrm{align}}$. This rate gives the reference curve in the figures, and capacity meets it exactly at the balanced fractions. For $K\le3$ these are finitely many isolated points. Between two consecutive ones the configuration splits, and direct alignment on each component gives a horizontal and then a rising segment below the reference curve.

\subsection{Tame type: infinitely many linear pieces}
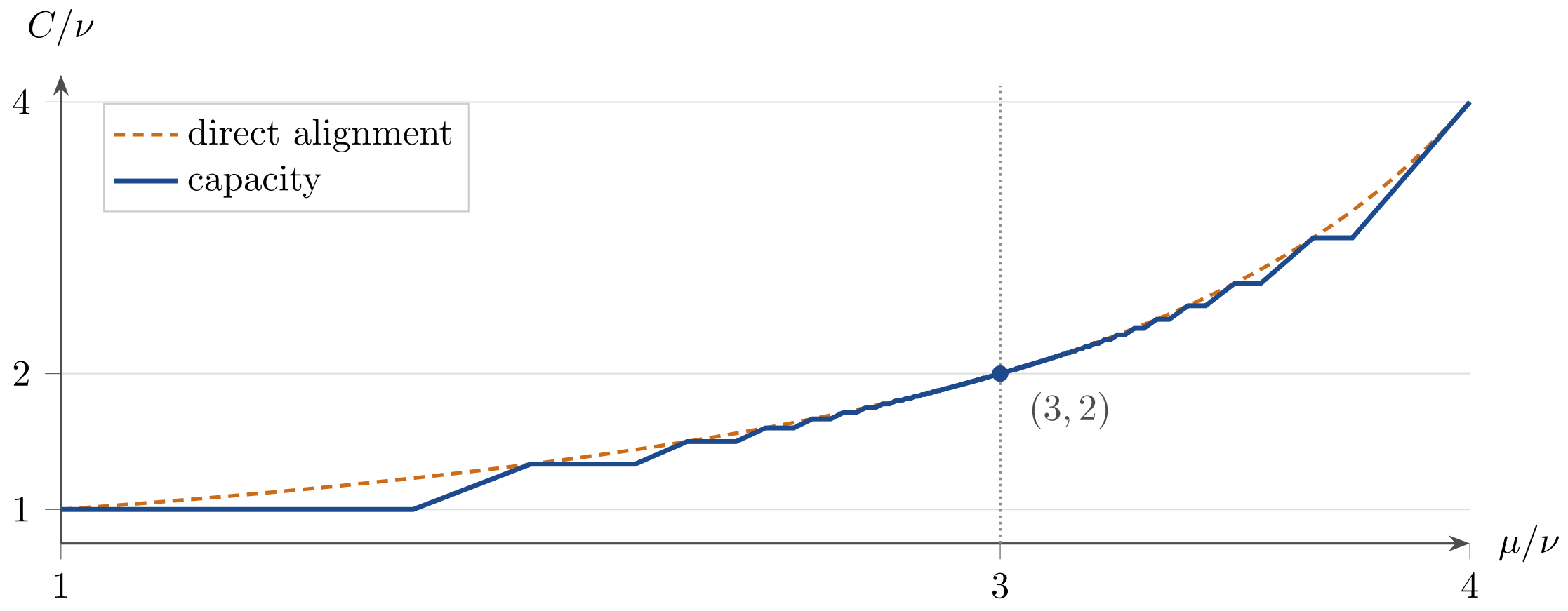
\begin{figure}[H]
\centering
\begin{tikzpicture}
\begin{axis}[width=.9\textwidth,height=6cm,capplot,xlabel={$\mu/\nu$},ylabel={$C/\nu$},
 xmin=1,xmax=4,ymin=.75,ymax=4.2,
 xtick={1,3,4},ytick={1,2,4}]
\addplot[alignline,domain=1:4,samples=250] {4/(5-x)};
\addlegendentry{direct alignment}
\addplot[capline] coordinates {(1,1) (1.002506266,1) (1.005025126,1) (1.0075,1) (1.007556675,1) (1.01010101,1) (1.012658228,1) (1.015,1) (1.015228426,1) (1.017811705,1) (1.020408163,1) (1.0225,1) (1.023017903,1) (1.025641026,1) (1.028277635,1) (1.03,1) (1.030927835,1) (1.033591731,1) (1.03626943,1) (1.0375,1) (1.038961039,1) (1.041666667,1) (1.044386423,1) (1.045,1) (1.047120419,1) (1.049868766,1) (1.0525,1) (1.052631579,1) (1.055408971,1) (1.058201058,1) (1.06,1) (1.061007958,1) (1.063829787,1) (1.066666667,1) (1.0675,1) (1.069518717,1) (1.072386059,1) (1.075,1) (1.075268817,1) (1.078167116,1) (1.081081081,1) (1.0825,1) (1.08401084,1) (1.086956522,1) (1.089918256,1) (1.09,1) (1.092896175,1) (1.095890411,1) (1.0975,1) (1.098901099,1) (1.101928375,1) (1.104972376,1) (1.105,1) (1.108033241,1) (1.111111111,1) (1.1125,1) (1.114206128,1) (1.117318436,1) (1.12,1) (1.120448179,1) (1.123595506,1) (1.126760563,1) (1.1275,1) (1.129943503,1) (1.133144476,1) (1.135,1) (1.136363636,1) (1.13960114,1) (1.1425,1) (1.142857143,1) (1.146131805,1) (1.149425287,1) (1.15,1) (1.152737752,1) (1.156069364,1) (1.1575,1) (1.15942029,1) (1.162790698,1) (1.165,1) (1.166180758,1) (1.169590643,1) (1.1725,1) (1.173020528,1) (1.176470588,1) (1.179941003,1) (1.18,1) (1.183431953,1) (1.18694362,1) (1.1875,1) (1.19047619,1) (1.194029851,1) (1.195,1) (1.19760479,1) (1.201201201,1) (1.2025,1) (1.204819277,1) (1.208459215,1) (1.21,1) (1.212121212,1) (1.215805471,1) (1.2175,1) (1.219512195,1) (1.22324159,1) (1.225,1) (1.226993865,1) (1.230769231,1) (1.2325,1) (1.234567901,1) (1.238390093,1) (1.24,1) (1.242236025,1) (1.246105919,1) (1.2475,1) (1.25,1) (1.253918495,1) (1.255,1) (1.257861635,1) (1.261829653,1) (1.2625,1) (1.265822785,1) (1.26984127,1) (1.27,1) (1.27388535,1) (1.2775,1) (1.277955272,1) (1.282051282,1) (1.285,1) (1.286173633,1) (1.290322581,1) (1.2925,1) (1.294498382,1) (1.298701299,1) (1.3,1) (1.302931596,1) (1.307189542,1) (1.3075,1) (1.31147541,1) (1.315,1) (1.315789474,1) (1.320132013,1) (1.3225,1) (1.324503311,1) (1.328903654,1) (1.33,1) (1.333333333,1) (1.3375,1) (1.337792642,1) (1.342281879,1) (1.345,1) (1.346801347,1) (1.351351351,1) (1.3525,1) (1.355932203,1) (1.36,1) (1.360544218,1) (1.365187713,1) (1.3675,1) (1.369863014,1) (1.374570447,1) (1.375,1) (1.379310345,1) (1.3825,1) (1.384083045,1) (1.388888889,1) (1.39,1) (1.393728223,1) (1.3975,1) (1.398601399,1) (1.403508772,1) (1.405,1) (1.408450704,1) (1.4125,1) (1.413427562,1) (1.418439716,1) (1.42,1) (1.423487544,1) (1.4275,1) (1.428571429,1) (1.433691756,1) (1.435,1) (1.438848921,1) (1.4425,1) (1.444043321,1) (1.449275362,1) (1.45,1) (1.454545455,1) (1.4575,1) (1.459854015,1) (1.465,1) (1.465201465,1) (1.470588235,1) (1.4725,1) (1.47601476,1) (1.48,1) (1.481481481,1) (1.486988848,1) (1.4875,1) (1.492537313,1) (1.495,1) (1.498127341,1) (1.5025,1) (1.503759398,1) (1.509433962,1) (1.51,1) (1.515151515,1) (1.5175,1) (1.520912548,1) (1.525,1) (1.526717557,1) (1.5325,1) (1.53256705,1) (1.538461538,1) (1.54,1) (1.544401544,1) (1.5475,1) (1.550387597,1) (1.555,1) (1.556420233,1) (1.5625,1) (1.568627451,1) (1.57,1) (1.57480315,1) (1.5775,1) (1.581027668,1) (1.585,1) (1.587301587,1) (1.5925,1) (1.593625498,1) (1.6,1) (1.606425703,1) (1.6075,1) (1.612903226,1) (1.615,1) (1.619433198,1) (1.6225,1) (1.62601626,1) (1.63,1) (1.632653061,1) (1.6375,1) (1.639344262,1) (1.645,1) (1.646090535,1) (1.6525,1) (1.652892562,1) (1.659751037,1) (1.66,1) (1.666666667,1) (1.6675,1) (1.673640167,1) (1.675,1) (1.680672269,1) (1.6825,1) (1.687763713,1) (1.69,1) (1.694915254,1) (1.6975,1) (1.70212766,1) (1.705,1) (1.709401709,1) (1.7125,1) (1.716738197,1) (1.72,1) (1.724137931,1) (1.7275,1) (1.731601732,1) (1.735,1) (1.739130435,1) (1.7425,1) (1.746724891,1) (1.75,1) (1.75,1) (1.754385965,1.005847953) (1.7575,1.01) (1.762114537,1.016152717) (1.765,1.02) (1.769911504,1.026548673) (1.7725,1.03) (1.777777778,1.037037037) (1.78,1.04) (1.785714286,1.047619048) (1.7875,1.05) (1.793721973,1.058295964) (1.795,1.06) (1.801801802,1.069069069) (1.8025,1.07) (1.809954751,1.079939668) (1.81,1.08) (1.8175,1.09) (1.818181818,1.090909091) (1.825,1.1) (1.826484018,1.101978691) (1.8325,1.11) (1.834862385,1.113149847) (1.84,1.12) (1.843317972,1.124423963) (1.8475,1.13) (1.851851852,1.135802469) (1.855,1.14) (1.860465116,1.147286822) (1.8625,1.15) (1.869158879,1.158878505) (1.87,1.16) (1.8775,1.17) (1.877934272,1.17057903) (1.885,1.18) (1.886792453,1.182389937) (1.8925,1.19) (1.895734597,1.194312796) (1.9,1.2) (1.904761905,1.206349206) (1.9075,1.21) (1.913875598,1.218500797) (1.915,1.22) (1.9225,1.23) (1.923076923,1.230769231) (1.93,1.24) (1.93236715,1.2431562) (1.9375,1.25) (1.941747573,1.25566343) (1.945,1.26) (1.951219512,1.268292683) (1.9525,1.27) (1.96,1.28) (1.960784314,1.281045752) (1.9675,1.29) (1.97044335,1.293924466) (1.975,1.3) (1.98019802,1.306930693) (1.9825,1.31) (1.99,1.32) (1.990049751,1.320066335) (1.9975,1.33) (2,1.333333333) (2.005,1.333333333) (2.010050251,1.333333333) (2.0125,1.333333333) (2.02,1.333333333) (2.02020202,1.333333333) (2.0275,1.333333333) (2.030456853,1.333333333) (2.035,1.333333333) (2.040816327,1.333333333) (2.0425,1.333333333) (2.05,1.333333333) (2.051282051,1.333333333) (2.0575,1.333333333) (2.06185567,1.333333333) (2.065,1.333333333) (2.0725,1.333333333) (2.07253886,1.333333333) (2.08,1.333333333) (2.083333333,1.333333333) (2.0875,1.333333333) (2.094240838,1.333333333) (2.095,1.333333333) (2.1025,1.333333333) (2.105263158,1.333333333) (2.11,1.333333333) (2.116402116,1.333333333) (2.1175,1.333333333) (2.125,1.333333333) (2.127659574,1.333333333) (2.1325,1.333333333) (2.139037433,1.333333333) (2.14,1.333333333) (2.1475,1.333333333) (2.150537634,1.333333333) (2.155,1.333333333) (2.162162162,1.333333333) (2.1625,1.333333333) (2.17,1.333333333) (2.173913043,1.333333333) (2.1775,1.333333333) (2.185,1.333333333) (2.18579235,1.333333333) (2.1925,1.333333333) (2.197802198,1.333333333) (2.2,1.333333333) (2.2075,1.333333333) (2.209944751,1.333333333) (2.215,1.333333333) (2.222222222,1.333333333) (2.2225,1.33375) (2.23,1.345) (2.234636872,1.351955307) (2.2375,1.35625) (2.245,1.3675) (2.247191011,1.370786517) (2.2525,1.37875) (2.259887006,1.389830508) (2.26,1.39) (2.2675,1.40125) (2.272727273,1.409090909) (2.275,1.4125) (2.2825,1.42375) (2.285714286,1.428571429) (2.29,1.435) (2.2975,1.44625) (2.298850575,1.448275862) (2.305,1.4575) (2.312138728,1.468208092) (2.3125,1.46875) (2.32,1.48) (2.325581395,1.488372093) (2.3275,1.49125) (2.333333333,1.5) (2.335,1.5) (2.339181287,1.5) (2.3425,1.5) (2.35,1.5) (2.352941176,1.5) (2.3575,1.5) (2.365,1.5) (2.366863905,1.5) (2.3725,1.5) (2.38,1.5) (2.380952381,1.5) (2.3875,1.5) (2.395,1.5) (2.395209581,1.5) (2.4025,1.5) (2.409638554,1.5) (2.41,1.5) (2.4175,1.5) (2.424242424,1.5) (2.425,1.5) (2.4325,1.5) (2.4375,1.5) (2.43902439,1.502439024) (2.44,1.504) (2.4475,1.516) (2.45398773,1.526380368) (2.455,1.528) (2.4625,1.54) (2.469135802,1.550617284) (2.47,1.552) (2.4775,1.564) (2.48447205,1.57515528) (2.485,1.576) (2.4925,1.588) (2.5,1.6) (2.5075,1.6) (2.515,1.6) (2.51572327,1.6) (2.5225,1.6) (2.53,1.6) (2.53164557,1.6) (2.5375,1.6) (2.545,1.6) (2.547770701,1.6) (2.5525,1.6) (2.56,1.6) (2.564102564,1.606837607) (2.5675,1.6125) (2.575,1.625) (2.580645161,1.634408602) (2.5825,1.6375) (2.59,1.65) (2.597402597,1.662337662) (2.5975,1.6625) (2.6,1.666666667) (2.605,1.666666667) (2.6125,1.666666667) (2.614379085,1.666666667) (2.62,1.666666667) (2.6275,1.666666667) (2.631578947,1.666666667) (2.635,1.666666667) (2.638888889,1.666666667) (2.6425,1.672857143) (2.649006623,1.684011353) (2.65,1.685714286) (2.6575,1.698571429) (2.665,1.711428571) (2.666666667,1.714285714) (2.6725,1.714285714) (2.68,1.714285714) (2.684563758,1.714285714) (2.6875,1.714285714) (2.693877551,1.714285714) (2.695,1.71625) (2.7025,1.729375) (2.702702703,1.72972973) (2.71,1.7425) (2.714285715,1.75) (2.7175,1.75) (2.721088435,1.75) (2.725,1.75) (2.7325,1.75) (2.734375,1.75) (2.739726027,1.759512938) (2.74,1.76) (2.7475,1.773333333) (2.75,1.777777778) (2.755,1.777777778) (2.75862069,1.777777778) (2.7625,1.777777778) (2.765432098,1.777777778) (2.77,1.786) (2.7775,1.7995) (2.777777778,1.8) (2.785,1.8) (2.79,1.8) (2.7925,1.804545455) (2.797202797,1.813095995) (2.8,1.818181818) (2.8,1.818181818) (2.8075,1.818181818) (2.809917356,1.818181819) (2.815,1.8275) (2.816901408,1.830985915) (2.818181818,1.833333333) (2.8225,1.833333333) (2.826388889,1.833333333) (2.83,1.84) (2.833333333,1.846153846) (2.836879433,1.846153846) (2.8375,1.846153846) (2.840236687,1.846153847) (2.845,1.855) (2.846153846,1.857142856) (2.852040816,1.857142857) (2.8525,1.858) (2.857142857,1.866666667) (2.86,1.866666667) (2.862222222,1.866666667) (2.866666667,1.875) (2.8675,1.875) (2.87109375,1.875) (2.875,1.88235294) (2.875,1.882352941) (2.877697842,1.882352941) (2.878892734,1.882352942) (2.882352941,1.888888889) (2.8825,1.888888889) (2.885802469,1.888888889) (2.888888889,1.894736841) (2.89,1.894736842) (2.891966759,1.894736842) (2.894736842,1.899999999) (2.8975,1.9) (2.8975,1.9) (2.898550725,1.90200138) (2.9,1.904761905) (2.902494331,1.904761905) (2.904761905,1.909090909) (2.905,1.909090909) (2.907024794,1.90909091) (2.909090909,1.913043478) (2.911153119,1.913043478) (2.9125,1.915625) (2.913043478,1.916666666) (2.914930556,1.916666667) (2.916666666,1.919999999) (2.9184,1.92) (2.919708029,1.922515441) (2.92,1.923076923) (2.92,1.923076923) (2.921597633,1.923076923) (2.923076923,1.925925925) (2.924554184,1.925925926) (2.925925926,1.928571429) (2.927295919,1.928571429) (2.9275,1.928965517) (2.928571429,1.931034483) (2.929845422,1.931034483) (2.931034483,1.933333333) (2.932222222,1.933333333) (2.933333333,1.935483871) (2.934443289,1.935483872) (2.935,1.9365625) (2.935483871,1.937499999) (2.936523438,1.937500001) (2.9375,1.939393939) (2.938475666,1.939393939) (2.939393939,1.94117647) (2.940311419,1.941176471) (2.941176471,1.942857143) (2.942040816,1.942857143) (2.9425,1.94375) (2.942857143,1.944444444) (2.943672839,1.944444444) (2.944444445,1.945945946) (2.945215485,1.945945946) (2.945945946,1.947368421) (2.9466759,1.947368421) (2.947368421,1.948717948) (2.948060487,1.948717949) (2.948717949,1.95) (2.949375,1.95) (2.95,1.951219512) (2.95,1.951219512) (2.950624628,1.951219513) (2.951219512,1.952380952) (2.951814059,1.952380953) (2.952380953,1.953488372) (2.95294754,1.953488373) (2.953488372,1.954545454) (2.954028926,1.954545455) (2.954545454,1.955555555) (2.955061729,1.955555556) (2.955555555,1.956521738) (2.956049149,1.956521739) (2.956521739,1.957446809) (2.956994115,1.957446809) (2.957446809,1.958333333) (2.9575,1.958333333) (2.957899306,1.958333334) (2.958333333,1.959183673) (2.95876718,1.959183674) (2.959183673,1.96) (2.9596,1.96) (2.96,1.960784314) (2.960399846,1.960784314) (2.960784313,1.961538461) (2.961168639,1.961538462) (2.961538461,1.96226415) (2.961908152,1.962264151) (2.962264151,1.962962963) (2.962620028,1.962962963) (2.962962963,1.963636364) (2.963305785,1.963636364) (2.963636363,1.964285714) (2.963966837,1.964285715) (2.964285714,1.964912281) (2.964604494,1.964912281) (2.964912281,1.965517241) (2.965,1.965517241) (2.965219976,1.965517241) (2.965517242,1.966101695) (2.965814421,1.966101695) (2.966101695,1.966666667) (2.966388889,1.966666667) (2.966666667,1.967213115) (2.9725,1.97277027) (2.98,1.98019802) (2.985074627,1.985185185) (2.9875,1.98757764) (2.995,1.995012469) (3,2) (3.0025,2.002503129) (3.007518797,2.00754717) (3.01,2.010050251) (3.0175,2.017610619) (3.025,2.025316456) (3.03030303,2.030769231) (3.032258065,2.032786885) (3.0325,2.032786885) (3.032518141,2.032786886) (3.032786886,2.033333333) (3.033055556,2.033333334) (3.033333333,2.033898305) (3.033611031,2.033898305) (3.033898306,2.034482759) (3.034185493,2.034482759) (3.034482759,2.035087719) (3.034779932,2.035087719) (3.035087719,2.035714285) (3.035395408,2.035714286) (3.035714286,2.036363636) (3.036033058,2.036363636) (3.036363636,2.037037036) (3.036694102,2.037037038) (3.037037037,2.037735849) (3.03737985,2.037735849) (3.037735849,2.038461538) (3.038091716,2.038461539) (3.038461538,2.039215685) (3.038831219,2.039215687) (3.039215686,2.039999999) (3.0396,2.04) (3.04,2.040816327) (3.04,2.040816327) (3.040399833,2.040816327) (3.040816327,2.041666667) (3.041232639,2.041666667) (3.041666667,2.042553191) (3.042100498,2.042553191) (3.042553191,2.04347826) (3.043005671,2.043478261) (3.043478261,2.044444444) (3.043950617,2.044444444) (3.044444444,2.045454545) (3.044938016,2.045454545) (3.045454545,2.046511627) (3.045970795,2.046511628) (3.046511628,2.047619048) (3.047052154,2.047619048) (3.0475,2.048536585) (3.047619048,2.048780488) (3.048185604,2.048780488) (3.048780488,2.05) (3.049375,2.05) (3.05,2.05128205) (3.050624589,2.051282052) (3.051282052,2.052631579) (3.051939059,2.05263158) (3.052631579,2.054054054) (3.053323594,2.054054054) (3.053435115,2.054283291) (3.054054054,2.055555556) (3.05478395,2.055555556) (3.055,2.056) (3.055555555,2.057142857) (3.056326531,2.057142858) (3.057142858,2.058823529) (3.057958477,2.058823529) (3.05882353,2.060606061) (3.059687787,2.060606061) (3.06060606,2.062499999) (3.061523438,2.062500001) (3.0625,2.064516129) (3.0625,2.064516129) (3.063475547,2.06451613) (3.064516129,2.066666666) (3.065555555,2.066666667) (3.066666667,2.068965517) (3.067776457,2.068965518) (3.068965518,2.071428571) (3.07,2.071428571) (3.070153062,2.071428572) (3.071428572,2.074074074) (3.072702332,2.074074074) (3.074074074,2.076923077) (3.075443787,2.076923077) (3.076923077,2.08) (3.0775,2.08) (3.0784,2.080000001) (3.08,2.083333333) (3.081597222,2.083333333) (3.083333334,2.086956522) (3.085,2.086956522) (3.085066162,2.086956522) (3.086956521,2.09090909) (3.088842975,2.090909091) (3.090909091,2.095238095) (3.0925,2.095238095) (3.092970521,2.095238095) (3.095238095,2.1) (3.0975,2.1) (3.1,2.105263157) (3.1,2.105263158) (3.100775194,2.105263158) (3.102493075,2.105263158) (3.105263158,2.111111111) (3.1075,2.111111111) (3.108024692,2.111111112) (3.111111111,2.117647059) (3.114186851,2.117647059) (3.115,2.119375) (3.117647059,2.125) (3.12109375,2.125000001) (3.1225,2.128) (3.125,2.133333333) (3.128888889,2.133333334) (3.13,2.135714286) (3.133333333,2.142857142) (3.1375,2.142857143) (3.137755102,2.142857143) (3.142857143,2.153846153) (3.145,2.153846154) (3.147928994,2.153846155) (3.149606299,2.157480315) (3.1525,2.16375) (3.153846154,2.166666667) (3.159722222,2.166666667) (3.16,2.167272727) (3.166666667,2.181818181) (3.1675,2.181818182) (3.173553719,2.181818182) (3.174603175,2.184126984) (3.175,2.185) (3.181818182,2.2) (3.1825,2.2) (3.19,2.2) (3.19,2.200000001) (3.1975,2.216666667) (3.2,2.222222222) (3.205,2.222222222) (3.209876544,2.222222223) (3.2125,2.228125) (3.22,2.245) (3.222222222,2.25) (3.225806452,2.25) (3.2275,2.25) (3.234375,2.25) (3.235,2.251428571) (3.2425,2.268571429) (3.25,2.285714286) (3.25,2.285714286) (3.25203252,2.285714286) (3.2575,2.285714286) (3.265,2.285714286) (3.265306122,2.285714286) (3.2725,2.3025) (3.278688525,2.316939891) (3.28,2.32) (3.285714286,2.333333333) (3.2875,2.333333333) (3.295,2.333333333) (3.3025,2.333333333) (3.305555556,2.333333333) (3.305785124,2.333884298) (3.31,2.344) (3.3175,2.362) (3.325,2.38) (3.3325,2.398) (3.333333333,2.4) (3.34,2.4) (3.3475,2.4) (3.355,2.4) (3.36,2.400000001) (3.361344538,2.403361345) (3.3625,2.40625) (3.37,2.425) (3.3775,2.44375) (3.385,2.4625) (3.389830508,2.474576271) (3.3925,2.48125) (3.4,2.499999999) (3.4,2.5) (3.4075,2.5) (3.415,2.5) (3.418803419,2.5) (3.4225,2.5) (3.43,2.5) (3.4375,2.5) (3.4375,2.5) (3.445,2.52) (3.448275862,2.528735632) (3.4525,2.54) (3.46,2.56) (3.4675,2.58) (3.475,2.6) (3.47826087,2.608695652) (3.4825,2.62) (3.49,2.64) (3.4975,2.66) (3.5,2.666666667) (3.505,2.666666667) (3.50877193,2.666666667) (3.5125,2.666666667) (3.52,2.666666667) (3.5275,2.666666667) (3.535,2.666666667) (3.539823009,2.666666667) (3.5425,2.666666667) (3.55,2.666666667) (3.555555556,2.666666667) (3.5575,2.6725) (3.565,2.695) (3.571428571,2.714285714) (3.5725,2.7175) (3.58,2.74) (3.5875,2.7625) (3.595,2.785) (3.6025,2.8075) (3.603603604,2.810810811) (3.61,2.83) (3.6175,2.8525) (3.625,2.875) (3.6325,2.8975) (3.636363636,2.909090909) (3.64,2.92) (3.6475,2.9425) (3.655,2.965) (3.6625,2.9875) (3.666666667,3) (3.669724771,3) (3.67,3) (3.6775,3) (3.685,3) (3.6925,3) (3.7,3) (3.703703704,3) (3.7075,3) (3.715,3) (3.7225,3) (3.73,3) (3.7375,3) (3.738317757,3) (3.745,3) (3.75,3) (3.7525,3.01) (3.76,3.04) (3.7675,3.07) (3.773584906,3.094339623) (3.775,3.1) (3.7825,3.13) (3.79,3.16) (3.7975,3.19) (3.805,3.22) (3.80952381,3.238095238) (3.8125,3.25) (3.82,3.28) (3.8275,3.31) (3.835,3.34) (3.8425,3.37) (3.846153846,3.384615385) (3.85,3.4) (3.8575,3.43) (3.865,3.46) (3.8725,3.49) (3.88,3.52) (3.883495146,3.533980583) (3.8875,3.55) (3.895,3.58) (3.9025,3.61) (3.91,3.64) (3.9175,3.67) (3.921568627,3.68627451) (3.925,3.7) (3.9325,3.73) (3.94,3.76) (3.9475,3.79) (3.955,3.82) (3.96039604,3.841584158) (3.9625,3.85) (3.97,3.88) (3.9775,3.91) (3.985,3.94) (3.9925,3.97) (4,4)};
\addlegendentry{capacity}
\draw[black!45,densely dotted,line width=.7pt] (axis cs:3,.75) -- (axis cs:3,4.12);
\addplot[only marks,mark=*,capblue,mark size=2pt,forget plot] coordinates {(3,2)};
\node[font=\small,anchor=north west,black!70] at (axis cs:3.03,1.97) {$(3,2)$};
\end{axis}
\end{tikzpicture}
\caption{Tame type: four users. Infinitely many horizontal and rising line segments approach $(\mu/\nu,C/\nu)=(3,2)$ (dot). The dotted vertical line marks $\mu/\nu=3$.}
\label{fig:tame}
\end{figure}

For $K=4$, the balanced fractions are infinitely many isolated points accumulating at $\nu/\kappa=1/2$, so the split regimes give infinitely many linear pieces. Their horizontal and rising segments become shorter as they approach $(\mu/\nu,C/\nu)=(3,2)$; see Figure~\ref{fig:tame}. Finitely many split regimes no longer cover all data-to-demand ratios, but a single recursion generates the entire sequence.

\subsection{Wild type: a curved part with linear pieces outside}
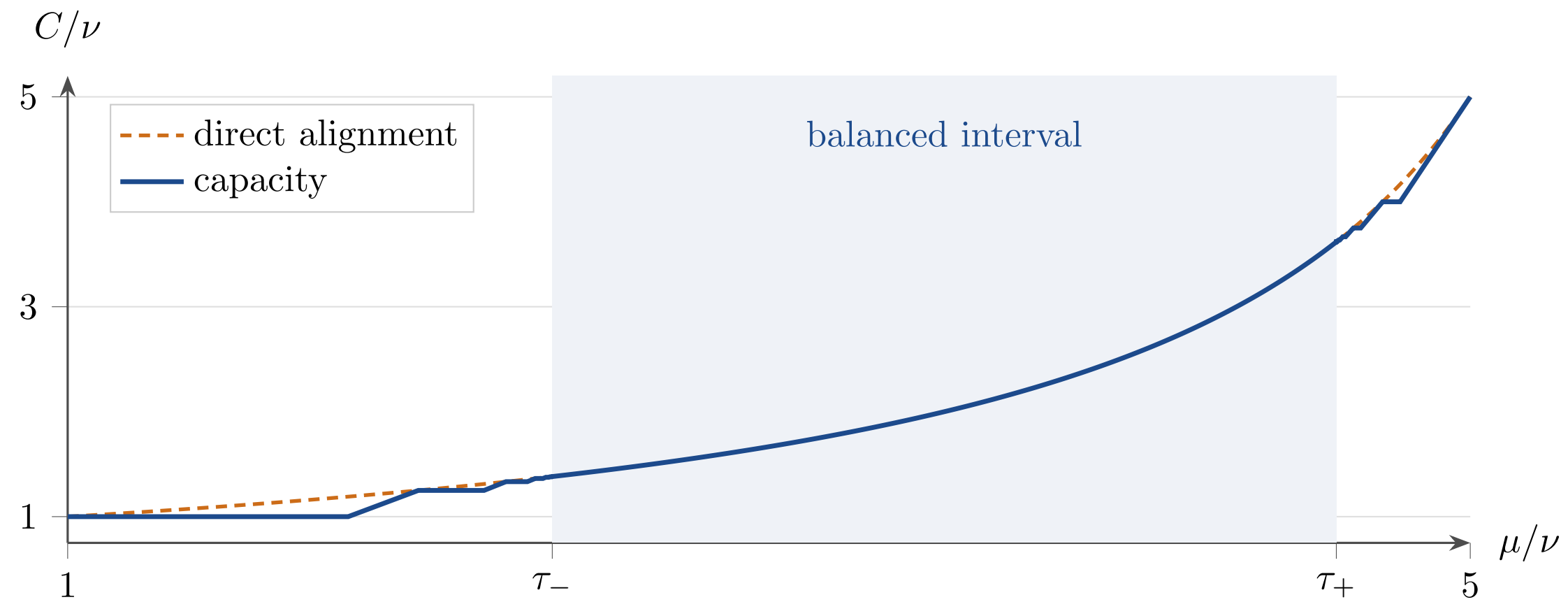
\begin{figure}[H]
\centering
\begin{tikzpicture}
\begin{axis}[width=.9\textwidth,height=6cm,capplot,xlabel={$\mu/\nu$},ylabel={$C/\nu$},
 xmin=1,xmax=5,ymin=.75,ymax=5.2,
 xtick={1,2.381966011,4.618033989,5},xticklabels={$1$,$\tau_-$,$\tau_+$,$5$},ytick={1,3,5}]
\path[fill=capblue!7] (axis cs:2.381966011,.75) rectangle (axis cs:4.618033989,5.2);
\addplot[alignline,domain=1:5,samples=250] {5/(6-x)};
\addlegendentry{direct alignment}
\addplot[capline] coordinates {(1,1) (1.002673797,1) (1.00536193,1) (1.008064516,1) (1.01,1) (1.010781671,1) (1.013513513,1) (1.016260163,1) (1.019021739,1) (1.02,1) (1.021798365,1) (1.024590164,1) (1.02739726,1) (1.03,1) (1.03021978,1) (1.033057851,1) (1.035911602,1) (1.038781163,1) (1.04,1) (1.041666667,1) (1.044568245,1) (1.047486033,1) (1.05,1) (1.050420168,1) (1.053370787,1) (1.056338028,1) (1.059322034,1) (1.06,1) (1.062322946,1) (1.065340909,1) (1.068376068,1) (1.07,1) (1.071428571,1) (1.074498567,1) (1.077586207,1) (1.08,1) (1.080691643,1) (1.083815029,1) (1.086956522,1) (1.09,1) (1.090116279,1) (1.093294461,1) (1.096491228,1) (1.099706745,1) (1.1,1) (1.102941176,1) (1.10619469,1) (1.109467456,1) (1.11,1) (1.112759644,1) (1.116071429,1) (1.119402985,1) (1.12,1) (1.122754491,1) (1.126126126,1) (1.129518072,1) (1.13,1) (1.132930514,1) (1.136363636,1) (1.139817629,1) (1.14,1) (1.143292683,1) (1.146788991,1) (1.15,1) (1.150306749,1) (1.153846154,1) (1.157407407,1) (1.16,1) (1.160990712,1) (1.164596273,1) (1.168224299,1) (1.17,1) (1.171875,1) (1.175548589,1) (1.179245283,1) (1.18,1) (1.1829653,1) (1.186708861,1) (1.19,1) (1.19047619,1) (1.194267516,1) (1.198083067,1) (1.2,1) (1.201923077,1) (1.205787781,1) (1.209677419,1) (1.21,1) (1.213592233,1) (1.217532468,1) (1.22,1) (1.221498371,1) (1.225490196,1) (1.229508197,1) (1.23,1) (1.233552632,1) (1.237623762,1) (1.24,1) (1.241721854,1) (1.245847176,1) (1.25,1) (1.254180602,1) (1.258389262,1) (1.26,1) (1.262626263,1) (1.266891892,1) (1.27,1) (1.271186441,1) (1.275510204,1) (1.279863481,1) (1.28,1) (1.284246575,1) (1.288659794,1) (1.29,1) (1.293103448,1) (1.297577855,1) (1.3,1) (1.302083333,1) (1.306620209,1) (1.31,1) (1.311188811,1) (1.315789474,1) (1.32,1) (1.320422535,1) (1.325088339,1) (1.329787234,1) (1.33,1) (1.334519573,1) (1.339285714,1) (1.34,1) (1.344086022,1) (1.348920863,1) (1.35,1) (1.353790614,1) (1.358695652,1) (1.36,1) (1.363636364,1) (1.368613139,1) (1.37,1) (1.373626374,1) (1.378676471,1) (1.38,1) (1.383763838,1) (1.388888889,1) (1.39,1) (1.394052045,1) (1.399253731,1) (1.4,1) (1.404494382,1) (1.409774436,1) (1.41,1) (1.41509434,1) (1.42,1) (1.420454545,1) (1.425855513,1) (1.43,1) (1.43129771,1) (1.436781609,1) (1.44,1) (1.442307692,1) (1.447876448,1) (1.45,1) (1.453488372,1) (1.459143969,1) (1.46,1) (1.46484375,1) (1.47,1) (1.470588235,1) (1.476377953,1) (1.48,1) (1.482213439,1) (1.488095238,1) (1.49,1) (1.494023904,1) (1.5,1) (1.5,1) (1.506024096,1) (1.51,1) (1.512096774,1) (1.518218623,1) (1.52,1) (1.524390244,1) (1.53,1) (1.530612245,1) (1.536885246,1) (1.54,1) (1.543209877,1) (1.549586777,1) (1.55,1) (1.556016597,1) (1.56,1) (1.5625,1) (1.569037657,1) (1.57,1) (1.575630252,1) (1.58,1) (1.582278481,1) (1.588983051,1) (1.59,1) (1.595744681,1) (1.6,1) (1.602564103,1) (1.60944206,1) (1.61,1) (1.61637931,1) (1.62,1) (1.623376623,1) (1.63,1) (1.630434783,1) (1.637554585,1) (1.64,1) (1.644736842,1) (1.65,1) (1.651982379,1) (1.659292035,1) (1.66,1) (1.666666667,1) (1.67,1) (1.674107143,1) (1.68,1) (1.68161435,1) (1.689189189,1) (1.69,1) (1.696832579,1) (1.7,1) (1.704545454,1) (1.71,1) (1.712328767,1) (1.72,1) (1.720183486,1) (1.728110599,1) (1.73,1) (1.736111111,1) (1.74,1) (1.744186047,1) (1.75,1) (1.752336448,1) (1.76,1) (1.76056338,1) (1.768867925,1) (1.77,1) (1.777251185,1) (1.78,1) (1.785714286,1) (1.79,1) (1.794258373,1) (1.8,1) (1.8,1) (1.802884615,1.003605769) (1.81,1.0125) (1.811594203,1.014492754) (1.82,1.025) (1.82038835,1.025485437) (1.829268293,1.036585366) (1.83,1.0375) (1.838235294,1.047794118) (1.84,1.05) (1.847290641,1.059113301) (1.85,1.0625) (1.856435643,1.070544554) (1.86,1.075) (1.865671642,1.082089552) (1.87,1.0875) (1.875,1.09375) (1.88,1.1) (1.88442211,1.105527638) (1.89,1.1125) (1.893939394,1.117424242) (1.9,1.125) (1.9035533,1.129441625) (1.91,1.1375) (1.913265306,1.141581633) (1.92,1.15) (1.923076923,1.153846154) (1.93,1.1625) (1.932989691,1.166237114) (1.94,1.175) (1.943005181,1.178756477) (1.95,1.1875) (1.953125,1.19140625) (1.96,1.2) (1.963350785,1.204188482) (1.97,1.2125) (1.97368421,1.217105263) (1.98,1.225) (1.984126984,1.23015873) (1.99,1.2375) (1.994680851,1.243351064) (2,1.25) (2.005347593,1.25) (2.01,1.25) (2.016129032,1.25) (2.02,1.25) (2.027027027,1.25) (2.03,1.25) (2.038043478,1.25) (2.04,1.25) (2.049180328,1.25) (2.05,1.25) (2.06,1.25) (2.060439561,1.25) (2.07,1.25) (2.071823204,1.25) (2.08,1.25) (2.083333333,1.25) (2.09,1.25) (2.094972067,1.25) (2.1,1.25) (2.106741573,1.25) (2.11,1.25) (2.118644068,1.25) (2.12,1.25) (2.13,1.25) (2.130681818,1.25) (2.14,1.25) (2.142857143,1.25) (2.15,1.25) (2.155172414,1.25) (2.16,1.25) (2.167630058,1.25) (2.17,1.25) (2.18,1.25) (2.180232558,1.25) (2.1875,1.25) (2.19,1.253333333) (2.192982456,1.257309942) (2.2,1.266666667) (2.205882353,1.274509804) (2.21,1.28) (2.218934911,1.291913215) (2.22,1.293333333) (2.23,1.306666667) (2.232142857,1.30952381) (2.24,1.32) (2.245508982,1.32734531) (2.25,1.333333333) (2.25,1.333333333) (2.259036144,1.333333333) (2.26,1.333333333) (2.27,1.333333333) (2.272727273,1.333333333) (2.28,1.333333333) (2.286585366,1.333333333) (2.29,1.333333333) (2.3,1.333333333) (2.300613497,1.333333333) (2.31,1.333333333) (2.311111111,1.333333333) (2.314814815,1.338383838) (2.32,1.345454545) (2.329192547,1.357989836) (2.33,1.359090909) (2.333333333,1.363636363) (2.34,1.363636364) (2.34375,1.363636364) (2.35,1.363636364) (2.355371901,1.363636364) (2.358490566,1.367924528) (2.36,1.37) (2.363636364,1.375) (2.37,1.375) (2.371875,1.375) (2.373417722,1.377127892) (2.375,1.379310345) (2.378121284,1.379310345) (2.379310345,1.380952381) (2.38,1.380952381) (2.380498866,1.380952381) (2.380952381,1.381578947) (2.381405817,1.381578947) (2.381578947,1.381818182) (2.381752066,1.381818182) (2.381818182,1.381909548) (2.381884296,1.381909548) (2.381909548,1.381944444) (2.381934799,1.381944444) (2.381944444,1.381957773) (2.38195409,1.381957774) (2.381957774,1.381962865) (2.381961458,1.381962865) (2.381962865,1.381964809) (2.381964272,1.381964809) (2.38196481,1.381965552) (2.381965347,1.381965552) (2.381965552,1.381965836) (2.381965758,1.381965836) (2.381965836,1.381965944) (2.381965914,1.381965944) (2.381965944,1.381965986) (2.381965974,1.381965986) (2.381965986,1.381966001) (2.381965997,1.381966001) (2.381966002,1.381966008) (2.381966006,1.381966008) (2.381966007,1.38196601) (2.381966009,1.38196601) (2.38196601,1.38196601) (2.381966011,1.381966011) (2.381966011,1.381966011) (2.388535032,1.384479718) (2.39,1.385041551) (2.4,1.388888889) (2.403846154,1.390374332) (2.41,1.39275766) (2.419354839,1.396396396) (2.42,1.396648045) (2.43,1.400560224) (2.435064935,1.402550091) (2.44,1.404494382) (2.45,1.408450704) (2.450980392,1.408839779) (2.46,1.412429379) (2.467105263,1.415270019) (2.47,1.416430595) (2.48,1.420454545) (2.483443708,1.421845574) (2.49,1.424501425) (2.5,1.428571429) (2.51,1.432664756) (2.516778524,1.435452794) (2.52,1.436781609) (2.53,1.44092219) (2.533783784,1.442495127) (2.54,1.445086705) (2.55,1.449275362) (2.551020408,1.449704142) (2.56,1.453488372) (2.568493151,1.457085828) (2.57,1.457725948) (2.58,1.461988304) (2.586206896,1.464646465) (2.59,1.46627566) (2.6,1.470588235) (2.604166667,1.472392638) (2.61,1.474926254) (2.62,1.479289941) (2.622377623,1.480331263) (2.63,1.483679525) (2.64,1.488095238) (2.64084507,1.488469602) (2.65,1.492537313) (2.659574468,1.496815287) (2.66,1.497005988) (2.67,1.501501502) (2.678571429,1.505376344) (2.68,1.506024096) (2.69,1.510574018) (2.697841726,1.51416122) (2.7,1.515151515) (2.71,1.519756839) (2.717391304,1.523178808) (2.72,1.524390244) (2.73,1.529051988) (2.737226278,1.532438479) (2.74,1.533742331) (2.75,1.538461538) (2.757352941,1.541950113) (2.76,1.543209877) (2.77,1.547987616) (2.777777778,1.551724138) (2.78,1.552795031) (2.79,1.557632399) (2.798507463,1.561771562) (2.8,1.5625) (2.81,1.567398119) (2.819548872,1.572104019) (2.82,1.572327044) (2.83,1.577287066) (2.84,1.582278481) (2.840909091,1.582733813) (2.85,1.587301587) (2.86,1.592356688) (2.86259542,1.593673966) (2.87,1.597444089) (2.88,1.602564103) (2.884615384,1.604938271) (2.89,1.607717042) (2.9,1.612903226) (2.906976744,1.616541353) (2.91,1.618122977) (2.92,1.623376623) (2.9296875,1.628498728) (2.93,1.628664495) (2.94,1.633986928) (2.95,1.639344262) (2.952755905,1.640826873) (2.96,1.644736842) (2.97,1.650165017) (2.976190476,1.653543307) (2.98,1.655629139) (2.99,1.661129568) (3,1.666666667) (3,1.666666667) (3.01,1.672240803) (3.02,1.677852349) (3.024193548,1.680216802) (3.03,1.683501684) (3.04,1.689189189) (3.048780488,1.694214876) (3.05,1.694915254) (3.06,1.700680272) (3.07,1.706484642) (3.073770492,1.708683474) (3.08,1.712328767) (3.09,1.718213058) (3.099173553,1.723646723) (3.1,1.724137931) (3.11,1.730103806) (3.12,1.736111111) (3.125,1.739130435) (3.13,1.742160279) (3.14,1.748251748) (3.15,1.754385965) (3.151260505,1.755162242) (3.16,1.76056338) (3.17,1.766784452) (3.177966101,1.771771772) (3.18,1.773049645) (3.19,1.779359431) (3.2,1.785714286) (3.205128205,1.788990826) (3.21,1.792114695) (3.22,1.798561151) (3.23,1.805054152) (3.232758621,1.806853583) (3.24,1.811594203) (3.25,1.818181818) (3.26,1.824817518) (3.260869565,1.825396825) (3.27,1.831501832) (3.28,1.838235294) (3.289473684,1.844660194) (3.29,1.84501845) (3.3,1.851851852) (3.31,1.858736059) (3.318584071,1.864686469) (3.32,1.865671642) (3.33,1.872659176) (3.34,1.879699248) (3.348214285,1.885521885) (3.35,1.886792453) (3.36,1.893939394) (3.37,1.901140684) (3.378378378,1.907216495) (3.38,1.908396947) (3.39,1.915708812) (3.4,1.923076923) (3.409090909,1.929824562) (3.41,1.930501931) (3.42,1.937984496) (3.43,1.945525292) (3.44,1.953125) (3.440366972,1.953405018) (3.45,1.960784314) (3.46,1.968503937) (3.47,1.976284585) (3.472222222,1.978021978) (3.48,1.984126984) (3.49,1.992031873) (3.5,2) (3.504672898,2.003745319) (3.51,2.008032129) (3.52,2.016129032) (3.53,2.024291498) (3.537735849,2.030651341) (3.54,2.032520325) (3.55,2.040816327) (3.56,2.049180328) (3.57,2.057613169) (3.571428571,2.058823529) (3.58,2.066115702) (3.59,2.074688797) (3.6,2.083333333) (3.605769231,2.088353414) (3.61,2.092050209) (3.62,2.100840336) (3.63,2.109704641) (3.64,2.118644068) (3.640776699,2.119341563) (3.65,2.127659574) (3.66,2.136752137) (3.67,2.145922747) (3.676470588,2.151898734) (3.68,2.155172414) (3.69,2.164502165) (3.7,2.173913043) (3.71,2.183406114) (3.712871288,2.186147187) (3.72,2.192982456) (3.73,2.202643172) (3.74,2.212389381) (3.75,2.222222222) (3.75,2.222222222) (3.76,2.232142857) (3.77,2.242152466) (3.78,2.252252252) (3.787878788,2.260273973) (3.79,2.262443439) (3.8,2.272727273) (3.81,2.283105023) (3.82,2.293577982) (3.826530613,2.300469484) (3.83,2.304147465) (3.84,2.314814815) (3.85,2.325581395) (3.86,2.336448598) (3.865979381,2.342995169) (3.87,2.34741784) (3.88,2.358490566) (3.89,2.369668246) (3.9,2.380952381) (3.90625,2.388059701) (3.91,2.392344498) (3.92,2.403846154) (3.93,2.415458937) (3.94,2.427184466) (3.947368422,2.435897437) (3.95,2.43902439) (3.96,2.450980392) (3.97,2.463054187) (3.98,2.475247525) (3.989361702,2.486772486) (3.99,2.487562189) (4,2.5) (4.01,2.512562814) (4.02,2.525252525) (4.03,2.538071066) (4.032258065,2.540983607) (4.04,2.551020408) (4.05,2.564102564) (4.06,2.577319588) (4.07,2.590673575) (4.076086957,2.598870057) (4.08,2.604166667) (4.09,2.617801047) (4.1,2.631578947) (4.11,2.645502646) (4.12,2.659574468) (4.12087912,2.660818713) (4.13,2.673796791) (4.14,2.688172043) (4.15,2.702702703) (4.16,2.717391304) (4.166666667,2.727272727) (4.17,2.732240437) (4.18,2.747252747) (4.19,2.762430939) (4.2,2.777777778) (4.21,2.793296089) (4.213483147,2.798742139) (4.22,2.808988764) (4.23,2.824858757) (4.24,2.840909091) (4.25,2.857142857) (4.26,2.873563218) (4.261363636,2.875816992) (4.27,2.89017341) (4.28,2.906976744) (4.29,2.923976608) (4.3,2.941176471) (4.31,2.958579882) (4.310344828,2.959183673) (4.32,2.976190476) (4.33,2.994011976) (4.34,3.012048193) (4.35,3.03030303) (4.36,3.048780488) (4.360465117,3.049645391) (4.37,3.067484663) (4.38,3.086419753) (4.39,3.105590062) (4.4,3.125) (4.41,3.144654088) (4.411764705,3.148148147) (4.42,3.164556962) (4.43,3.184713376) (4.44,3.205128205) (4.45,3.225806452) (4.46,3.246753247) (4.464285714,3.255813953) (4.47,3.267973856) (4.48,3.289473684) (4.49,3.311258278) (4.5,3.333333333) (4.51,3.355704698) (4.51807229,3.373983741) (4.52,3.378378378) (4.53,3.401360544) (4.54,3.424657534) (4.55,3.448275862) (4.56,3.472222222) (4.57,3.496503497) (4.573170731,3.504273503) (4.58,3.521126761) (4.59,3.546099291) (4.6,3.571428571) (4.61,3.597122302) (4.618033988,3.618033986) (4.61803399,3.618033992) (4.618033992,3.618033997) (4.618033994,3.618033999) (4.618033999,3.618034014) (4.618034003,3.618034014) (4.618034013,3.618034053) (4.618034026,3.618034058) (4.618034056,3.618034164) (4.618034086,3.618034165) (4.618034165,3.618034448) (4.618034242,3.618034448) (4.618034449,3.618035191) (4.618034653,3.618035191) (4.618035191,3.618037135) (4.618035728,3.618037135) (4.618037136,3.618042226) (4.618038543,3.618042229) (4.618042226,3.618055555) (4.618045911,3.618055559) (4.618055555,3.618090451) (4.618065201,3.618090455) (4.618090452,3.618181817) (4.618115703,3.61818182) (4.618181818,3.61842105) (4.618247923,3.618421056) (4.618421053,3.619047619) (4.618594104,3.619047619) (4.619047618,3.620689652) (4.619500594,3.620689655) (4.62,3.6225) (4.620689656,3.625) (4.621875001,3.625000002) (4.625,3.636363636) (4.628099173,3.636363636) (4.62962963,3.641975309) (4.63,3.643333333) (4.636363637,3.666666667) (4.64,3.666666667) (4.644444444,3.666666667) (4.65,3.6875) (4.66,3.725) (4.666666666,3.749999999) (4.67,3.75) (4.68,3.75) (4.687500001,3.750000003) (4.69,3.76) (4.7,3.8) (4.71,3.84) (4.72,3.88) (4.73,3.92) (4.74,3.96) (4.746835442,3.987341769) (4.75,3.999999999) (4.75,4) (4.76,4) (4.77,4) (4.78,4) (4.79,4) (4.8,4) (4.800000001,4.000000004) (4.807692308,4.038461538) (4.81,4.05) (4.82,4.1) (4.83,4.15) (4.84,4.2) (4.85,4.25) (4.86,4.3) (4.87,4.35) (4.870129871,4.350649355) (4.88,4.4) (4.89,4.45) (4.9,4.5) (4.91,4.55) (4.92,4.6) (4.93,4.65) (4.934210526,4.671052628) (4.94,4.7) (4.95,4.75) (4.96,4.8) (4.97,4.85) (4.98,4.9) (4.99,4.95) (5,5)};
\addlegendentry{capacity}
\node[font=\small,capblue] at (axis cs:3.5,4.65) {balanced interval};
\end{axis}
\end{tikzpicture}
\caption{Wild type: five users. $C/\nu$ equals $R_{\mathrm{align}}/\nu$ on the shaded balanced interval $[\tau_-,\tau_+]$. Outside it, horizontal segments alternate with rising line segments, becoming shorter as they approach the two endpoints.}
\label{fig:wild}
\end{figure}

For $K\ge5$, the single ratio approached by the linear pieces is replaced by an interval $\tau_-\le\tau\le\tau_+$. Here $\tau_-$ and $\tau_+$, given in Theorem~\ref{thm:main}, are the lower and upper data-to-demand ratios where the curved part begins and ends. On this whole interval the generic configuration is balanced, and direct alignment is optimal. We call it the \emph{balanced interval}. Outside it, split regimes again give infinitely many linear pieces, becoming shorter as they approach its two endpoints. Figure~\ref{fig:wild} illustrates five users. Although the configurations themselves are wild to classify, the generic capacity has a closed form, since it depends only on the regime and the component dimensions.

\subsection{Converse and connections}\label{sec:techniques}
\emph{Converse.} The converse bounds the sum of the users' remaining observation entropies in terms of the remaining data entropy. The same balanced and split configurations that guide the scheme give the required entropy bounds (Section~\ref{sec:geo-converse}). Communication limits how much observation entropy can be removed, converting these bounds into lower bounds on the sum-rate (Section~\ref{sec:conv}).

\emph{Connections.} Piecewise linear and curved regimes also arise in interference alignment, through subspace alignment chains~\cite{WangGouJafar,BreslerCartwrightTse} and the counting-versus-decomposition picture of genie chains~\cite{GenieChains}. Section~\ref{sec:connections} discusses this connection and shows that the same geometry gives the sum-capacity of a many-to-one interference channel.

\section{Problem Statement}\label{sec:model}

\subsection{Codes and capacity}
User $k\in[K]=\{1,\ldots,K\}$ observes $\bz_k=\mathbf H_k\bw$, and the server wants $\by=\mathbf D\bw$, where the data are $\bw\in\Fq^\mu$ and $\mathbf H_k,\mathbf D\in\Fq^{\nu\times\mu}$. Throughout, vectors are columns, and a tuple of vectors or scalars, such as $(\bz_1,\ldots,\bz_K)$, denotes the column vector obtained by stacking them. Assume $\mathbf H_1,\ldots,\mathbf H_K,\mathbf D$ have full row rank and $[\mathbf H_1^\top\ \cdots\ \mathbf H_K^\top]_{\mu\times K\nu}$ has rank $\mu$, so $1\le\nu\le\mu\le K\nu$. The matrices are known to everyone and fixed across instances. Symmetry refers to the common observation and demand dimension $\nu$. Logarithms and entropies are in $q$-ary units.

A code of block length $L$ operates on $\bw^L=(\bw_1,\ldots,\bw_L)\in\Fq^{\mu L}$. Write $\mathbf H_k^{[L]}=\operatorname{diag}(\mathbf H_k,\ldots,\mathbf H_k)$ with $L$ blocks, and define $\mathbf D^{[L]}$ similarly. User $k$ sends $X_k=\phi_k(\bz_k^L)\in\mathcal X_k$, where $\bz_k^L=\mathbf H_k^{[L]}\bw^L$. The server outputs $\psi(X_1,\ldots,X_K)$ as its estimate of $\by^L=\mathbf D^{[L]}\bw^L$. The encoders $\phi_k$ and decoder $\psi$ are arbitrary deterministic functions. The encoding and decoding functions may depend on the known matrices $\mathbf H_1,\ldots,\mathbf H_K,\mathbf D$.

User $k$ sends $N_k=\log_q|\mathcal X_k|$ symbols per block, at rate $R_k=N_k/L$, and the sum-rate is $R_\Sigma=\sum_kR_k$. For fixed $\mathbf H_1,\ldots,\mathbf H_K,\mathbf D$, a sum-rate $R$ is achievable if codes with $L\to\infty$ satisfy $\limsup_{L\to\infty}R_\Sigma\le R$ and $\Pr\{\psi(X_1,\ldots,X_K)\ne\by^L\}\to0$, where the data are uniform on $\Fq^{\mu L}$. The capacity $C_q$ is the infimum of achievable sum-rates; dependence on the fixed matrices is suppressed. A code has \emph{zero error} if $\psi(X_1,\ldots,X_K)=\by^L$ for every data input.

\subsection{Generic capacity}\label{sec:generic}
Fix $(K,\mu,\nu)$ and draw all entries of $\mathbf H_1,\ldots,\mathbf H_K,\mathbf D$ independently and uniformly from $\Fq$. Condition on the rank assumptions above, which hold with probability $1-O(1/q)$. The matrices are drawn once and remain fixed during coding. Thus $C_q$ is random only through this initial draw.

We study $q=p^n\to\infty$, allowing $p$, $n$, or both to grow. The \emph{generic capacity} $C$ is the limit of $C_q$ in probability along every such sequence: for every $\epsilon>0$,
\[
\Pr\{|C_q-C|>\epsilon\}\longrightarrow0.
\]
Here the probability is over the draw of $\mathbf H_1,\ldots,\mathbf H_K,\mathbf D$. Unless another limit is stated, asymptotic notation refers to $q\to\infty$ with $(K,\mu,\nu)$ fixed: $o(1)$ tends to zero, and $O(1/q)$ is bounded by a dimension-dependent constant times $1/q$.

\section{Main result}\label{sec:capacity}
For $K=1$, feasibility forces $\mu=\nu$. For $K=2$, the capacity is given by~\eqref{eq:twouserintro} and Figure~\ref{fig:problem}(b). We now give the capacity for $K\ge3$.
\Needspace{6\baselineskip}
\begin{theorem}[Generic capacity]\label{thm:main}
For fixed $K\ge3$ and $1\le\nu\le\mu\le K\nu$, let $\tau=\mu/\nu$ be the data-to-demand ratio. Then $C/\nu$ depends only on $\tau$ and is given below.
\begin{enumerate}
\item For $K=3$,
\[
\renewcommand{\arraystretch}{1.3}
\begin{array}{c|cccccc}
\tau&[1,5/3]&[5/3,2]&[2,9/4]&[9/4,5/2]&[5/2,8/3]&[8/3,3]\\ \hline
C/\nu&1&(3\tau-3)/2&3/2&2\tau-3&2&3\tau-6
\end{array}
\]
\item For $K\ge4$, let $\tau_\pm=(K+2\pm\sqrt{K(K-4)})/2$, and define the sequence
\[
\beta_0=0,\qquad \beta_{j+1}=\frac1{K(1-\beta_j)},\qquad j\ge0.
\]
Consider the lower pairs and the upper pairs
\[
(\beta_-,\beta_+)=(\beta_j,\beta_{j+1}),\ j\ge1,\qquad\text{and}\qquad
(\beta_-,\beta_+)=(1-\beta_{j+1},1-\beta_j),\ j\ge0.
\]
For each pair, define its left endpoint, changeover, and right endpoint by
\[
\tau_{\mathrm L}=K+1-\frac1{\beta_-},\qquad
\tau_\star=K-\frac{1-\beta_-}{\beta_+},\qquad
\tau_{\mathrm R}=K+1-\frac1{\beta_+}.
\]
The intervals $[\tau_{\mathrm L},\tau_{\mathrm R}]$ of the lower pairs cover $1\le\tau<\tau_-$, and those of the upper pairs cover $\tau_+<\tau\le K$.

Then $C/\nu$ is
\begin{equation}\label{eq:main}
\boxed{
\frac C\nu=\begin{cases}
\displaystyle\frac{K}{K+1-\tau},
&\tau_-\le\tau\le\tau_+,\\[5pt]
K\beta_-,
&\tau_{\mathrm L}\le\tau\le\tau_\star,\\[3pt]
K\beta_+\tau-K(K\beta_+-1),
&\tau_\star\le\tau\le\tau_{\mathrm R}.
\end{cases}}
\end{equation}
\end{enumerate}
\end{theorem}

For $K=3$, three horizontal segments alternate with three rising segments (Figure~\ref{fig:finite}). For $K\ge4$, every lower or upper pair gives a horizontal segment from $\tau_{\mathrm L}$ to $\tau_\star$ followed by a rising segment from $\tau_\star$ to $\tau_{\mathrm R}$, and increasing $j$ gives infinitely many pieces. For $K\ge4$, the sequence $\beta_j$ increases to
\[
\beta_\infty=\frac{1-\sqrt{1-4/K}}2,
\]
the smaller root of $K\beta(1-\beta)=1$, because the map $\beta\mapsto1/(K(1-\beta))$ is increasing and fixes $\beta_\infty$; for $K\le3$, the map has no fixed point below $1$, and the sequence reaches $1$ after finitely many steps. For $K=4$, $\beta_j=j/(2j+2)$ and $\beta_\infty=1/2$; the balanced interval $[\tau_-,\tau_+]$ is the single point $\tau_-=\tau_+=3$, and the pieces accumulate there from both sides (Figure~\ref{fig:tame}). For $K\ge5$, the balanced interval has positive length; the intervals of the lower and upper pairs accumulate at $\tau_-$ and $\tau_+$, between which $C/\nu=K/(K+1-\tau)$ is curved (Figure~\ref{fig:wild} for $K=5$).

The first row of~\eqref{eq:main} is the balanced regime, and each lower or upper pair gives one split regime, between its endpoints (Section~\ref{sec:regimes}). At the endpoints, $C/\nu=R_{\mathrm{align}}/\nu=K/(K+1-\tau)$.

Asymptotic set alignment approaches capacity with $L=1$: for every fixed $\epsilon>0$, it gives sum-rate at most $C+\epsilon$ with probability tending to one as the field size grows.

\section{Set alignment in the relation space}\label{sec:relations}
There are $\kappa=(K+1)\nu-\mu$ independent equations connecting the observations and demand, written as
\begin{equation}\label{eq:constraint}
\sum_{k=1}^K\mathbf A_k\bz_k=\mathbf B\by,
\qquad \mathbf A_k,\mathbf B\in\Fq^{\kappa\times\nu}.
\end{equation}
We call the matrices $\mathbf A_k$ the \emph{user channels} and $\mathbf B$ the \emph{desired channel}, by analogy with wireless channels; they are given and fixed throughout coding. User $k$ contributes in $\mathcal A_k=\col(\mathbf A_k)$ and the demand in $\mathcal B=\col(\mathbf B)$. Lemma~\ref{lem:constraints} in Appendix~\ref{app:reduction} justifies the independent uniform channel representation.

\subsection{Uncertainty sets and the conditions for set alignment}\label{sec:ia-conditions}
The encoder $\phi_k$ partitions the observations into \emph{bins}: each observation belongs to exactly one bin, and user $k$ sends the index of its bin. A bin is therefore the set of observations left possible once its index is received, the user's remaining local uncertainty. We design the bins around one finite set $\mathcal Z_k$, called the \emph{uncertainty set}: each bin index tells the server a vector $\boldsymbol s\in\Fq^\nu$ such that the observation is $\boldsymbol s$ plus an unknown element of $\mathcal Z_k$. After receiving the index, the server thus knows the observation up to an element of $\mathcal Z_k$. We call the set $\boldsymbol s+\mathcal Z_k$ of these possible observations a \emph{shift} of $\mathcal Z_k$.

With $L=1$, choose a \emph{precoding matrix} $\mathbf V_k$ over $\Fq$ and a finite \emph{coefficient alphabet} $\mathcal T_k$ whose elements, the \emph{coefficient arrays} $\boldsymbol t_k$, are vectors over $\Fq$, and write $\mathcal Z_k=\mathbf V_k\mathcal T_k=\{\mathbf V_k\boldsymbol t_k:\boldsymbol t_k\in\mathcal T_k\}$; in general, a matrix times a set, or a subspace, denotes the set of all products. Two observations in the same bin then differ by $\mathbf V_k(\boldsymbol t_k-\boldsymbol t_k')$.

Write $\mathcal T_k-\mathcal T_k=\{\boldsymbol t_k-\boldsymbol t_k':\boldsymbol t_k,\boldsymbol t_k'\in\mathcal T_k\}$; a sum of sets contains all sums obtained by choosing one element from each set. We place the combined observation differences in a finite \emph{interference set} $\mathcal I\subseteq\Fq^\kappa$. Write $\Delta\by=\by-\by'$, $\Delta\bz_k=\bz_k-\bz_k'$, and $\Delta\boldsymbol t_k=\boldsymbol t_k-\boldsymbol t_k'$ for differences of demands, observations, and coefficient arrays.

In Lemma~\ref{lem:packing} below, the first two conditions ensure that equal messages imply equal demands. The third determines the communication saving from each coefficient alphabet: a bin index saves $\log_q|\mathcal Z_k|$ symbols, and requiring different coefficient arrays to give different observation values, $|\mathcal Z_k|=|\mathcal T_k|$, lets this saving be computed from $|\mathcal T_k|$, which the constructions control.
\Needspace{16\baselineskip}
\begin{lemma}[Set alignment gives computation codes]\label{lem:packing}
For fixed channels, let $\mathbf V_k$ be precoding matrices and $\mathcal T_k$ nonempty finite coefficient alphabets, with uncertainty sets $\mathcal Z_k=\mathbf V_k\mathcal T_k$. Suppose that, for a finite set $\mathcal I\subseteq\Fq^\kappa$, the following three conditions hold.
\begin{enumerate}
\item \emph{Interference alignment.} \begin{equation}\label{eq:ia-alignment}
\sum_{k=1}^K\mathbf A_k\mathbf V_k(\mathcal T_k-\mathcal T_k)
\subseteq\mathcal I.
\end{equation}

\item \emph{Desired-signal independence.} \begin{equation}\label{eq:separation}
\mathbf B\Delta\by\notin\mathcal I
\qquad\text{for every }\Delta\by\in\Fq^\nu\setminus\{\boldsymbol0\}.
\end{equation}

\item \emph{Distinct uncertainty values.} \begin{equation}\label{eq:distinct}
|\mathcal Z_k|=|\mathcal T_k|\qquad\text{for every }k\in[K].
\end{equation}
\end{enumerate}
There is then a zero-error computation code with $L=1$ and
\begin{equation}\label{eq:packing-cost}
R_\Sigma\le K\nu-\sum_k\log_q|\mathcal T_k|+o(1).
\end{equation}

\end{lemma}

The constructions below establish the lemma's conditions on channel events whose probability tends to one.

To see the rate in~\eqref{eq:packing-cost}, user $k$ has $q^\nu$ possible observations, and each shift of $\mathcal Z_k$ contains $|\mathcal Z_k|=|\mathcal T_k|$ values by~\eqref{eq:distinct}. If the shifts partition the observation space, their number is exactly $q^\nu/|\mathcal T_k|$. For general sets, a covering by shifts and an assignment of each observation to one containing shift give bins with only a vanishing overhead in rate. User $k$'s rate is therefore at most $\nu-\log_q|\mathcal T_k|+o(1)$. Summing over users gives~\eqref{eq:packing-cost}: each user's uncertainty set $\mathcal Z_k$ saves communication. The exact covering bound and the decoding proof are in Appendix~\ref{app:reduction}; decoding uses the subtraction~\eqref{eq:subtract} of the relation equations for two data inputs.

We will choose the uncertainty sets $\mathcal Z_k$ and the interference set $\mathcal I$ to depend only on the user channels $\mathbf A_k$, so they are independent of the desired channel $\mathbf B$. Lemma~\ref{lem:separation} in Appendix~\ref{app:reduction} shows that desired-signal independence~\eqref{eq:separation} holds with probability tending to one whenever
\[
|\mathcal I|=o(q^{\kappa-\nu}).
\]
The $\nu$ demand coordinates leave room for this many combined observation differences in the $\kappa$ relation coordinates.

\subsection{Comparison with subspace alignment}\label{sec:set-comparison}
Table~\ref{tab:wireless} compares the design choices and decoding conditions. The wireless column refers to subspace alignment; structured wireless codes, such as those of real interference alignment~\cite{Motahari}, can also use finite alphabets.
\begin{table}[H]
\centering
\small
\renewcommand{\arraystretch}{1.18}
\begin{tabular}{@{}p{0.19\textwidth}p{0.35\textwidth}p{0.39\textwidth}@{}}
\toprule
& Wireless subspace alignment & Distributed computation here\\
\midrule
Given channels & Desired and interfering channel matrices & $\mathbf B$ and $\mathbf A_k$ in the relation equations\\
Design choices & Precoding matrices & Precoding matrices $\mathbf V_k$ and finite alphabets $\mathcal T_k$\\
Aligned interference & Vectors share few directions & Many local choices give few combined observation differences in $\mathcal I$,~\eqref{eq:ia-alignment}\\
Decoding condition & Desired signals are linearly independent of the aligned interference & No nonzero demand difference belongs to $\mathcal I$,~\eqref{eq:separation}\\
Performance gain & More data streams fit beside interference & Larger uncertainty sets require fewer bin indices,~\eqref{eq:packing-cost}\\
\bottomrule
\end{tabular}
\caption{Subspace alignment and set alignment.}
\label{tab:wireless}
\end{table}

\subsection{A first example: interval alphabets}\label{sec:interval-example}
The coefficient alphabet $\mathcal T_k$ is a chosen subset of the field-vector space on which $\mathbf V_k$ acts, rather than the entire space. The next scalar example uses consecutive field elements, so the allowed observation differences form a small set.
\begin{example}[Set alignment with interval alphabets]\label{ex:interval}
Let $(K,\mu,\nu)=(3,2,1)$, so $\kappa=2$. For this example, take $q=p>2$ and perform all arithmetic over $\Fp$. We first bring the generic observations and demand into a convenient coordinate form.

\emph{Given channels.} The three observation rows and the demand row specify four lines in the two-dimensional data space. With probability $1-O(1/p)$, all four lines are distinct. Choose data coordinates along the demand and the first observation. Both entries of each remaining observation are nonzero. Rescaling the coordinates and the scalar observations then gives
\[
y=w_1,\qquad z_1=w_2,\qquad
z_2=w_1+w_2,\qquad z_3=b w_1+w_2,
\]
where $b\notin\{0,1\}$ because the four lines are distinct. These are invertible changes of basis, so they preserve the computation problem and its rates. Conditional on the first three lines (the demand and observations $1,2$), the fourth is uniform among the remaining lines. Thus $b$ is uniform on $\Fp\setminus\{0,1\}$ and is fixed throughout coding, as in the four-line problem~\cite{FourLine}; it is a cross-ratio parameter for the four lines.

Subtracting the shared $w_2$ gives two independent relation equations: $z_2-z_1=y$ and $z_3-z_1=b y$. Together, these equations give the relation-space representation:
\[
\underbrace{\begin{bmatrix}-z_1\\-z_1\end{bmatrix}}_{\mathbf A_1z_1}
+\underbrace{\begin{bmatrix}z_2\\0\end{bmatrix}}_{\mathbf A_2z_2}
+\underbrace{\begin{bmatrix}0\\z_3\end{bmatrix}}_{\mathbf A_3z_3}
=\underbrace{\begin{bmatrix}y\\b y\end{bmatrix}}_{\mathbf B y}.
\]
Hence the given channel columns are
\[
\mathbf A_1=\begin{bmatrix}-1\\-1\end{bmatrix},\quad
\mathbf A_2=\begin{bmatrix}1\\0\end{bmatrix},\quad
\mathbf A_3=\begin{bmatrix}0\\1\end{bmatrix},\quad
\mathbf B=\begin{bmatrix}1\\b\end{bmatrix}.
\]
In these coordinates, the user channels $\mathbf A_k$ are fixed and only the desired channel $\mathbf B$ contains the random parameter $b$. This lets us choose the uncertainty sets $\mathcal Z_k$ independently of $b$, verify alignment, and then use its randomness to prove desired-signal independence.

\emph{Uncertainty sets.} Set $\sigma=\lfloor p^{1/2-\epsilon}\rfloor$ for fixed $0<\epsilon<1/2$. Choose $\mathbf V_k=[1]$ and $\mathcal T_k=\{0,\ldots,\sigma-1\}\subseteq\Fp$, with all arithmetic over $\Fp$. Then $\mathcal Z_k=\mathbf V_k\mathcal T_k=\mathcal T_k$.

\emph{Interference alignment.} Each coefficient difference $\Delta t_{k}$ lies in $\{-(\sigma-1),\ldots,\sigma-1\}$. Each relation coordinate below is a difference of two such coefficient differences, so it lies in $\mathcal I_0=\{-2(\sigma-1),\ldots,2(\sigma-1)\}\subseteq\Fp$. Thus the combined observation difference satisfies
\[
\sum_{k=1}^3\mathbf A_k\mathbf V_k \Delta t_{k}
=\begin{bmatrix}\Delta t_{2}-\Delta t_{1}\\\Delta t_{3}-\Delta t_{1}\end{bmatrix}
\in\mathcal I_0^2.
\]
This verifies~\eqref{eq:ia-alignment} directly with $\mathcal I=\mathcal I_0^2$. Three coefficient alphabets $\mathcal T_k$ of size $\sigma$ produce at most $(4\sigma-3)^2=O(\sigma^2)$ combined observation differences.

\emph{Desired-signal independence.} Here a demand difference is a scalar $\Delta y\in\Fp$. For~\eqref{eq:separation}, we require
\[
\mathbf B \Delta y=\begin{bmatrix}\Delta y\\b \Delta y\end{bmatrix}
\notin\mathcal I=\mathcal I_0^2
\qquad\text{whenever }\Delta y\ne0.
\]
Failure would require both $\Delta y\in\mathcal I_0$ and $b \Delta y\in\mathcal I_0$.
For each fixed nonzero $\Delta y$, the value $b \Delta y$ is uniform on $\Fp\setminus\{0,\Delta y\}$. Summing over the possible nonzero $\Delta y\in\mathcal I_0$ bounds the probability of any such change by $\frac{|\mathcal I_0|^2}{p-2}=O(p^{-2\epsilon})$. Thus~\eqref{eq:separation} holds with probability $1-O(p^{-2\epsilon})$, and Lemma~\ref{lem:packing} applies to every such fixed $b$.

\emph{Distinct uncertainty values.} Since $\mathcal Z_k=\mathcal T_k$, we have $|\mathcal Z_k|=|\mathcal T_k|=\sigma$, which is~\eqref{eq:distinct}.

\emph{Rate.} Instead of the covering in the proof of Lemma~\ref{lem:packing}, we use explicit bins. Represent each observation $z_k\in\Fp$ by its integer value in $\{0,\ldots,p-1\}$, and let user $k$ send the bin index $\lfloor z_k/\sigma\rfloor$. Bin $i$ is $\{i\sigma,\ldots,i\sigma+\sigma-1\}\cap\{0,\ldots,p-1\}$, so from its index, the server knows the observation up to an element of $\mathcal Z_k$, with $\boldsymbol s=i\sigma$. Two observations with the same index therefore differ by an element of $\mathcal Z_k-\mathcal Z_k$, which is the only property of the bins that the decoding argument of Lemma~\ref{lem:packing} uses (Appendix~\ref{app:reduction}); with~\eqref{eq:ia-alignment} and~\eqref{eq:separation}, the indices determine $y$ with zero error. Each user sends one of $\lceil p/\sigma\rceil$ indices, so
\[
R_\Sigma=3\log_p\lceil p/\sigma\rceil
=3-3\log_p\sigma+o(1)=\frac32+3\epsilon+o(1).
\]
These message alphabets are the same for every $b$ satisfying~\eqref{eq:separation}. Letting $p\to\infty$ and then $\epsilon\to0$ gives sum-rate $3/2$.
\end{example}

\Needspace{10\baselineskip}
\section{Asymptotic set alignment}\label{sec:alignment-tools}

The monomial precoding matrices follow the asymptotic alignment design of Cadambe and Jafar~\cite{CJ}: multiplication by a channel entry increases an exponent by one, so all interfering columns fit in a slightly enlarged monomial matrix. Here the coefficient alphabets also control the number of distinct combined observation differences. Bounded coefficient alphabets on channel monomials also underlie real interference alignment~\cite{Motahari}, which aligns integer constellations in fractional dimensions, and asymptotic alignment with fixed channels also gives the generic capacity of linear computation broadcast~\cite{YaoJafar}. We apply Lemma~\ref{lem:packing} by checking its three conditions: interference alignment~\eqref{eq:ia-alignment}, desired-signal independence~\eqref{eq:separation}, and distinct uncertainty values~\eqref{eq:distinct}.

\begin{example}[Asymptotic set alignment]\label{ex:count}
Let $(K,\mu,\nu)=(5,3,1)$, so $\kappa=3$. For this example, take $q=p$, with all arithmetic over $\Fp$.

\emph{Given channels.} Generically, a change of relation coordinates makes the first three user channels the standard basis, leaving the remaining columns independent and uniform:
\[
\begin{bmatrix}z_1\\z_2\\z_3\end{bmatrix}+\ba_4z_4+\ba_5z_5=\bb\,y,
\qquad \ba_4,\ba_5,\bb\in\Fp^3.
\]
Here $\mathbf A_1,\mathbf A_2,\mathbf A_3$ are the standard basis columns, $\mathbf A_4=\ba_4$, $\mathbf A_5=\ba_5$, and $\mathbf B=\bb$. Write $a_{k,i}$ for entry $i$ of $\ba_k$. We design the uncertainty sets $\mathcal Z_k$ from the six user-channel entries, independently of $\bb$.

\emph{Uncertainty sets.} Choose integers $\eta\ge2$ and $2\le\sigma<p/2$, to be specified below. For each choice of six scalar exponents $j_1,\ldots,j_6\in\{0,\ldots,\eta-1\}$, form one scalar monomial. Collect all $\eta^6$ monomials, in any fixed order, as the entries of the precoding row:
\[
\boldsymbol v_\eta^\top=
\bigl[a_{4,1}^{j_1}a_{4,2}^{j_2}a_{4,3}^{j_3}a_{5,1}^{j_4}a_{5,2}^{j_5}a_{5,3}^{j_6}\bigr]_
{0\le j_1,\ldots,j_6<\eta}\in\Fp^{1\times\eta^6}.
\]
Thus $\boldsymbol v_\eta^\top$ is the entire row, with one scalar entry for each choice of $j_1,\ldots,j_6$. Each user uses $\mathbf V_k=\boldsymbol v_\eta^\top\in\Fp^{1\times\eta^6}$ and a column of coefficients in the same order:
\[
\boldsymbol t_k=[t_{k,j_1,\ldots,j_6}]_{0\le j_1,\ldots,j_6<\eta}\in\mathcal T_k
=\mathcal T_0=\{0,\ldots,\sigma-1\}^{\eta^6}\subseteq\Fp^{\eta^6}.
\]
There are $\sigma^{\eta^6}$ allowed coefficient arrays. The row times this column gives one scalar uncertainty value:
\[
\boldsymbol v_\eta^\top\boldsymbol t_k
=\sum_{0\le j_1,\ldots,j_6<\eta}
(\boldsymbol v_\eta^\top)_{j_1,\ldots,j_6}\,t_{k,j_1,\ldots,j_6}\in\Fp,
\qquad
\mathcal Z_0=\boldsymbol v_\eta^\top\mathcal T_k\subseteq\Fp.
\]

\emph{Interference alignment.} Multiplication by a user-channel entry increases one exponent by one, so the resulting monomials fit in $\{0,\ldots,\eta\}^6$. Adding coefficient differences enlarges their coefficient range by only a constant factor; here each relation coordinate has three signals. Let $\mathcal I_0\subseteq\Fp$ contain all monomial sums with these enlarged exponents and coefficients in $\{-3\sigma,\ldots,3\sigma\}$. Then
\[
\sum_{k=1}^5\mathbf A_k\mathbf V_k(\mathcal T_k-\mathcal T_k)
\subseteq\mathcal I=\mathcal I_0^3\subseteq\Fp^3,
\qquad
|\mathcal I_0|\le(6\sigma+1)^{(\eta+1)^6}.
\]
Here $\mathbf A_k\mathbf V_k\in\Fp^{3\times\eta^6}$ maps each coefficient-difference column to a three-dimensional relation-space change, verifying~\eqref{eq:ia-alignment} for every choice of the user-channel entries. Allowing each of the six exponents to range from $0$ to $\eta$, instead of $0$ to $\eta-1$, increases the number of monomials from $\eta^6$ to $(\eta+1)^6$; their ratio $(1+1/\eta)^6$ approaches $1$.

To choose the alphabet sizes, fix $0<\epsilon<2/3$. Choose a sufficiently large fixed $\eta$ so that
\[
\left(\frac23-\epsilon\right)\left(1+\frac1\eta\right)^6
<\frac23-\frac\epsilon2,
\qquad
\sigma=\left\lfloor p^{(2/3-\epsilon)/\eta^6}\right\rfloor.
\]
The interference count gives, for every choice of the user-channel entries,
\[
\log_p|\mathcal I_0|
\le(\eta+1)^6\log_p(6\sigma+1)
=\left(\frac23-\epsilon\right)\left(1+\frac1\eta\right)^6+o(1)
\le\frac23-\frac\epsilon2
\]
for sufficiently large $p$. The strict inequality in the choice of $\eta$ absorbs the last $o(1)$. Thus the combined observation differences occupy at most $|\mathcal I|\le p^{2-3\epsilon/2}$ values in the three-dimensional relation space, leaving room for the one-dimensional demand.

\emph{Desired-signal independence.} The set $\mathcal I$ depends only on the user-channel entries, and $|\mathcal I|=o(p^2)=o(q^{\kappa-\nu})$. By Lemma~\ref{lem:separation}, as noted after Lemma~\ref{lem:packing}, condition~\eqref{eq:separation} holds with probability tending to one: the aligned uncertainty leaves enough space for the independently generated demand.

\emph{Distinct uncertainty values.} For~\eqref{eq:distinct}, different arrays in $\mathcal T_k$ must give different scalar values under $\boldsymbol v_\eta^\top$. Fix a nonzero difference $\boldsymbol t-\boldsymbol t'\in\mathcal T_k-\mathcal T_k$, independently of the user-channel entries. Its product with $\mathbf V_k$ is a nonzero polynomial of degree at most $6(\eta-1)$, so it vanishes with probability at most $6(\eta-1)/p$ by the Schwartz--Zippel lemma. Counting at most $(2\sigma-1)^{\eta^6}$ difference arrays gives
\[
\Pr(\text{two different arrays give the same value})
\le\frac{6(\eta-1)(2\sigma-1)^{\eta^6}}p
=O(p^{-1/3-\epsilon}).
\]
Indeed, $(2\sigma-1)^{\eta^6}\le2^{\eta^6}\sigma^{\eta^6}\le2^{\eta^6}p^{2/3-\epsilon}$ by the choice of $\sigma$, and $6(\eta-1)2^{\eta^6}$ is constant as $p\to\infty$ because $\eta$ is fixed. With probability tending to one, \eqref{eq:distinct} holds for every user, since all use the same row and alphabet.

\emph{Rate.} All three conditions of Lemma~\ref{lem:packing} hold with probability $1-o(1)$. The coefficient alphabet has $|\mathcal T_k|=\sigma^{\eta^6}$, so $\log_p|\mathcal T_k|=\eta^6\log_p\sigma=2/3-\epsilon+o(1)$, and~\eqref{eq:packing-cost} gives $R_\Sigma\le5-5(2/3-\epsilon)+o(1)=5/3+5\epsilon+o(1)$. Letting $p\to\infty$ and then $\epsilon\to0$ gives $5/3$.
\end{example}

Appendix~\ref{app:alignment} extends this construction to arbitrary prime powers $q=p^n$, treating bounded and large $n$ separately. Section~\ref{sec:ach} applies the scheme to the whole relation space or to one component at a time, and a component may need only part of the demand. The next lemma therefore separates the dimensions. Let $\kappa_0$ be the relation dimension, $\nu_0$ the observation dimension at each user, and $0\le\nu_{\mathrm d}\le\kappa_0$ the demand dimension, with $1\le\nu_0\le\kappa_0$. The user channels may be the original channels or the component blocks of Corollary~\ref{lem:component-coordinates}, whose entries are ratios of polynomials of bounded degree in the original independent user-channel entries; the desired channel, which multiplies the demand, is independent and uniform given them.

\begin{lemma}[Rate of asymptotic set alignment]\label{lem:alignment-rate}
For every fixed $\epsilon>0$, with probability tending to one, asymptotic set alignment gives a zero-error code with $L=1$ and
\begin{equation}\label{eq:unsplit-general}
R_\Sigma\le K\nu_0\frac{\nu_{\mathrm d}}{\kappa_0}+\epsilon.
\end{equation}
\end{lemma}

The alignment of the user signals through $\mathbf A_k$ is unchanged; only the space reserved for demand differences now has $\nu_{\mathrm d}$ dimensions. Appendix~\ref{app:alignment} gives the proof over arbitrary prime powers.

\section{Geometry of the relation space}\label{sec:geometry}
We treat the user subspaces $\mathcal A_1,\ldots,\mathcal A_K$ as a \emph{configuration} $(\Fq^\kappa;\mathcal A_1,\ldots,\mathcal A_K)$ of $K$ subspaces of dimension $\nu$ in the relation space $\Fq^\kappa$; its properties depend only on the matrices $\mathbf A_1,\ldots,\mathbf A_K$. This section establishes the geometric properties used by both achievability and converse: we first describe the balanced and split configurations, then write the relation equations by component and derive entropy bounds.

\subsection{Balance and the two regimes}\label{sec:geo-curve}
\subsubsection{Balance}\label{sec:balance}
Each user occupies a $\nu$-dimensional subspace $\mathcal A_k$ of the $\kappa$-dimensional relation space, so on average the users occupy $K\nu/\kappa$ dimensions per dimension of $\Fq^\kappa$. We write $\mathcal U\le\mathcal V$ when $\mathcal U$ is a subspace of $\mathcal V$. For a subspace $\mathcal W\le\Fq^\kappa$, the intersection $\mathcal W\cap\mathcal A_k$ consists of the directions contributed by user $k$ that lie in $\mathcal W$. We call the configuration $(\Fq^\kappa;\mathcal A_1,\ldots,\mathcal A_K)$ balanced if no subspace contains more than its proportional share of the users' total observation dimensions:
\begin{equation}\label{eq:balance}
\sum_k\dim(\mathcal W\cap\mathcal A_k)\le \frac{K\nu}{\kappa}\dim(\mathcal W)
\qquad\text{for every subspace }\mathcal W\le\Fq^\kappa.
\end{equation}
Equality holds for $\mathcal W=\{\boldsymbol0\}$ and for $\mathcal W=\Fq^\kappa$, since the user subspaces have total dimension $K\nu$. Balance says that no intermediate subspace does better than these two extremes. Three small configurations illustrate the definition; each is generic for its dimensions.

\emph{Three lines in a plane} ($K=3$, $\nu=1$, $\kappa=2$). A line contains at most one of the three user lines, and $1\le(3/2)\cdot1$; the whole plane contains all three, and $3=(3/2)\cdot2$. The configuration is balanced.

\emph{Three planes in $\Fq^3$} ($K=3$, $\nu=2$, $\kappa=3$), so $K\nu/\kappa=2$. Generic planes meet pairwise in three lines that span $\Fq^3$; in a basis along these lines, the planes are $\col[\boldsymbol e_1\ \boldsymbol e_2]$, $\col[\boldsymbol e_1\ \boldsymbol e_3]$, and $\col[\boldsymbol e_2\ \boldsymbol e_3]$. A line lies in at most two of them, and $2\le2$; a plane meets them in at most $2+1+1=4$ dimensions, and $4\le2\cdot2$; the whole space contains $6=2\cdot3$. The configuration is balanced.

\emph{Three planes in $\Fq^5$} ($K=3$, $\nu=2$, $\kappa=5$), so $K\nu/\kappa=6/5$. Since $2+2<5$, two generic planes have no overlap. Since $2+2+2>5$, there are vectors $\ba_k\in\mathcal A_k$, unique up to a common scaling, with $\ba_1+\ba_2+\ba_3=\boldsymbol0$; all three are nonzero because no two planes overlap. Their span $\mathcal W$ is a plane containing one line of each user, so $\sum_k\dim(\mathcal W\cap\mathcal A_k)=3>(6/5)\cdot2$, and the configuration is not balanced. The plane $\mathcal W$ with its three user lines is a copy of three lines in a plane. The remaining direction of each user plane gives three independent lines, which together with $\mathcal W$ fill $\Fq^5$. Thus the configuration splits into a dense part, where the users occupy half of each dimension, and a sparse part, where they occupy one third; Section~\ref{sec:balanced-fractions} makes such splits precise.

Balance also has an equivalent form on the null space of $\mathbf A=[\mathbf A_1\ \cdots\ \mathbf A_K]_{\kappa\times K\nu}$, the tuples $(\bz_1,\ldots,\bz_K)$ with $\sum_k\mathbf A_k\bz_k=\boldsymbol0$. Write $\boldsymbol\Pi_k=[\mathbf 0\ \cdots\ \mathbf 0\ \mathbf I_\nu\ \mathbf 0\ \cdots\ \mathbf 0]_{\nu\times K\nu}$, with the identity $\mathbf I_\nu$ in block $k$, so that $\boldsymbol\Pi_k(\bz_1,\ldots,\bz_K)=\bz_k$ picks out user $k$'s entry. This form is used by the converse (Section~\ref{sec:geo-converse}) and by the reduction in Appendix~\ref{app:elementary}.
\begin{lemma}[Balance on the null space]\label{lem:null-balance}
Let each $\mathbf A_k$ have full column rank, let $\mathbf A$ have full row rank, and let $\beta=\nu/\kappa>1/K$. The configuration is balanced if and only if
\begin{equation}\label{eq:null-balance}
\sum_k\dim(\boldsymbol\Pi_k\mathcal U)\ge\frac{K\beta}{K\beta-1}\dim(\mathcal U)\qquad\text{for every }\mathcal U\le\nullspace(\mathbf A).
\end{equation}
\end{lemma}
The factor $K\beta/(K\beta-1)=K\nu/(K\nu-\kappa)$ in~\eqref{eq:null-balance} is the ratio for the whole configuration: the users have $K\nu$ dimensions in total, and the null space has dimension $K\nu-\kappa$. Balance says that no subspace $\mathcal W$ holds more user dimensions per dimension than the whole relation space; dually, the lemma says that no part of the null space has fewer user dimensions per dimension than the whole null space. The proof is in Appendix~\ref{app:null-balance}.

\subsubsection{Balanced and split fractions}\label{sec:balanced-fractions}
A configuration \emph{splits} if the relation space is a direct sum $\Fq^\kappa=\mathcal E_-\oplus\mathcal E_+$ of two nonzero subspaces, that is, $\mathcal E_-+\mathcal E_+=\Fq^\kappa$ and $\mathcal E_-\cap\mathcal E_+=\{\boldsymbol0\}$, such that every user subspace splits accordingly, $\mathcal A_k=(\mathcal A_k\cap\mathcal E_-)\oplus(\mathcal A_k\cap\mathcal E_+)$. Each part $\mathcal E_\pm$ gives a \emph{component} $(\mathcal E_\pm;\mathcal A_1\cap\mathcal E_\pm,\ldots,\mathcal A_K\cap\mathcal E_\pm)$, again a configuration of $K$ subspaces of equal dimension, now in the smaller space $\mathcal E_\pm$. In bases adapted to the split, all user channels are block diagonal with matching blocks. A component is \emph{balanced} if it satisfies~\eqref{eq:balance} with $\mathcal E_\pm$ in place of $\Fq^\kappa$ and $\mathcal A_k\cap\mathcal E_\pm$ in place of $\mathcal A_k$. For the three planes in $\Fq^5$ of Section~\ref{sec:balance}, $\mathcal E_+=\mathcal W$ and $\mathcal E_-$ is spanned by the three independent lines.

For generic channels, the per-user fraction $\nu/\kappa$ alone decides which of two regimes occurs: the configuration is balanced, or it splits into two balanced components. We call these the balanced and split regimes, and the corresponding fractions balanced and split. The balanced fractions form the set
\[
\mathcal S_K=\begin{cases}
\{0,1/2,1\},&K=2,\\
\{0,1/3,1/2,2/3,1\},&K=3,\\
\{\beta_j,1-\beta_j:j\ge0\}\cup\{1/2\},&K=4,\\
\{\beta_j,1-\beta_j:j\ge0\}\cup[\beta_\infty,1-\beta_\infty],&K\ge5,
\end{cases}
\]
where $\beta_j$ is the sequence in Theorem~\ref{thm:main} and $\beta_\infty$ its limit (Section~\ref{sec:capacity}). The elements of $\mathcal S_K$ are particular values of the per-user fraction $\beta=\nu/\kappa$: by Lemma~\ref{lem:geometry} below, a generic configuration is balanced when $\nu/\kappa\in\mathcal S_K$, and otherwise splits into components whose fractions are the two neighboring elements of $\mathcal S_K$. So Lemma~\ref{lem:null-balance} applies with $\beta=\nu/\kappa$ in the first case and, component by component, with $\beta=\beta_\pm$ in the second. The following lemma is the main geometric result: it divides the possible fractions $\nu/\kappa$ into these two regimes.

\begin{lemma}[Balanced and split fractions]\label{lem:geometry}
Fix $K\ge2$ and $\nu\le\kappa\le K\nu$, and let the user channels be independent and uniform. Along every sequence of prime powers, the following holds with probability tending to one.
\begin{enumerate}
\item \emph{Balanced fractions.} If $\nu/\kappa\in\mathcal S_K$, the configuration is balanced.
\item \emph{Split fractions.} Let $\beta_-<\beta_+$ be consecutive elements of $\mathcal S_K$. If $\beta_-<\nu/\kappa<\beta_+$, there are subspaces $\mathcal E_-,\mathcal E_+\le\Fq^\kappa$ such that, for every user $k$,
\[
\Fq^\kappa=\mathcal E_-\oplus\mathcal E_+,\qquad
\mathcal A_k=(\mathcal A_k\cap\mathcal E_-)\oplus(\mathcal A_k\cap\mathcal E_+),\qquad
\dim(\mathcal A_k\cap\mathcal E_\pm)=\beta_\pm\dim(\mathcal E_\pm),
\]
and both components are balanced. Their dimensions $\kappa_\pm=\dim(\mathcal E_\pm)$ satisfy
\begin{equation}\label{eq:blocks}
\kappa_-+\kappa_+=\kappa,\qquad
\beta_-\kappa_-+\beta_+\kappa_+=\nu,\qquad\text{so}\qquad
\kappa_- =\frac{\beta_+\kappa-\nu}{\beta_+-\beta_-},\quad
\kappa_+ =\frac{\nu-\beta_-\kappa}{\beta_+-\beta_-}.
\end{equation}
\end{enumerate}
\end{lemma}

Appendices~\ref{app:central}--\ref{app:geometry-proof} prove the lemma. The split fractions between two consecutive elements of $\mathcal S_K$ form one split regime; we call the components on $\mathcal E_-$ and $\mathcal E_+$ the \emph{sparse} and \emph{dense} components.

\begin{example}[A split configuration]\label{ex:blocks}
Let $(K,\mu,\nu)=(3,7,3)$, so $\kappa=(K+1)\nu-\mu=5$ and $\nu/\kappa=3/5$, between the consecutive elements $\beta_-=1/2$ and $\beta_+=2/3$ of $\mathcal S_3$. By~\eqref{eq:blocks}, $\kappa_-=2$ and $\kappa_+=3$, with $\beta_-\kappa_-=1$ and $\beta_+\kappa_+=2$ dimensions per user. Three generic $3$-dimensional subspaces of $\Fq^5$ split into two balanced components with these dimensions, each a generic configuration of its own dimensions. In suitable bases, they are the two balanced configurations of Section~\ref{sec:balance}:
\begin{itemize}
\item \emph{Sparse component: three lines in a plane.} In $\Fq^2$, the users occupy $\col(\boldsymbol e_1)$, $\col(\boldsymbol e_2)$, and $\col(\boldsymbol e_1+\boldsymbol e_2)$.
\item \emph{Dense component: three planes in $\Fq^3$.} In $\Fq^3$, the users occupy $\col[\boldsymbol e_1\ \boldsymbol e_2]$, $\col[\boldsymbol e_1\ \boldsymbol e_3]$, and $\col[\boldsymbol e_2\ \boldsymbol e_3]$.
\end{itemize}
Each user's $3$-dimensional subspace is the direct sum of its line and its plane. The three planes in $\Fq^5$ of Section~\ref{sec:balance} split instead between $1/3$ and $1/2$, so there three lines in a plane form the dense component; here, between $1/2$ and $2/3$, they form the sparse one.
\end{example}

The balanced fractions arise in two ways.

\emph{The balanced interval.} A configuration of $K$ subspaces of dimension $\nu$ in $\Fq^\kappa$ has $K\nu(\kappa-\nu)$ parameters, while changes of basis of the relation space have $\kappa^2$ entries. When
\[
K\nu(\kappa-\nu)\ge\kappa^2,\qquad\text{equivalently}\qquad K\beta(1-\beta)\ge1\ \text{ with }\beta=\nu/\kappa,
\]
this count suggests the quadratic range of balanced fractions. Lemma~\ref{lem:central} in Appendix~\ref{app:central} makes this heuristic rigorous by a counting argument: in this range, the configuration is balanced with probability $1-O(1/q)$, uniformly in the characteristic. Since $\beta(1-\beta)\le1/4$, the condition has no solutions for $K\le3$, gives only $\beta=1/2$ for $K=4$, and gives the interval $[\beta_\infty,1-\beta_\infty]$ for $K\ge5$.

\emph{Isolated fractions.} The other balanced fractions are built from the three base configurations of Appendix~\ref{app:elementary}, with fractions $\beta_0=0$, $\beta_1=1/K$, and $\beta_2=1/(K-1)$: lifting by the reverse alignment construction~\eqref{eq:reverse-dims} sends $\beta_j$ to $\beta_{j+2}$ and preserves balance (Corollary~\ref{lem:relations-balance}), which gives every $\beta_j$, and orthogonal complements send $\beta$ to $1-\beta$ and preserve balance (Lemma~\ref{lem:sum-perp}), which gives every $1-\beta_j$. A limit of the recursion satisfies $\beta=1/(K(1-\beta))$, that is $K\beta(1-\beta)=1$, so the isolated fractions accumulate exactly at the endpoints of the balanced interval.

\subsubsection{The two regimes along the data-to-demand ratio}\label{sec:regimes}
Since $\kappa/\nu=K+1-\tau$, the per-user fraction $\beta=\nu/\kappa=1/(K+1-\tau)$ increases with $\tau$, and conversely a fraction $\beta$ corresponds to $\tau=K+1-1/\beta$. The balanced interval $[\beta_\infty,1-\beta_\infty]$ of fractions therefore corresponds to $\tau_-\le\tau\le\tau_+$, with
\[
\tau_-=K+1-\frac1{\beta_\infty}=\frac{K+2-\sqrt{K(K-4)}}2,\qquad
\tau_+=K+1-\frac1{1-\beta_\infty}=\frac{K+2+\sqrt{K(K-4)}}2,
\]
since $1/\beta_\infty$ and $1/(1-\beta_\infty)$ are the two roots of $x^2-Kx+K=0$, namely $(K\pm\sqrt{K(K-4)})/2$. We also call $[\tau_-,\tau_+]$ the balanced interval. Likewise, an isolated balanced fraction $\beta$ corresponds to a single point, and the split regime between consecutive $\beta_-<\beta_+$ corresponds to $\tau_{\mathrm L}<\tau<\tau_{\mathrm R}$, where $\tau_{\mathrm L}$ and $\tau_{\mathrm R}$ are the points corresponding to $\beta_-$ and $\beta_+$, the endpoints in Theorem~\ref{thm:main}. Within a split regime, the component dimensions~\eqref{eq:blocks} vary linearly: as $\nu/\kappa$ moves from $\beta_-$ to $\beta_+$, the sparse component shrinks from all of $\Fq^\kappa$ to nothing and the dense component grows, and at each endpoint one component vanishes and the configuration is balanced.

\subsection{Relation equations by component}\label{sec:shared-coordinates}
The achievability uses the decomposition of Lemma~\ref{lem:geometry} in matrix form, in which the relation equations separate by component. Superscripts $-$ and $+$ denote the sparse and dense parts.

\begin{corollary}[Relation equations by component]\label{lem:component-coordinates}
In the split regime, after changes of basis adapted to the decomposition of Lemma~\ref{lem:geometry}, each observation splits as $\bz_k=(\bz_k^-,\bz_k^+)$ with $\bz_k^\pm\in\Fq^{\beta_\pm\kappa_\pm}$, and the relation equations become
\begin{equation}\label{eq:component-equations}
\sum_k\mathbf A_k^-\bz_k^-=\mathbf B^-\by,\qquad
\sum_k\mathbf A_k^+\bz_k^+=\mathbf B^+\by.
\end{equation}
\end{corollary}
\begin{proof}
In bases of $\Fq^\kappa$ and of each observation space adapted to $\Fq^\kappa=\mathcal E_-\oplus\mathcal E_+$, each user channel is block diagonal, $\mathbf A_k=\operatorname{diag}(\mathbf A_k^-,\mathbf A_k^+)$ with $\mathbf A_k^\pm\in\Fq^{\kappa_\pm\times\beta_\pm\kappa_\pm}$. Splitting $\mathbf B$ into the matching row blocks $\mathbf B^-$ and $\mathbf B^+$ separates $\sum_k\mathbf A_k\bz_k=\mathbf B\by$ into~\eqref{eq:component-equations}.
\end{proof}
The two systems in~\eqref{eq:component-equations} share only the demand $\by$. The changes of basis are rational functions of the user-channel entries with nonzero denominators (proof of Lemma~\ref{lem:outer}), so they depend only on the user channels. Hence $\mathbf B^-$ and $\mathbf B^+$ remain independent and uniform conditional on the user channels (Lemma~\ref{lem:constraints}).

\subsection{Entropy bounds for the converse}\label{sec:geo-converse}
The converse uses the geometry through entropy bounds on what the users still observe once the demand is fixed. With $\mathbf A$ and $\boldsymbol\Pi_k$ as in Lemma~\ref{lem:null-balance}, fixing the demand and subtracting a compatible observation tuple leaves a vector in $\nullspace(\mathbf A)$, so we represent the data left unspecified by a fixed demand, the \emph{residual data}, by $\bwr$, written in the users' observation coordinates, in $\mathcal U_{\mathrm{res}}=\nullspace(\mathbf A)$, where the subscript stands for ``residual''; $\boldsymbol\Pi_k\bwr$ is user $k$'s remaining observation. The residual space $\mathcal U_{\mathrm{res}}$ has dimension $K\nu-\kappa=\mu-\nu$.

\begin{example}[Residual data for five users]\label{ex:five-residual}
In Example~\ref{ex:count}, $(K,\mu,\nu)=(5,3,1)$, and the observations $z_1,\ldots,z_5$ and the demand $y$ satisfy the relation equation $(z_1,z_2,z_3)+\ba_4z_4+\ba_5z_5=\bb\,y$.
Fix the demand $y$, and take two observation tuples $(z_1,\ldots,z_5)$ and $(z_1',\ldots,z_5')$ that are both consistent with it. Subtracting their relation equations cancels $\bb\,y$, so the difference $\bwr=(\Delta z_1,\ldots,\Delta z_5)$, with $\Delta z_k=z_k-z_k'$, satisfies
\[
(\Delta z_1,\Delta z_2,\Delta z_3)+\ba_4\Delta z_4+\ba_5\Delta z_5=\boldsymbol0,
\qquad\text{that is,}\qquad
\Delta z_i=-a_{4,i}\Delta z_4-a_{5,i}\Delta z_5,\quad i=1,2,3.
\]
Conversely, every such $\bwr$ arises in this way. Hence the observation tuples consistent with a fixed demand differ by elements of $\mathcal U_{\mathrm{res}}=\nullspace(\mathbf A)$, which has dimension $\mu-\nu=2$ with free coordinates $(\Delta z_4,\Delta z_5)$, and user $k$'s remaining observation is the coordinate $\boldsymbol\Pi_k\bwr=\Delta z_k$.
\end{example}

The converse rests on entropy bounds of the form
\[
\sum_kH(\boldsymbol\Pi_k\bwr)\ge\zeta H(\bwr)+\gamma_\zeta
\qquad\text{for every distribution of }\bwr\in\mathcal U_{\mathrm{res}},
\]
a lower bound on the sum of the users' remaining observation entropies by a multiple $\zeta$, the \emph{entropy factor}, of the remaining data entropy, with a constant term $\gamma_\zeta$. These bounds depend only on the user subspaces. Section~\ref{sec:conv-tools} reduces them to subspace dimensions and gives the entropy factor of a balanced configuration; Section~\ref{sec:conv-split} treats the split regime.

\subsubsection{Entropy bounds from subspace dimensions}\label{sec:conv-tools}
\begin{example}[Entropy factor for five users]\label{ex:five-entropy}
In Example~\ref{ex:five-residual}, each $\Delta z_k$ is a linear function of $(\Delta z_4,\Delta z_5)$, and for each pair $i\ne j$ the $2\times2$ matrix taking $(\Delta z_4,\Delta z_5)$ to $(\Delta z_i,\Delta z_j)$ is nonsingular with probability $1-O(1/q)$, so every pair of remaining observations determines $\bwr$: $H(\boldsymbol\Pi_i\bwr,\boldsymbol\Pi_j\bwr)=H(\bwr)$. Pairwise subadditivity gives
\begin{align*}
4\sum_{k=1}^5H(\boldsymbol\Pi_k\bwr)
&=\sum_{1\le i<j\le5}\bigl[H(\boldsymbol\Pi_i\bwr)+H(\boldsymbol\Pi_j\bwr)\bigr]\\
&\ge\sum_{1\le i<j\le5}H(\boldsymbol\Pi_i\bwr,\boldsymbol\Pi_j\bwr)
=10H(\bwr),\\
\Longrightarrow\quad\sum_{k=1}^5H(\boldsymbol\Pi_k\bwr)
&\ge\frac52H(\bwr).
\end{align*}
\end{example}

In general, the pairwise argument used above is replaced by Lemma~\ref{lem:BL} and Corollary~\ref{lem:balanced-entropy}. Lemma~\ref{lem:BL} reduces every entropy bound to subspace dimensions: the minimum over distributions of the total local entropy minus a multiple of the joint entropy equals the corresponding minimum over subspaces $\mathcal U$. Unlike the pairwise argument, its proof uses the additive structure of $\Fq$, through sums of independent copies (Remark~\ref{rem:entropy}).

\begin{lemma}[Entropy bounds from subspace dimensions]\label{lem:BL}
For any configuration, with its residual space $\mathcal U_{\mathrm{res}}$ and matrices $\boldsymbol\Pi_k$, and any $\zeta\in\mathbb R$, every random vector $\bwr$ taking values in $\mathcal U_{\mathrm{res}}$ satisfies
\begin{equation}\label{eq:entropy}
\sum_kH(\boldsymbol\Pi_k\bwr)-\zeta H(\bwr)\ge
\min_{\mathcal U\le\mathcal U_{\mathrm{res}}}\left\{\sum_k\dim(\boldsymbol\Pi_k\mathcal U)-\zeta\dim(\mathcal U)\right\},
\end{equation}
with equality for $\bwr$ uniform on a minimizing subspace.
\end{lemma}
\begin{proof}
Write $\operatorname{ent}(\bwr)=\sum_kH(\boldsymbol\Pi_k\bwr)-\zeta H(\bwr)$ and let $\operatorname{ent}_\star$ be its minimum over distributions. This minimum exists because the set of distributions on the finite alphabet is compact and entropy is continuous. Among minimizers choose $\bwr$ with largest entropy, and let $\bwr'$ be an independent copy.

Fix any nonzero $c\in\Fq$ and let $\bwr_{\mathrm{sum}}=\bwr+c \bwr'$. Write $\mathbb E_{\bwr_{\mathrm{sum}}} \operatorname{ent}(\bwr'\mid \bwr_{\mathrm{sum}})$ for the average of $\operatorname{ent}$ evaluated on the conditional distributions given each value of $\bwr_{\mathrm{sum}}$. The chain rule, invertibility of $(\bwr,\bwr')\mapsto(\bwr_{\mathrm{sum}},\bwr')$, and conditioning give
\begin{align*}
2\operatorname{ent}_\star
&=\sum_k\{H(\boldsymbol\Pi_k \bwr_{\mathrm{sum}})+H(\boldsymbol\Pi_k\bwr'\mid\boldsymbol\Pi_k \bwr_{\mathrm{sum}})\}
-\zeta\{H(\bwr_{\mathrm{sum}})+H(\bwr'\mid \bwr_{\mathrm{sum}})\}\\
&\ge \operatorname{ent}(\bwr_{\mathrm{sum}})+\mathbb E_{\bwr_{\mathrm{sum}}} \operatorname{ent}(\bwr'\mid \bwr_{\mathrm{sum}})\ge2\operatorname{ent}_\star.
\end{align*}
Hence $\bwr_{\mathrm{sum}}$ is also a minimizer. Maximality of $H(\bwr)$ and independence imply
\[
H(\bwr)\ge H(\bwr_{\mathrm{sum}})\ge H(\bwr_{\mathrm{sum}}\mid \bwr')=H(\bwr),
\]
so $\bwr_{\mathrm{sum}}$ and $\bwr'$ are independent. Consequently the distribution of $\bwr+c\bwr'$ is the same for every fixed $\bwr'$ in the support of $\bwr$, the set of values with positive probability. The distribution is invariant under every $c(\bwr'-\bwr'')$, for every nonzero $c$.

Let $\mathcal U$ be the $\Fq$-span of these support differences. The distribution of $\bwr$ is invariant under $\mathcal U$, and its support is $\{\bwr_0+\boldsymbol x:\boldsymbol x\in\mathcal U\}$ for any $\bwr_0$ in it, so it is uniform on this set. Since $\bwr$ and $\bwr'$ take values in the subspace $\mathcal U_{\mathrm{res}}$, so do their differences, and $\mathcal U\le\mathcal U_{\mathrm{res}}$. The objective of this uniform distribution is $\sum_k\dim(\boldsymbol\Pi_k\mathcal U)-\zeta\dim(\mathcal U)$, proving the lower bound; a uniform distribution on a minimizing subspace gives equality.
\end{proof}

The subspace minimum in~\eqref{eq:entropy} is the constant $\gamma_\zeta$ of an entropy bound $\sum_kH(\boldsymbol\Pi_k\bwr)\ge\zeta H(\bwr)+\gamma_\zeta$. Thus every entropy bound reduces to the geometry of the relation space.

\begin{remark}[Connections with entropy inequalities]\label{rem:entropy}
Lemma~\ref{lem:BL} is a finite-field form of the discrete Brascamp--Lieb inequality, for which an optimizer can be chosen uniform on a subgroup~\cite{Christ,LauNair}; here it is uniform on an $\Fq$-subspace, and the proof follows the doubling method of~\cite{LauNair}. In the Euclidean setting, the validity of such inequalities is characterized by a scaling condition together with dimension conditions over all subspaces~\cite{BCCT}, which for configurations of subspaces amount to semistability~\cite{ChindrisDerksenIMRN}, the counterpart of balance~\eqref{eq:balance}. Quiver methods have also been applied to other entropy inequalities~\cite{ChindrisDerksen}.
\end{remark}

For a balanced configuration, Lemma~\ref{lem:null-balance} gives the entropy factor directly, with constant term $\gamma_\zeta=0$; Example~\ref{ex:five-entropy} is the case $\beta=1/3$, $\zeta_\beta=5/2$. Consider a configuration, or a component, with per-user fraction $\beta>1/K$ and channels of full rank as in Lemma~\ref{lem:null-balance}; its residual data $\bwr_\beta$ take values in $\nullspace(\mathbf A)$ for its own $\mathbf A$.
\Needspace{8\baselineskip}
\begin{corollary}[Entropy forced by a balanced component]\label{lem:balanced-entropy}
For a balanced component and every distribution of its residual data $\bwr_{\beta}$,
\begin{equation}\label{eq:block-entropy}
\sum_kH(\boldsymbol\Pi_k\bwr_{\beta})\ge\zeta_{\beta} H(\bwr_{\beta}),\qquad
\zeta_{\beta}=\frac{K\beta}{K\beta-1}.
\end{equation}
\end{corollary}
\begin{proof}
Combine Lemmas~\ref{lem:null-balance} and~\ref{lem:BL}: by Lemma~\ref{lem:null-balance}, applied to the component, $\sum_k\dim(\boldsymbol\Pi_k\mathcal U)\ge\zeta_\beta\dim(\mathcal U)$ for every subspace $\mathcal U$ of its null space, and Lemma~\ref{lem:BL} converts this dimension inequality into~\eqref{eq:block-entropy}.
\end{proof}

\subsubsection{Entropy bounds in the split regime}\label{sec:conv-split}
Fix one split regime: consecutive elements $\beta_-<\beta_+$ of $\mathcal S_K$, and $\beta_-<\nu/\kappa<\beta_+$. The residual space separates by component. With the blocks of Corollary~\ref{lem:component-coordinates}, set $\mathbf A^\pm=[\mathbf A_1^\pm\ \cdots\ \mathbf A_K^\pm]_{\kappa_\pm\times K\beta_\pm\kappa_\pm}$, and let $\boldsymbol\Pi_k^\pm$ select user $k$'s block of coordinates in that component, as $\boldsymbol\Pi_k$ does in Lemma~\ref{lem:null-balance}.

\begin{corollary}[Residual data]\label{lem:residual}
In the split regime, in the coordinates of Corollary~\ref{lem:component-coordinates}, the space of observation differences for a fixed demand is
\begin{equation}\label{eq:component-residuals}
\mathcal U_{\mathrm{res}}=\mathcal U_{\mathrm{res},-}\oplus\mathcal U_{\mathrm{res},+},\qquad
\mathcal U_{\mathrm{res},\pm}=\nullspace(\mathbf A^\pm),\qquad
\dim(\mathcal U_{\mathrm{res},\pm})=(K\beta_\pm-1)\kappa_\pm,
\end{equation}
and $\boldsymbol\Pi_k\bwr=(\boldsymbol\Pi_k^-\bwr_-,\boldsymbol\Pi_k^+\bwr_+)$ for $\bwr=(\bwr_-,\bwr_+)\in\mathcal U_{\mathrm{res}}$.
\end{corollary}
\begin{proof}
This is Corollary~\ref{lem:component-coordinates} with the demand fixed. Subtracting two observation tuples consistent with the same demand sets both right sides of~\eqref{eq:component-equations} to zero, and $\mathbf A^\pm$ act on separate coordinates, so the null space and each user's projection split as stated. In each component, the users' coordinates have total dimension $K\beta_\pm\kappa_\pm$, and the $\kappa_\pm$ relation equations are independent, so the residual space has dimension $K\beta_\pm\kappa_\pm-\kappa_\pm=(K\beta_\pm-1)\kappa_\pm$.
\end{proof}

\begin{example}[Residual data for the split configuration]\label{ex:split-residual}
In Example~\ref{ex:blocks}, fix the demand and subtract two consistent observation tuples. By Corollary~\ref{lem:residual}, the difference splits into a sparse and a dense part.
\begin{itemize}
\item \emph{Sparse part.} The users' lines are spanned by $\boldsymbol e_1$, $\boldsymbol e_2$, and $\boldsymbol e_1+\boldsymbol e_2$, so their sparse differences $\Delta z_1^-,\Delta z_2^-,\Delta z_3^-$ satisfy $\Delta z_1^-\boldsymbol e_1+\Delta z_2^-\boldsymbol e_2+\Delta z_3^-(\boldsymbol e_1+\boldsymbol e_2)=\boldsymbol0$, that is, $\Delta z_1^-=\Delta z_2^-=-\Delta z_3^-$. The sparse residual space has dimension $(K\beta_--1)\kappa_-=1$; write $\bwr_-=\Delta z_1^-$, so each user's sparse difference is $\bwr_-$ or $-\bwr_-$.
\item \emph{Dense part.} The users' planes are $\col[\boldsymbol e_1\ \boldsymbol e_2]$, $\col[\boldsymbol e_1\ \boldsymbol e_3]$, and $\col[\boldsymbol e_2\ \boldsymbol e_3]$. Their dense differences cancel coordinate by coordinate: the $\boldsymbol e_1$-coordinates of users $1$ and $2$ are opposite, as are the $\boldsymbol e_2$-coordinates of users $1$ and $3$ and the $\boldsymbol e_3$-coordinates of users $2$ and $3$. The dense residual space has dimension $(K\beta_+-1)\kappa_+=3$, with coordinates $\bwr_+=(\bwr_{+,1},\bwr_{+,2},\bwr_{+,3})$, each observed by exactly one pair of users.
\end{itemize}
Signs do not change entropies, so we drop them: each $\boldsymbol\Pi_k$ then selects $3$ of the $4$ residual coordinates,
\[
\boldsymbol\Pi_1\bwr=(\bwr_-,\bwr_{+,1},\bwr_{+,2}),\qquad
\boldsymbol\Pi_2\bwr=(\bwr_-,\bwr_{+,1},\bwr_{+,3}),\qquad
\boldsymbol\Pi_3\bwr=(\bwr_-,\bwr_{+,2},\bwr_{+,3}).
\]
\end{example}

By Lemma~\ref{lem:geometry}, the two components are balanced, with different entropy factors. Suppose first that $\beta_->1/K$; the endpoint $\beta_-=1/K$ is treated in Lemma~\ref{lem:split-entropy}. Corollary~\ref{lem:balanced-entropy} gives the factor $\zeta_-=K\beta_-/(K\beta_--1)$ for the sparse component and the smaller factor $\zeta_+=K\beta_+/(K\beta_+-1)$ for the dense component. Applying the dense-component factor to all residual entropy gives the \emph{dense-factor bound}; applying the sparse-component factor and subtracting a correction bounded by the dense component's dimension gives the \emph{sparse-factor bound}.

\begin{example}[Entropy bounds for the split configuration]\label{ex:split-entropy}
In Example~\ref{ex:split-residual}, where $\zeta_-=3$ and $\zeta_+=2$, for the three parts $\bwr_{+,1},\bwr_{+,2},\bwr_{+,3}$, the pairs $\{1,2\}$, $\{1,3\}$, and $\{2,3\}$ cover each coordinate twice. Conditional Shearer's inequality given $\bwr_-$ gives
\begin{align}
\sum_kH(\boldsymbol\Pi_k \bwr)
&=3H(\bwr_-)+H(\bwr_{+,1},\bwr_{+,2}\mid \bwr_-)+H(\bwr_{+,1},\bwr_{+,3}\mid \bwr_-)+H(\bwr_{+,2},\bwr_{+,3}\mid \bwr_-)\notag\\
&\ge3H(\bwr_-)+2H(\bwr_{+,1},\bwr_{+,2},\bwr_{+,3}\mid \bwr_-)\notag\\
&=2H(\bwr)+H(\bwr_-).\label{eq:two-multiplicities}
\end{align}
\emph{Dense-factor bound.} Dropping $H(\bwr_-)\ge0$ in~\eqref{eq:two-multiplicities} immediately gives
\[
\sum_kH(\boldsymbol\Pi_k\bwr)\ge2H(\bwr).
\]

\emph{Sparse-factor bound.} The dense residual data have dimension $3$, so $H(\bwr_-)=H(\bwr)-H(\bwr_{+,1},\bwr_{+,2},\bwr_{+,3}\mid\bwr_-)\ge H(\bwr)-3$. Substituting this lower bound into~\eqref{eq:two-multiplicities} gives
\[
\sum_kH(\boldsymbol\Pi_k\bwr)\ge3H(\bwr)-3.
\]
\end{example}

The general case follows the same steps, with conditional Shearer's inequality replaced by Corollary~\ref{lem:balanced-entropy} on each component.
\begin{lemma}[Entropy bounds in the split regime]\label{lem:split-entropy}
In a split regime with $\beta_->1/K$, every distribution of $\bwr\in\mathcal U_{\mathrm{res}}$ satisfies
\begin{equation}\label{eq:uncertainty-budget}
\begin{aligned}
\sum_kH(\boldsymbol\Pi_k\bwr)&\ge\zeta_+H(\bwr),\\
\sum_kH(\boldsymbol\Pi_k\bwr)&\ge\zeta_-H(\bwr)-(\zeta_--\zeta_+)(K\beta_+-1)\kappa_+.
\end{aligned}
\end{equation}
If $\beta_-=1/K$, the first inequality holds.
\end{lemma}
\begin{proof}
Let $\beta_->1/K$, so $\zeta_->\zeta_+$. By Corollary~\ref{lem:residual}, $\boldsymbol\Pi_k\bwr=(\boldsymbol\Pi_k^-\bwr_-,\boldsymbol\Pi_k^+\bwr_+)$. The chain rule, and conditioning on all of $\bwr_-$ rather than only on $\boldsymbol\Pi_k^-\bwr_-$, give for each $k$
\[
H(\boldsymbol\Pi_k\bwr)=H(\boldsymbol\Pi_k^-\bwr_-)+H(\boldsymbol\Pi_k^+\bwr_+\mid\boldsymbol\Pi_k^-\bwr_-)
\ge H(\boldsymbol\Pi_k^-\bwr_-)+H(\boldsymbol\Pi_k^+\bwr_+\mid\bwr_-).
\]
Sum over $k$. Corollary~\ref{lem:balanced-entropy} for the sparse component gives $\sum_kH(\boldsymbol\Pi_k^-\bwr_-)\ge\zeta_-H(\bwr_-)$. For each value of $\bwr_-$, Corollary~\ref{lem:balanced-entropy} for the dense component applies to the conditional distribution of $\bwr_+$; taking the expectation over $\bwr_-$ gives $\sum_kH(\boldsymbol\Pi_k^+\bwr_+\mid\bwr_-)\ge\zeta_+H(\bwr_+\mid\bwr_-)$. Hence
\[
\sum_kH(\boldsymbol\Pi_k\bwr)\ge\zeta_-H(\bwr_-)+\zeta_+H(\bwr_+\mid\bwr_-).
\]
With $H(\bwr)=H(\bwr_-)+H(\bwr_+\mid\bwr_-)$, the right side can be written in two ways:
\[
\zeta_+H(\bwr)+(\zeta_--\zeta_+)H(\bwr_-)
\qquad\text{and}\qquad
\zeta_-H(\bwr)-(\zeta_--\zeta_+)H(\bwr_+\mid\bwr_-).
\]
In the first, $(\zeta_--\zeta_+)H(\bwr_-)\ge0$, which gives the dense-factor bound. In the second, $H(\bwr_+\mid\bwr_-)\le\dim(\mathcal U_{\mathrm{res},+})=(K\beta_+-1)\kappa_+$, which gives the sparse-factor bound.

If $\beta_-=1/K$, the sparse residual space has dimension $(K\beta_--1)\kappa_-=0$, so $\bwr=\bwr_+$, and Corollary~\ref{lem:balanced-entropy} applied to the dense component gives the first inequality.
\end{proof}

\section{Proof of Theorem~\ref{thm:main}}\label{sec:proof}
The proof follows the two regimes of Lemma~\ref{lem:geometry}. The achievability proof applies Lemma~\ref{lem:alignment-rate} to the whole relation space or, by Corollary~\ref{lem:component-coordinates}, to each component (Section~\ref{sec:ach}), and the converse converts the entropy bounds of Corollary~\ref{lem:balanced-entropy} and Lemma~\ref{lem:split-entropy} into lower bounds on achievable sum-rates (Section~\ref{sec:conv}). Section~\ref{sec:conv-completion} combines the two.

\subsection{Achievability}\label{sec:ach}

\emph{Balanced regime.} At a balanced fraction, Lemma~\ref{lem:alignment-rate} with $\kappa_0=\kappa$ and $\nu_0=\nu_{\mathrm d}=\nu$ shows that $R_{\mathrm{align}}=K\nu^2/\kappa$ in~\eqref{eq:align-reference} is achievable, the first row of~\eqref{eq:main}.

\emph{Split regime.} Fix a split regime $\beta_-<\nu/\kappa<\beta_+$, that is, $\tau_{\mathrm L}<\tau<\tau_{\mathrm R}$ (Section~\ref{sec:regimes}). By Lemma~\ref{lem:geometry} and Corollary~\ref{lem:component-coordinates}, the configuration is two balanced components that share the demand, and direct alignment on a component with per-user fraction $\beta$ costs $K\beta$ per demand dimension (Lemma~\ref{lem:alignment-rate}). The scheme uses the cheaper sparse component first, for as many demand dimensions as it can supply, at most $\kappa_-$, and the dense component for the rest.

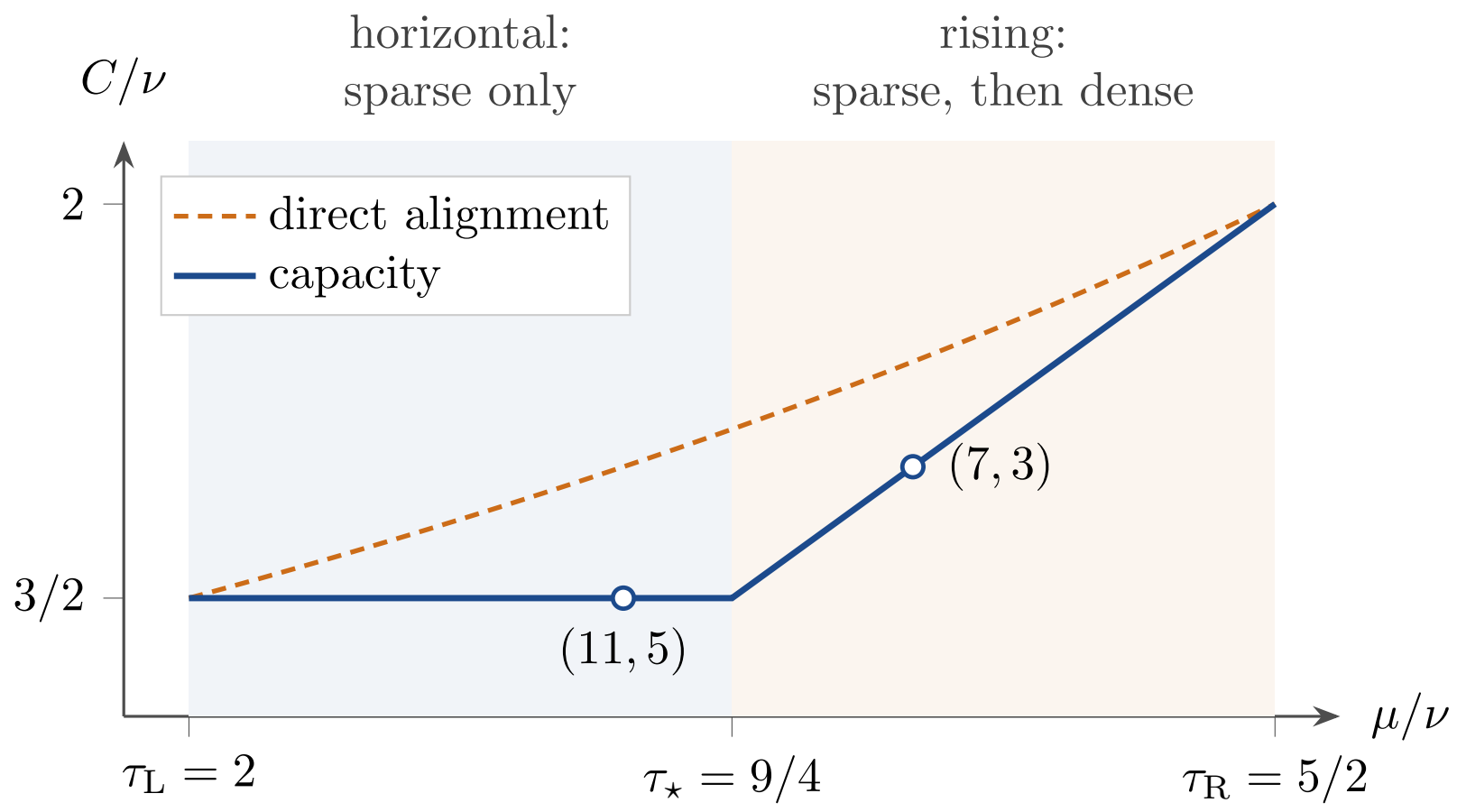
\begin{figure}[H]
\centering
\begin{tikzpicture}
\begin{axis}[capplot,ymajorgrids=false,width=10.5cm,height=5.8cm,xmin=1.97,xmax=2.53,ymin=1.35,ymax=2.08,
xtick={2,2.25,2.5},xticklabels={{$\tau_{\mathrm L}=2$},{$\tau_\star=9/4$},{$\tau_{\mathrm R}=5/2$}},
ytick={1.5,2},yticklabels={$3/2$,$2$},xlabel={$\mu/\nu$},ylabel={$C/\nu$}]
\path[fill=capblue!6] (axis cs:2,1.35) rectangle (axis cs:2.25,2.08);
\path[fill=alignorange!7] (axis cs:2.25,1.35) rectangle (axis cs:2.5,2.08);
\addplot[alignline,domain=2:2.5,samples=60]{3/(4-x)};
\addlegendentry{direct alignment}
\addplot[capline] coordinates {(2,1.5) (2.25,1.5) (2.5,2)};
\addlegendentry{capacity}
\node[font=\small,align=center,anchor=south,black!75] at (axis cs:2.125,2.09) {horizontal:\\ sparse only};
\node[font=\small,align=center,anchor=south,black!75] at (axis cs:2.375,2.09) {rising:\\ sparse, then dense};
\draw[capblue,fill=white,line width=.9pt] (axis cs:2.2,1.5) circle (2.2pt) node[below=2pt,font=\small,black] {$(11,5)$};
\draw[capblue,fill=white,line width=.9pt] (axis cs:2.3333,1.6667) circle (2.2pt) node[right=3pt,font=\small,black] {$(7,3)$};
\end{axis}
\end{tikzpicture}
\caption{The split regime $K=3$, $(\beta_-,\beta_+)=(1/2,2/3)$. The horizontal and rising segments are shaded; circles mark the instances $(\mu,\nu)=(11,5)$ and $(7,3)$ of Example~\ref{ex:two-step}.}
\label{fig:snapshot}
\end{figure}

\begin{example}[Two split instances]\label{ex:two-step}
Let $K=3$, and consider $(\mu,\nu)=(11,5)$ and $(7,3)$, which lie in the split regime of Figure~\ref{fig:snapshot}, with $(\beta_-,\beta_+)=(1/2,2/3)$; a demand dimension costs $3/2$ in the sparse component and $2$ in the dense one. By~\eqref{eq:blocks}, $(\kappa_-,\kappa_+)=(6,3)$ for $(11,5)$ and $(2,3)$ for $(7,3)$.
\begin{itemize}
\item $(\mu,\nu)=(11,5)$: the sparse component computes all $5$ demand dimensions, at sum-rate $5\cdot3/2=15/2$, on the horizontal segment.
\item $(\mu,\nu)=(7,3)$: the sparse component computes $2$ demand dimensions and the dense component $1$, at sum-rate $2\cdot3/2+1\cdot2=5$, on the rising segment.
\end{itemize}
\end{example}

In general, by~\eqref{eq:blocks} and $\kappa/\nu=K+1-\tau$,
\begin{equation}\label{eq:sparse-room}
\frac{\kappa_-}\nu=\frac{\beta_+(K+1-\tau)-1}{\beta_+-\beta_-},
\end{equation}
which decreases in $\tau$ and equals $1$ at the changeover $\tau_\star$. So each split regime breaks at $\tau_\star$ into a horizontal and a rising segment, as in Figure~\ref{fig:snapshot}.
\begin{itemize}
\item \emph{Horizontal segment, $\tau_{\mathrm L}<\tau\le\tau_\star$, $\kappa_-\ge\nu$.} The sparse component computes the whole demand: apply Lemma~\ref{lem:alignment-rate} to the sparse equations with $\kappa_0=\kappa_-$, $\nu_0=\beta_-\kappa_-$, $\nu_{\mathrm d}=\nu$, and desired channel $\mathbf B^-$; nothing is sent for the dense component. The sum-rate is $K\beta_-\nu$, so $R/\nu=K\beta_-$, the horizontal row of~\eqref{eq:main}.
\item \emph{Rising segment, $\tau_\star\le\tau<\tau_{\mathrm R}$, $\kappa_-\le\nu$.} The sparse component computes $\kappa_-$ demand dimensions and the dense component the remaining $\nu-\kappa_-$. Change demand coordinates, depending only on $\mathbf B^-$, to $\by=(\by_1,\by_2)$ with $\by_1=\mathbf B^-\by$, generically $\kappa_-$ independent demand coordinates. The users send all sparse observations, so the server decodes $\by_1$, and the dense equations become $\sum_k\mathbf A_k^+\bz_k^+=\mathbf B^+_1\by_1+\mathbf B^+_2\by_2$. The dense step works because the term $\mathbf B^+_1\by_1$ is known to the server once $\by_1$ is decoded and cancels when two inputs with the same messages are compared, so the dense encoders need not know $\by_1$ (Appendix~\ref{app:alignment}), and because $\mathbf B^+_2$, with $\nu-\kappa_-$ columns, remains independent and uniform given $\mathbf B^-$. Hence Lemma~\ref{lem:alignment-rate} applies to the dense equations with $\kappa_0=\kappa_+$, $\nu_0=\beta_+\kappa_+$, desired channel $\mathbf B^+_2$, and $\nu_{\mathrm d}=\nu-\kappa_-$. The sum-rate is $K\beta_-\kappa_-+K\beta_+(\nu-\kappa_-)$, and~\eqref{eq:sparse-room} gives $R/\nu=K\beta_+-K(\beta_+-\beta_-)\kappa_-/\nu=K\beta_+\tau-K(K\beta_+-1)$, the rising row of~\eqref{eq:main}.
\end{itemize}
Together with direct alignment at the balanced fractions, these rates achieve every row of~\eqref{eq:main}.

\subsection{Converse}\label{sec:conv}\label{sec:entropy-to-rate}
Let the data be uniform, and use the residual data $\bwr\in\mathcal U_{\mathrm{res}}$ and the matrices $\boldsymbol\Pi_k$ of Section~\ref{sec:geo-converse}. Here $H(\bwr)$ measures how much data remain unspecified once the demand is fixed, and $H(\boldsymbol\Pi_k\bwr)$ measures how much of those data user $k$ still observes. After the messages are fixed as well, communication has reduced both quantities. A lower bound on total observation entropy in terms of data entropy limits how much the encoders can remove. The following lemma performs this conversion for every bound of this form.
\begin{lemma}[An entropy bound gives a sum-rate lower bound]\label{lem:entropy-rate}
For fixed channels, suppose that for some $\zeta>1$ and constant $\gamma_\zeta$, every distribution of $\bwr$ in $\mathcal U_{\mathrm{res}}$ satisfies $\sum_kH(\boldsymbol\Pi_k\bwr)\ge\zeta H(\bwr)+\gamma_\zeta$. Then every achievable sum-rate $R$ for these fixed channels satisfies
\begin{equation}\label{eq:hidden}
R\ge\frac{\zeta\mu-K\nu+\gamma_\zeta}{\zeta-1}.
\end{equation}

\end{lemma}
\begin{proof}
The hypothesis of the lemma concerns a single vector $\bwr\in\mathcal U_{\mathrm{res}}$; a code of block length $L$ produces residual data $\bwr^L=(\bwr_1,\ldots,\bwr_L)\in\mathcal U_{\mathrm{res}}^L$, one vector per instance, and the hypothesis extends to such blocks. Write $\bwr_{<\ell}=(\bwr_1,\ldots,\bwr_{\ell-1})$, and let $\boldsymbol\Pi_k^{[L]}$ apply $\boldsymbol\Pi_k$ separately to each instance. The chain rule and conditioning give, for every distribution of $\bwr^L$ in $\mathcal U_{\mathrm{res}}^L$,
\begin{equation}\label{eq:block-bound}
\sum_kH(\boldsymbol\Pi_k^{[L]}\bwr^L)
\ge\sum_{\ell,k}H(\boldsymbol\Pi_k\bwr_\ell\mid \bwr_{<\ell})
\ge\zeta H(\bwr^L)+L\gamma_\zeta.
\end{equation}
Write $X_{[K]}=(X_1,\ldots,X_K)$ for the message tuple. The observations $\bz_k^L=\mathbf H_k^{[L]}\bw^L$ satisfy the relation equations~\eqref{eq:constraint} in every instance. Hence, as in Section~\ref{sec:geo-converse}, given the messages and the demand, $\mathbf H_k^{[L]}\bw^L=\bz_{k,0}^L+\boldsymbol\Pi_k^{[L]}\bwr^L$ for every $k$, for a fixed tuple $\bz_0^L=(\bz_{1,0}^L,\ldots,\bz_{K,0}^L)$ consistent with them and some $\bwr^L\in\mathcal U_{\mathrm{res}}^L$. Moreover, $\bw^L$ and $\bwr^L$ determine each other, because $[\mathbf H_1^\top\ \cdots\ \mathbf H_K^\top]$ has rank $\mu$, so the observations determine the data. Hence the conditional entropies of $\mathbf H_k^{[L]}\bw^L$ and $\bw^L$ equal those of $\boldsymbol\Pi_k^{[L]}\bwr^L$ and $\bwr^L$, and~\eqref{eq:block-bound} for the conditional distribution of $\bwr^L$ gives
\begin{equation}\label{eq:cond-bound}
\sum_kH(\mathbf H_k^{[L]}\bw^L\mid X_{[K]},\by^L)\ge\zeta H(\bw^L\mid X_{[K]},\by^L)+L\gamma_\zeta.
\end{equation}
The two sides are bounded through the messages. Since each message is a function of its user's observation, uniform data give
\begin{align}
H(\bw^L\mid X_{[K]},\by^L)
&\ge\mu L-\sum_kH(X_k)-H(\by^L\mid X_{[K]}),\label{eq:data-left}\\
\sum_kH(\mathbf H_k^{[L]}\bw^L\mid X_{[K]},\by^L)
&\le\sum_kH(\mathbf H_k^{[L]}\bw^L\mid X_k)
=K\nu L-\sum_kH(X_k).\label{eq:obs-left}
\end{align}
Bounding the left side of~\eqref{eq:cond-bound} from above by~\eqref{eq:obs-left} and its right side from below by~\eqref{eq:data-left}, and rearranging,
\[
(\zeta-1)\sum_kH(X_k)
\ge L(\zeta\mu-K\nu+\gamma_\zeta)-\zeta H(\by^L\mid X_{[K]}).
\]
Use $\sum_kH(X_k)\le LR_\Sigma$ for each code, where $\limsup_{L\to\infty}R_\Sigma\le R$. For the fixed field and matrices, Fano's inequality gives $H(\by^L\mid X_{[K]})/L\to0$ as $L\to\infty$ under vanishing error. Dividing by $L$ and taking the limit proves~\eqref{eq:hidden}.
\end{proof}

\emph{Balanced regime.} By Lemma~\ref{lem:geometry} the whole configuration is balanced, and Corollary~\ref{lem:balanced-entropy} with $\beta=\nu/\kappa$ gives the entropy factor $\zeta_\beta=K\nu/(K\nu-\kappa)=K\nu/(\mu-\nu)$, with $\gamma_\zeta=0$. Lemma~\ref{lem:entropy-rate} then shows that every achievable sum-rate satisfies $R\ge(\zeta_\beta\mu-K\nu)/(\zeta_\beta-1)=K\nu^2/\kappa$, the rate of direct alignment in Section~\ref{sec:ach}. The case $\mu=\nu$ follows from the demand bound $R\ge\nu$, since the server must recover $\nu$ demand symbols per instance.

\emph{Split regime.} Lemma~\ref{lem:split-entropy} gives the sparse-factor and dense-factor bounds, and both hold throughout the regime when $\beta_->1/K$; when $\beta_-=1/K$, the demand bound $R\ge\nu$ replaces the sparse-factor bound. Each, converted by Lemma~\ref{lem:entropy-rate}, gives a lower bound that matches the scheme on one segment of Figure~\ref{fig:snapshot}. The dense factor of Corollary~\ref{lem:balanced-entropy} is $\zeta_+=K\beta_+/(K\beta_+-1)$; when $\beta_->1/K$, the sparse factor is $\zeta_-=K\beta_-/(K\beta_--1)$. Each such factor satisfies
\[
(K\beta-1)(\zeta_\beta-1)=1,\qquad (K\beta-1)\zeta_\beta=K\beta.
\]
\begin{itemize}
\item \emph{Horizontal segment.} The sparse-factor bound has $\zeta=\zeta_-$ and $\gamma_\zeta=-(\zeta_--\zeta_+)(K\beta_+-1)\kappa_+$. Split $\mu=\nu+(K\beta_--1)\kappa_-+(K\beta_+-1)\kappa_+$, the demand plus the two residual dimensions of Corollary~\ref{lem:residual}. Then the numerator of~\eqref{eq:hidden} is
\[
\zeta_-\mu-K\nu+\gamma_\zeta
=\zeta_-\nu+\zeta_-(K\beta_--1)\kappa_-+\zeta_+(K\beta_+-1)\kappa_+-K\nu
=\zeta_-\nu,
\]
since $\zeta_\pm(K\beta_\pm-1)\kappa_\pm=K\beta_\pm\kappa_\pm$ and $K\beta_-\kappa_-+K\beta_+\kappa_+=K\nu$ by~\eqref{eq:blocks}. Hence $R\ge\zeta_-\nu/(\zeta_--1)=K\beta_-\nu$, the horizontal row of Theorem~\ref{thm:main}.
\item \emph{Rising segment.} The dense-factor bound has $\zeta=\zeta_+$ and $\gamma_\zeta=0$. Multiplying the numerator and denominator of~\eqref{eq:hidden} by $K\beta_+-1$ gives a denominator of $1$ and transforms the term $\zeta_+\mu$ into $K\beta_+\mu$, so
\[
R\ge K\beta_+\mu-K(K\beta_+-1)\nu,
\]
which, with $\mu=\tau\nu$, is $\{K\beta_+\tau-K(K\beta_+-1)\}\nu$, the rising row of Theorem~\ref{thm:main}.
\end{itemize}

\subsection{Completing the proof}\label{sec:conv-completion}
Let $C_\star$ denote the claimed capacity in Theorem~\ref{thm:main}. For the lower bound, consider the event that the configuration is balanced or splits as stated in Lemma~\ref{lem:geometry}. This event has probability tending to one, and for every fixed choice of channels in it, the entropy bounds of Corollary~\ref{lem:balanced-entropy} or Lemma~\ref{lem:split-entropy}, converted by Lemma~\ref{lem:entropy-rate}, show that every achievable sum-rate is at least $C_\star$; hence $\Pr\{C_q<C_\star\}=o(1)$. For the upper bound, fix $\epsilon>0$. Section~\ref{sec:ach} applies Lemma~\ref{lem:alignment-rate}, whose alignment events, together with the event above, have probability tending to one, and on them the sum-rate is at most $C_\star+\epsilon$; hence $\Pr\{C_q>C_\star+\epsilon\}=o(1)$. For $K=3$, the consecutive fractions $1/3,1/2,2/3,1$ give the six intervals in the capacity formula; for $K\ge4$, the balanced regime gives the first row of~\eqref{eq:main} and each split regime its two linear rows. The two bounds give $C_q\to C_\star$ in probability, proving Theorem~\ref{thm:main}.

\section{Discussion}\label{sec:discussion}\label{sec:connections}
\emph{Interference alignment.} The separation between linear pieces and a curved central part has a precedent in interference alignment. Subspace alignment chains~\cite{WangGouJafar}, and the closely related alignment paths of~\cite{BreslerCartwrightTse}, explain the piecewise linear degrees-of-freedom curves of the three-user multiple-input multiple-output (MIMO) interference channel. Genie chains~\cite{GenieChains} compare a counting upper bound for one-shot linear alignment with a decomposition inner bound, and develop information-theoretic converses in specified regimes. The balanced and split regimes of Lemma~\ref{lem:geometry} play roles similar to the counting and decomposition pictures, and here they give the capacity in every regime. The two models and their regimes differ, however.

\emph{A many-to-one interference channel.} The same geometry gives the sum-capacity of a many-to-one interference channel~\cite{BreslerParekhTse}, in its finite-field form~\cite{KrishnamurthyJafar}. User $0$ transmits a desired vector $\by\in\Fq^\nu$, and each user $k\in[K]$ transmits its own independent uniform message encoded into $\bz_k\in\Fq^\nu$. With the channels of~\eqref{eq:constraint},
\begin{equation}\label{eq:ff-channel}
\text{receiver }0\text{ observes }\ \mathbf B\by+\sum_{k=1}^K\mathbf A_k\bz_k,
\qquad
\text{receiver }k\text{ observes }\ \bz_k,\quad k\in[K].
\end{equation}
All matrices are known to all terminals and remain fixed across uses. Over $L$ uses, the desired vector is uniform on $\Fq^{\nu L}$ and independent of the messages; receiver~$0$ must recover it and receiver $k$ its message, with error probabilities tending to zero. The objective is the \emph{interference sum-rate} $\sum_kR_k$, where $R_k$ is user $k$'s message rate in $q$-ary symbols per use, and $C_{\mathrm{int}}$ denotes its generic optimum: the limit in probability, as $q\to\infty$, of the largest achievable interference sum-rate, in the same way that $C$ is defined in Section~\ref{sec:generic}.
\begin{corollary}[Interference sum-capacity]\label{cor:channel}
For the parameters of Theorem~\ref{thm:main}, the channel~\eqref{eq:ff-channel} has generic interference sum-capacity $C_{\mathrm{int}}=K\nu-C$. It is approached by zero-error codes with $L=1$.
\end{corollary}
Appendix~\ref{app:channel} proves the corollary by reduction to computation. In computation, the server knows each observation up to an element of the uncertainty set $\mathcal Z_k$, and $\sum_k\log_q|\mathcal Z_k|$ is the communication saved; in the channel, the same sets are codebooks, and the same quantity is the interference sum-rate. For the converse, the entropy bounds of Sections~\ref{sec:conv-tools}--\ref{sec:conv-split} are properties of the configuration and apply unchanged; only Lemma~\ref{lem:entropy-rate} is replaced.

\appendix
\section{Proofs for set alignment}\label{app:reduction}

\subsection{Independent channel entries}
The generic matrices $\mathbf H_1,\ldots,\mathbf H_K,\mathbf D$ have joint row rank $\mu$ with probability $1-O(1/q)$. Their space of row relations therefore has dimension $(K+1)\nu-\mu=\kappa$. A basis of this space gives matrices $\mathbf A_k\in\Fq^{\kappa\times\nu}$ and $\mathbf B\in\Fq^{\kappa\times\nu}$ satisfying
\[
\sum_k\mathbf A_k\mathbf H_k=\mathbf B\mathbf D.
\]
Multiplying by $\bw$ gives $\sum_k\mathbf A_k\bz_k=\mathbf B\by$.

Invertible changes of basis of the data, any user's observation, or the demand preserve capacity: the encoders and decoder can incorporate the corresponding inverse. A change of basis of the row relations only changes coordinates in the relation space. With this in mind, the following lemma justifies using independent random channel entries.
\begin{lemma}[Independent channel entries]\label{lem:constraints}
Up to events of probability $O(1/q)$, the model with independent uniform entries in $\mathbf H_1,\ldots,\mathbf H_K,\mathbf D$ is equivalently described by $\sum_k\mathbf A_k\bz_k=\mathbf B\by$, with independent uniform entries in $\mathbf A_1,\ldots,\mathbf A_K,\mathbf B$. Each $\mathbf A_k$ and $\mathbf B$ has full column rank $\nu$, and $\mathbf A$ has full row rank $\kappa$.
\end{lemma}

\begin{proof}
Recall $\mathbf A=[\mathbf A_1\ \cdots\ \mathbf A_K]$. The standard equal-basis count~\cite[Prop.~1.7.2]{Stanley} transfers uniformity between full-rank matrices and their null spaces; thus choosing a uniform basis of the row relations gives a uniform full-row-rank $[\mathbf A\ {-\mathbf B}]$. All required rank failures have total probability $O(1/q)$ by the Schwartz--Zippel lemma applied to finitely many nonzero minor determinants.
\end{proof}

\subsection{Proof of Lemma~\ref{lem:packing}}
In Example~\ref{ex:interval}, the uncertainty sets are intervals of field elements, and the messages identify consecutive bins. The following proof covers arbitrary finite uncertainty sets by covering the observation space with shifts.
\begin{proof}[Proof of Lemma~\ref{lem:packing}]
Recall $\mathcal Z_k=\mathbf V_k\mathcal T_k$, with $|\mathcal Z_k|=|\mathcal T_k|$ by~\eqref{eq:distinct}. We first cover user $k$'s observation space by shifts of this uncertainty set, then turn the covering into disjoint bins.

\emph{Binning intuition.} The random-binning idea is to describe any one acceptable choice from a larger set~\cite[Lem.~6]{SunJafarCB}. For an observation $\bz_k$, a shift $\boldsymbol s+\mathcal Z_k$ is acceptable if it contains $\bz_k$. There are $q^\nu$ possible shifts, one for each $\boldsymbol s\in\Fq^\nu$, of which exactly $|\mathcal Z_k|$, those with $\boldsymbol s\in\bz_k-\mathcal Z_k$, are acceptable. Choose slightly more than $q^\nu/|\mathcal Z_k|$ shifts at random, known to the encoder and decoder. Since each shift is acceptable with probability $|\mathcal Z_k|/q^\nu$, an acceptable one is likely among them, and the encoder sends its index among the chosen shifts as the bin index. This takes about $\nu-\log_q|\mathcal Z_k|$ $q$-ary symbols. The finite covering bound below gives this rate with $o(1)$ overhead while covering every observation.

\emph{Bins and messages.} The finite covering bound~\cite[Lem.~11.13]{CsiszarKorner} guarantees a collection of at most $q^{\nu+o(1)}/|\mathcal Z_k|$ shifts of $\mathcal Z_k$ covering $\Fq^\nu$. Assign each observation to the bin of the first shift that contains it, and send the bin index; the server then knows the observation up to an element of $\mathcal Z_k$. Two observations giving the same index therefore differ by $\mathbf V_k(\boldsymbol t_k-\boldsymbol t_k')$ for some $\boldsymbol t_k,\boldsymbol t_k'\in\mathcal T_k$.

\emph{Decoding.} If two valid data inputs give the same bin index at every user, let $\Delta\by$ be their demand difference. Subtracting their relation equations and using~\eqref{eq:ia-alignment} gives
\[
\mathbf B\Delta\by
=\sum_k\mathbf A_k\mathbf V_k(\boldsymbol t_k-\boldsymbol t_k')
\in\mathcal I.
\]
Condition~\eqref{eq:separation} forces $\Delta\by=\boldsymbol0$. Thus every message tuple that can occur is compatible with only one demand value, which defines a zero-error decoder.

\emph{Rate.} The number of bins, at most the number of shifts, gives $R_k\le\nu-\log_q|\mathcal Z_k|+o(1)=\nu-\log_q|\mathcal T_k|+o(1)$. Summing over users gives~\eqref{eq:packing-cost}.
\end{proof}

\subsection{Desired-signal independence}
The following lemma gives desired-signal independence~\eqref{eq:separation} from a size bound on the interference set, for an arbitrary demand dimension and an interference set chosen independently of $\mathbf B$.
\begin{lemma}[A counting bound for desired-signal independence]\label{lem:separation}
Let $\mathcal I\subseteq\Fq^\kappa$ be chosen independently of a uniform matrix $\mathbf B\in\Fq^{\kappa\times\nu}$. If $|\mathcal I|=o(q^{\kappa-\nu})$, then desired-signal independence~\eqref{eq:separation} holds with probability $1-o(1)$.
\end{lemma}
\begin{proof}
For each fixed nonzero $\Delta\by$, the vector $\mathbf B\Delta\by$ is uniform in $\Fq^\kappa$. Indeed, after fixing the other entries in a row, the entry multiplying a nonzero coordinate of $\Delta\by$ makes that row product uniform, and the rows are independent. Conditional on $\mathcal I$, its probability of belonging to $\mathcal I$ is $|\mathcal I|/q^\kappa$. A union bound over the $q^\nu-1$ nonzero demand differences gives, for every $q$ and every $\mathcal I$,
\[
\Pr\nolimits_{\mathbf B}\bigl(\exists\,\Delta\by\ne\boldsymbol0:
\mathbf B\Delta\by\in\mathcal I\mid\mathcal I\bigr)
\le\frac{(q^\nu-1)|\mathcal I|}{q^\kappa}=o(1).
\]
The event on the left is exactly the failure of~\eqref{eq:separation}, so~\eqref{eq:separation} holds with probability $1-o(1)$.
\end{proof}

\section{Asymptotic set alignment over all field sequences}\label{app:alignment}
We prove Lemma~\ref{lem:alignment-rate}. Compared with Example~\ref{ex:count}, the user channels now have $\kappa_0$ rows and $\nu_0$ columns, the independent desired channel has $\nu_{\mathrm d}$ columns, and the field is an arbitrary $\Fq$ with $q=p^n$. The proof also covers the setting of Section~\ref{sec:ach}, after its changes of coordinates and with an extra term on the demand side that the server already knows, such as $\mathbf B^+_1\by_1$ in Section~\ref{sec:ach}; this term cancels when two inputs with the same messages are compared, so the decoding argument of Lemma~\ref{lem:packing} is unchanged. The monomial construction is the same; the additional steps are to handle the resulting channel entries and to choose coefficient alphabets for all $p$ and $n$. After a common construction, the proof treats bounded and large $n$ as two separate cases.
\begin{proof}[Proof of Lemma~\ref{lem:alignment-rate}]
If $\nu_{\mathrm d}=0$, send nothing. If $\nu_{\mathrm d}=\kappa_0$, send all observations; this works when the square desired channel is nonsingular, an event of probability $1-O(1/q)$. Otherwise, for the prescribed $\epsilon>0$, choose $\alpha$ such that
\begin{equation}\label{eq:alpha}
0<\alpha<1-\frac{\nu_{\mathrm d}}{\kappa_0},\qquad
K\nu_0(1-\alpha)<K\nu_0\frac{\nu_{\mathrm d}}{\kappa_0}+\frac\epsilon2.
\end{equation}

\emph{Polynomial channel entries.} In Example~\ref{ex:count}, the six random user-channel entries enter the channels directly. Here the changes of coordinates can instead produce ratios of polynomials in the original independent user-channel entries $a_1,\ldots,a_\chi$. Indeed, Gaussian elimination divides by nonzero entries or, equivalently, by determinants of nonsingular submatrices. Fix a choice of such submatrices. Multiplying the relation coordinates by a common nonzero denominator clears all these divisions and preserves the relation equation and its decoding conditions. The user-channel entries are then polynomials of degree at most a fixed $\omega_0$ in each variable. They are reductions modulo $p$ of integer polynomials whose coefficient magnitudes are bounded by the fixed system dimensions.

All these operations depend only on the user-channel entries, so the desired channel remains uniform conditional on them: multiplying a uniform matrix on the left by a fixed invertible matrix, and taking a block of its rows, both preserve uniformity. Only finitely many choices of the nonzero submatrices occur, and the estimates below hold for each choice under the original independent entries; summing over these choices gives the stated bounds.

\emph{Common construction.} Write $q=p^n$. For an integer $\eta\ge1$ and an alphabet $\mathcal T_0\subseteq\Fp^{\eta^\chi}$, chosen below, let $\boldsymbol v_\eta^\top\in\Fq^{1\times\eta^\chi}$ list the $\eta^\chi$ monomials in $a_1,\ldots,a_\chi$ with each exponent in $\{0,\ldots,\eta-1\}$, and let $\mathcal Z_0=\boldsymbol v_\eta^\top\mathcal T_0\subseteq\Fq$. Each user has $\nu_0$ coordinates; apply the same row separately to each:
\[
\mathbf V_k=\operatorname{diag}(\boldsymbol v_\eta^\top,\ldots,\boldsymbol v_\eta^\top)
\in\Fq^{\nu_0\times\nu_0\eta^\chi},\qquad
\mathcal T_k=\mathcal T_0^{\nu_0}\subseteq\Fp^{\nu_0\eta^\chi},\qquad
\mathbf V_k\mathcal T_k=\mathcal Z_0^{\nu_0}.
\]
The coefficients lie in $\Fp\subseteq\Fq$, while all monomial evaluations, sums, and products take place in $\Fq$. Whatever $\eta$ and $\mathcal T_0$ are, the three conditions of Lemma~\ref{lem:packing} reduce to the following.
\begin{itemize}
\item \emph{Interference alignment~\eqref{eq:ia-alignment}.} As in Example~\ref{ex:count}, multiplication only enlarges the monomial list: a channel polynomial increases each exponent by at most $\omega_0$, so every relation coordinate uses monomials with exponents in $\{0,\ldots,\eta+\omega_0-1\}^\chi$. Let $\mathcal I_0\subseteq\Fq$ contain their possible sums after adding all coefficient differences. Then $\sum_k\mathbf A_k\mathbf V_k(\mathcal T_k-\mathcal T_k)\subseteq\mathcal I=\mathcal I_0^{\kappa_0}$.
\item \emph{Desired-signal independence~\eqref{eq:separation}.} The set $\mathcal I$ depends only on the user-channel entries. If $\log_q|\mathcal I_0|$ stays below $1-\nu_{\mathrm d}/\kappa_0$ by a fixed positive margin, then $|\mathcal I_0|^{\kappa_0}=o(q^{\kappa_0-\nu_{\mathrm d}})$, and the counting bound in the proof of Lemma~\ref{lem:separation}, applied conditionally on the user-channel entries, gives failure probability at most $(q^{\nu_{\mathrm d}}-1)|\mathcal I_0|^{\kappa_0}/q^{\kappa_0}=o(1)$.
\item \emph{Distinct uncertainty values~\eqref{eq:distinct}.} It suffices that different arrays in $\mathcal T_0$ give different scalar values under $\boldsymbol v_\eta^\top$, since the same row is applied to every coordinate of every user. A nonzero difference array $\Delta\boldsymbol t\in\mathcal T_0-\mathcal T_0$, fixed independently of the user-channel entries, gives the value $\boldsymbol v_\eta^\top\Delta\boldsymbol t$, a polynomial in $a_1,\ldots,a_\chi$ whose coefficients are the entries of $\Delta\boldsymbol t$; it is therefore a nonzero polynomial, of degree at most $\chi(\eta-1)$, which vanishes with probability at most $\chi(\eta-1)/q$ by the Schwartz--Zippel lemma. A union bound over the difference arrays gives the failure probability.
\item \emph{Rate.} By Lemma~\ref{lem:packing}, $R_\Sigma\le K\nu_0-K\log_q|\mathcal T_k|+o(1)$, with $L=1$ and zero error whenever the three conditions hold.
\end{itemize}

\emph{The threshold $n_0$.} For $n\ge1$, let $\eta_n=\lfloor(\alpha n)^{1/\chi}\rfloor$. Then $\eta_n^\chi/n\le\alpha$, and both $\eta_n^\chi/n$ and $(\eta_n+\omega_0)^\chi/n$ tend to $\alpha$ as $n\to\infty$. Hence there is $n_0$ such that, for every $n>n_0$, $\eta_n\ge1$ and
\begin{equation}\label{eq:n0}
\frac{(\eta_n+\omega_0)^\chi}{n}<1-\frac{\nu_{\mathrm d}}{\kappa_0},\qquad
K\nu_0\Big(\alpha-\frac{\eta_n^\chi}{n}\Big)<\frac\epsilon4,
\end{equation}
In the first inequality the left side tends to $\alpha$, which is below $1-\nu_{\mathrm d}/\kappa_0$ by~\eqref{eq:alpha}, so the gap stays above a fixed positive margin, as desired-signal independence requires; the second inequality involves the fixed $\epsilon/4$ and needs no further margin. Every sequence $q\to\infty$ splits into at most two subsequences, with $n\le n_0$ and with $n>n_0$, so it suffices to treat the two cases separately.

\emph{Case~1: $n\le n_0$.} Then $q\to\infty$ forces $p\to\infty$, so long intervals of integers fit in $\Fp$ without wrapping around. Fix one $\eta$ with $\eta^\chi>n_0\alpha$ and $\alpha(1+\omega_0/\eta)^\chi<1-\nu_{\mathrm d}/\kappa_0$, possible by the first inequality in~\eqref{eq:alpha}, and use the interval alphabet of Example~\ref{ex:count}:
\[
\mathcal T_0=\{0,\ldots,\sigma-1\}^{\eta^\chi},\qquad\sigma=\lfloor q^{\alpha/\eta^\chi}\rfloor.
\]
Since $n\le n_0$ and $\eta^\chi>n_0\alpha$, we have $\sigma\le p^{n\alpha/\eta^\chi}$ with $n\alpha/\eta^\chi<1$, so $\sigma/p\to0$ and $2\sigma<p$ eventually.
\begin{itemize}
\item \emph{Interference alignment.} In each enlarged monomial, the coefficient is a sum of a fixed number of coefficient differences multiplied by bounded integer coefficients of the channel polynomials, so it has $O(\sigma)$ possible values; reduction modulo $p$ can only decrease this count. With at most $(\eta+\omega_0)^\chi$ monomials, $\log_q|\mathcal I_0|\le\alpha(\eta+\omega_0)^\chi/\eta^\chi+o(1)$, which by the choice of $\eta$ is below $1-\nu_{\mathrm d}/\kappa_0$ by a fixed positive margin. Desired-signal independence follows.
\item \emph{Distinct uncertainty values.} Since $2\sigma<p$, a nonzero integer difference array stays nonzero modulo $p$. There are at most $(2\sigma-1)^{\eta^\chi}$ such arrays, so the failure probability is at most $\chi\eta(2\sigma)^{\eta^\chi}/q\le\chi\eta2^{\eta^\chi}q^{\alpha-1}=o(1)$, since $\eta$ is fixed.
\item \emph{Rate.} Since $\eta$ is fixed and $q\to\infty$, rounding gives $\log_q\sigma=\alpha/\eta^\chi+o(1)$, so $\log_q|\mathcal T_k|=\nu_0\eta^\chi\log_q\sigma=\nu_0\alpha+o(1)$. Hence $R_\Sigma\le K\nu_0(1-\alpha)+o(1)<K\nu_0\nu_{\mathrm d}/\kappa_0+\epsilon/2+o(1)$ by the second inequality in~\eqref{eq:alpha}.
\end{itemize}

\emph{Case~2: $n>n_0$.} Then $p$ may stay bounded, so we cannot rely on long intervals. Take $\eta=\eta_n$ and use the whole prime field:
\[
\mathcal T_0=\Fp^{\eta^\chi}.
\]
\begin{itemize}
\item \emph{Interference alignment.} Multiplication and addition keep every coefficient in $\Fp$, so $\mathcal I_0$ lies in the $\Fp$-span of at most $(\eta+\omega_0)^\chi$ monomials: $|\mathcal I_0|\le p^{(\eta+\omega_0)^\chi}$ and $\log_q|\mathcal I_0|\le(\eta+\omega_0)^\chi/n$, which by~\eqref{eq:n0} is below $1-\nu_{\mathrm d}/\kappa_0$ by a fixed positive margin. Desired-signal independence follows.
\item \emph{Distinct uncertainty values.} There are fewer than $p^{\eta^\chi}$ nonzero difference arrays. Since $q=p^n$ and $\eta^\chi\le\alpha n$, we have $p^{\eta^\chi}=q^{\eta^\chi/n}\le q^\alpha$, so the failure probability is at most $\chi\eta p^{\eta^\chi}/q\le\chi\eta q^{\alpha-1}$. This is $o(1)$ because $\alpha<1$ and $\eta\le(\alpha n)^{1/\chi}\le(\alpha\log_2q)^{1/\chi}$ grows only like a power of $\log q$.
\item \emph{Rate.} Since $|\mathcal T_k|=p^{\nu_0\eta^\chi}$ and $\log_qp=1/n$, $\log_q|\mathcal T_k|=\nu_0\eta^\chi/n$. Writing $1-\eta^\chi/n=(1-\alpha)+(\alpha-\eta^\chi/n)$ gives $R_\Sigma\le K\nu_0(1-\alpha)+K\nu_0(\alpha-\eta^\chi/n)+o(1)<K\nu_0\nu_{\mathrm d}/\kappa_0+\epsilon/2+\epsilon/4+o(1)$, where the first term is bounded by the second inequality in~\eqref{eq:alpha} and the second, the rounding loss from $\eta^\chi\le\alpha n$, by the second inequality in~\eqref{eq:n0}.
\end{itemize}
Both constructions keep the given field $\Fq$; they differ only in the coefficient alphabet.

\emph{All field sequences.} In both cases the three conditions hold with probability $1-o(1)$, with error terms depending only on $q$, and the bin-index overhead in Lemma~\ref{lem:packing} is $o(1)$. Hence, along every sequence $q=p^n\to\infty$, with probability tending to one, the rate is at most $K\nu_0\nu_{\mathrm d}/\kappa_0+\epsilon$ once $q$ is large enough that the $o(1)$ terms are below $\epsilon/4$. This proves~\eqref{eq:unsplit-general}.
\end{proof}

\section{Proofs of Lemmas~\ref{lem:null-balance} and~\ref{lem:geometry}}\label{app:geometry}
This appendix proves Lemmas~\ref{lem:null-balance} and~\ref{lem:geometry} using only the user channels, with probability bounds uniform in the characteristic. Appendix~\ref{app:null-balance} proves Lemma~\ref{lem:null-balance}. For Lemma~\ref{lem:geometry}, Appendix~\ref{app:central} proves balance on the balanced interval, Appendix~\ref{app:elementary} introduces the tools of the reduction, Appendix~\ref{app:split} follows the reduction chain to split every configuration outside the balanced interval into two balanced components, and Appendix~\ref{app:geometry-proof} combines the steps.

\subsection{Balance on the null space}\label{app:null-balance}
\begin{proof}[Proof of Lemma~\ref{lem:null-balance}]
Write~\eqref{eq:null-balance} as $\dim(\mathcal U)\le(1-1/(K\beta))\sum_k\dim(\boldsymbol\Pi_k\mathcal U)$, and balance~\eqref{eq:balance} as $\sum_k\dim(\mathcal W\cap\mathcal A_k)\le K\beta\dim(\mathcal W)$. The proof of each implication, balance to~\eqref{eq:null-balance} and~\eqref{eq:null-balance} to balance, uses the following dimension count. For subspaces $\mathcal P_k\le\Fq^\nu$, let
\[
\mathcal U_{\mathcal P}=\{\bz\in\nullspace(\mathbf A):\boldsymbol\Pi_k\bz\in\mathcal P_k\text{ for every }k\},
\]
the null space of $\mathbf A$ restricted to $\mathcal P_1\times\cdots\times\mathcal P_K$. This restriction has column space $\sum_k\mathbf A_k\mathcal P_k$, and $\dim(\mathbf A_k\mathcal P_k)=\dim(\mathcal P_k)$ by full column rank, so
\[
\dim(\mathcal U_{\mathcal P})=\sum_k\dim(\mathcal P_k)-\dim\Big(\sum_k\mathbf A_k\mathcal P_k\Big).
\]

\emph{Balance implies~\eqref{eq:null-balance}.} Fix $\mathcal U\le\nullspace(\mathbf A)$, and let $\mathcal P_k=\boldsymbol\Pi_k\mathcal U$ and $\mathcal W=\sum_k\mathbf A_k\mathcal P_k$. Then $\mathcal U\le\mathcal U_{\mathcal P}$, and balance at $\mathcal W$, with $\mathbf A_k\mathcal P_k\le\mathcal W\cap\mathcal A_k$, gives $\dim(\mathcal W)\ge\sum_k\dim(\mathcal P_k)/(K\beta)$. Hence
\[
\dim(\mathcal U)\le\sum_k\dim(\mathcal P_k)-\dim(\mathcal W)\le\Big(1-\frac1{K\beta}\Big)\sum_k\dim(\boldsymbol\Pi_k\mathcal U),
\]
which is~\eqref{eq:null-balance}.

\emph{\eqref{eq:null-balance} implies balance.} Fix $\mathcal W\le\Fq^\kappa$, and let $\mathcal P_k=\{\bz\in\Fq^\nu:\mathbf A_k\bz\in\mathcal W\}$, so $\dim(\mathcal P_k)=\dim(\mathcal W\cap\mathcal A_k)$ and $\sum_k\mathbf A_k\mathcal P_k\le\mathcal W$. Apply~\eqref{eq:null-balance} to $\mathcal U=\mathcal U_{\mathcal P}$, for which $\boldsymbol\Pi_k\mathcal U_{\mathcal P}\le\mathcal P_k$:
\[
\Big(1-\frac1{K\beta}\Big)\sum_k\dim(\mathcal P_k)\ge\dim(\mathcal U_{\mathcal P})\ge\sum_k\dim(\mathcal P_k)-\dim(\mathcal W).
\]
Rearranging gives $\sum_k\dim(\mathcal W\cap\mathcal A_k)\le K\beta\dim(\mathcal W)$.
\end{proof}

\subsection{Balance on the balanced interval}\label{app:central}
The condition $K\beta(1-\beta)\ge1$ below, with per-user fraction $\beta=\nu/\kappa$, is the balanced interval $[\beta_\infty,1-\beta_\infty]$ of Section~\ref{sec:balanced-fractions}, which corresponds to $\tau_-\le\tau\le\tau_+$. The proof bounds, by counting subspaces, the probability that balance fails, and shows that it tends to zero in this range.
\begin{lemma}[A sufficient condition for generic balance]\label{lem:central}
If $\beta=\nu/\kappa$ satisfies $K\beta(1-\beta)\ge1$, independent uniform subspaces of dimension $\nu$ in $\Fq^\kappa$ satisfy~\eqref{eq:balance} with probability $1-O(1/q)$, uniformly in characteristic.
\end{lemma}
\begin{proof}
Balance~\eqref{eq:balance} fails exactly when, for some $1\le d\le\kappa-1$ and some integers $0\le r_k\le\min\{d,\nu\}$ with $\sum_kr_k>K\beta d$, a $d$-dimensional subspace $\mathcal W$ satisfies $\dim(\mathcal W\cap\mathcal A_k)\ge r_k$ for all $k$. There are finitely many such tuples $(d,r_1,\ldots,r_K)$, depending only on the dimensions, so it suffices to show that each has probability $O(1/q)$.

The standard Gaussian-binomial estimate~\cite[Prop.~1.7.2]{Stanley} gives $O(q^{d(\kappa-d)})$ subspaces of $\Fq^\kappa$ of dimension $d$, with constants depending only on the dimensions, uniformly in characteristic.

Fix a $d$-dimensional subspace $\mathcal W$. There are $O(q^{r(d-r)})$ subspaces of $\mathcal W$ of dimension $r$, and a uniform $\nu$-dimensional subspace contains each one with probability $O(q^{-r(\kappa-\nu)})$. A union bound gives
\[
\Pr\bigl(\dim(\mathcal W\cap\mathcal A_k)\ge r\bigr)=O\bigl(q^{-r(\kappa-\nu-d+r)}\bigr).
\]
By independence across users and a union bound over $\mathcal W$, the probability that some $d$-dimensional $\mathcal W$ satisfies $\dim(\mathcal W\cap\mathcal A_k)\ge r_k$ for all $k$ is $O(q^{-E})$, where
\[
E=\sum_k r_k(\kappa-\nu-d+r_k)-d(\kappa-d).
\]

Using $\nu=\beta\kappa$, direct expansion gives
\[
E=\sum_k(r_k-\beta d)^2
+\Bigl(\sum_kr_k-K\beta d\Bigr)\bigl[(1-\beta)(\kappa-d)+\beta d\bigr]
+d(\kappa-d)\bigl[K\beta(1-\beta)-1\bigr].
\]
Every term is nonnegative because $K\beta(1-\beta)\ge1$. Moreover, $\sum_kr_k-K\beta d>0$, and its coefficient is positive because $0<\beta<1$ and $0<d<\kappa$, since $d=0$ and $d=\kappa$ satisfy balance. Thus $E>0$, even at the boundary $K\beta(1-\beta)=1$, and since $E$ is an integer, $E\ge1$. So each tuple has probability $O(1/q)$, and balance holds with probability $1-O(1/q)$, uniformly in characteristic.
\end{proof}

\subsection{Base configurations and two balance-preserving transformations}\label{app:elementary}
We reduce a configuration to base configurations using alignments among the user subspaces and orthogonal complements, and then reverse these steps to obtain its decomposition. We first describe the base configurations, where the reduction stops, and show that configurations with $\kappa\ge(K-1)\nu$ are direct sums of them. For $\kappa<(K-1)\nu$, we then describe the alignment transformation, the main reduction step, a recursive change of basis in the spirit of subspace alignment chains~\cite{WangGouJafar} and genie chains~\cite{GenieChains}; by Corollary~\ref{lem:relations-balance}, a consequence of Lemma~\ref{lem:null-balance}, a configuration is balanced exactly when its alignment configuration is. Finally, Lemma~\ref{lem:sum-perp} shows that orthogonal complements and direct sums preserve balance.

\Needspace{5\baselineskip}
\emph{Base configurations.} The \emph{base configurations} are the following three, with per-user fractions $0$, $1/K$, and $1/(K-1)$.
\begin{itemize}
\item \emph{Empty:} the users occupy nothing in the space.
\item \emph{Independent:} $K$ independent user subspaces, each user occupying its own dimensions, with no overlap.
\item \emph{Simplex:} $K$ user subspaces of dimension $\nu$ in a space of dimension $(K-1)\nu$, of which any $K-1$ are independent.
\end{itemize}
All three are balanced. This is clear for the empty and independent ones. In the simplex one, any $K-1$ user subspaces are independent, so for each $k$ and every $\mathcal W$,
\[
\sum_{i\ne k}\dim(\mathcal W\cap\mathcal A_i)\le\dim(\mathcal W)
\quad\Longrightarrow\quad
\sum_i\dim(\mathcal W\cap\mathcal A_i)\le\frac{K}{K-1}\dim(\mathcal W),
\]
where the implication follows by summing over $k$.

\emph{Configurations with $\kappa\ge(K-1)\nu$.} These are generically direct sums of base configurations, as follows. An \emph{alignment} is a tuple $(\bz_1,\ldots,\bz_K)$ of local vectors whose signals cancel, $\sum_k\mathbf A_k\bz_k=\boldsymbol0$; the alignments form $\nullspace(\mathbf A)$. The decomposition follows from three properties, which hold generically.
\begin{itemize}
\item \emph{Only $K$-way alignments.} Since $(K-1)\nu\le\kappa$, any $K-1$ user subspaces are generically independent, so every nonzero alignment involves all $K$ users.
\item \emph{Dimension of the alignment space.} Generically, $\dim(\nullspace(\mathbf A))=\max\{K\nu-\kappa,0\}$. If $\kappa\ge K\nu$, the $K$ user subspaces are independent and span $K\nu$ dimensions, and a complement of dimension $\kappa-K\nu$ is empty.
\item \emph{Components, for $\kappa<K\nu$.} By the first item, every nonzero alignment has a nonzero share $\boldsymbol\Pi_k\bz$ for each user $k$, so user $k$'s part of the alignments, $\mathcal A_k^{\mathrm{al}}$, has dimension $K\nu-\kappa$, and a complement $\mathcal A_k^{\mathrm{ind}}$ in $\mathcal A_k$ has dimension $\kappa-(K-1)\nu$. Every alignment has its $k$th signal in $\mathcal A_k^{\mathrm{al}}$, so, since $\mathcal A_k=\mathcal A_k^{\mathrm{al}}\oplus\mathcal A_k^{\mathrm{ind}}$, its part in $\mathcal A_k^{\mathrm{ind}}$ is zero. Hence the parts $\mathcal A_k^{\mathrm{ind}}$ have no overlap with one another or with the span of the parts $\mathcal A_k^{\mathrm{al}}$, and they span an independent component of dimension $K(\kappa-(K-1)\nu)$. The parts $\mathcal A_k^{\mathrm{al}}$ carry all alignments: their $K(K\nu-\kappa)$ dimensions carry an alignment space of dimension $K\nu-\kappa$, so they span a simplex component of dimension $(K-1)(K\nu-\kappa)$. The two dimensions add up to $\kappa$.
\end{itemize}
Writing a configuration by its dimensions $(\kappa,\nu)$, this gives
\begin{equation}\label{eq:base-table}
\begin{aligned}
&\kappa\ge K\nu: && (\kappa,\nu)=(K\nu,\nu)\oplus(\kappa-K\nu,0),\\
&(K-1)\nu\le\kappa\le K\nu: && (\kappa,\nu)=\bigl(K(\kappa-(K-1)\nu),\ \kappa-(K-1)\nu\bigr)\\
& && \phantom{(\kappa,\nu)={}}\oplus\bigl((K-1)(K\nu-\kappa),\ K\nu-\kappa\bigr),
\end{aligned}
\end{equation}
where the first summand is independent, with per-user fraction $1/K$, and the second is empty, with fraction $0$, or simplex, with fraction $1/(K-1)$; all are balanced.

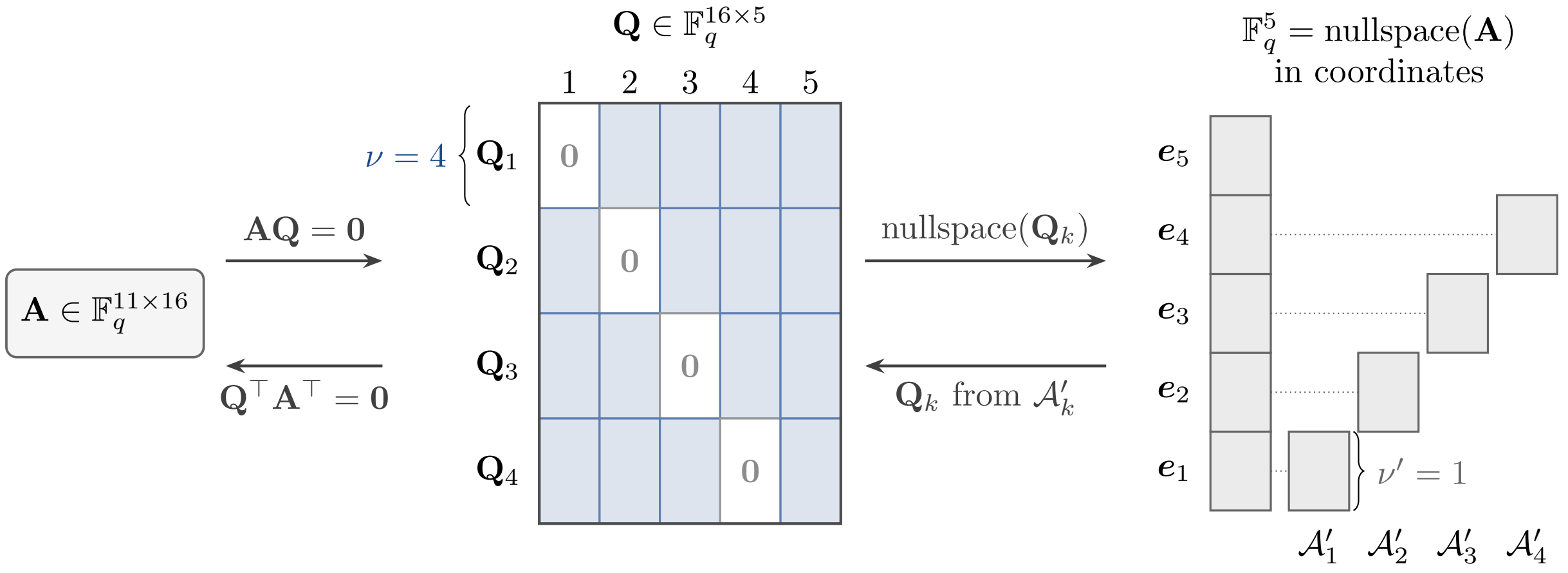
\begin{figure}[htbp]
\centering
\begin{tikzpicture}[x=0.62cm,y=0.27cm,>=Stealth,
  nz/.style={fill=capblue!15,draw=capblue!70,line width=.6pt},
  al/.style={fill=black!8,draw=black!60,line width=.6pt},
  zero/.style={fill=white,draw=black!40,line width=.6pt},
  box/.style={draw=black!60,fill=black!4,rounded corners=3pt,line width=.7pt,minimum height=0.9cm,align=center,inner sep=4pt,font=\small},
  lab/.style={font=\small}]
\node[box] (A) at (-7.2,8) {$\mathbf A\in\Fq^{11\times16}$};
\foreach \k in {1,...,4}{
  \pgfmathsetmacro{\yt}{16-4*(\k-1)}
  \foreach \j in {1,...,5}{
    \ifnum\j=\k
      \draw[zero] (\j-1,\yt-4) rectangle (\j,\yt);
      \node[lab,black!45] at (\j-0.5,\yt-2) {$\boldsymbol0$};
    \else
      \draw[nz] (\j-1,\yt-4) rectangle (\j,\yt);
    \fi
  }
  \node[lab,anchor=east] at (-0.1,\yt-2) {$\mathbf Q_{\k}$};
}
\draw[decorate,decoration={brace,amplitude=3pt}] (-1.15,12.1) -- (-1.15,15.9) node[midway,left=3pt,lab,capblue]{$\nu=4$};
\draw[black!70,line width=.8pt] (0,0) rectangle (5,16);
\foreach \j in {1,...,5}{\node[lab] at (\j-0.5,16.8) {$\j$};}
\node[lab,anchor=south] at (2.5,17.6) {$\mathbf Q\in\Fq^{16\times5}$};
\begin{scope}[xshift=6.9cm]
\node[lab,anchor=south,align=center] at (2.8,16.3) {$\Fq^5=\nullspace(\mathbf A)$\\in coordinates};
\foreach \r in {1,...,5}{
  \draw[al] (0,3*\r-2.5) rectangle (1,3*\r+0.5);
  \node[lab,anchor=east] at (-0.1,3*\r-1) {$\boldsymbol e_{\r}$};
}
\foreach \u in {1,...,4}{
  \pgfmathsetmacro{\x}{0.15+\u*1.15}
  \draw[black!60,densely dotted] (1,3*\u-1) -- (\x,3*\u-1);
  \draw[al] (\x,3*\u-2.5) rectangle (\x+1,3*\u+0.5);
  \node[lab,anchor=north] at (\x+0.5,0.3) {$\mathcal A_{\u}'$};
}
\end{scope}
\draw[decorate,decoration={brace,amplitude=3pt,mirror}] (13.5,0.5) -- (13.5,3.5) node[midway,right=3pt,lab,black!60]{$\nu'=1$};
\draw[-{Stealth[length=2mm]},line width=.8pt,black!75] (-5.2,10) -- node[lab,above]{$\mathbf A\mathbf Q=\mathbf 0$} (-2.6,10);
\draw[{Stealth[length=2mm]}-,line width=.8pt,black!75] (-5.2,6) -- node[lab,below]{$\mathbf Q^\top\mathbf A^\top=\mathbf 0$} (-2.6,6);
\draw[-{Stealth[length=2mm]},line width=.8pt,black!75] (5.4,10) -- node[lab,above]{$\nullspace(\mathbf Q_k)$} (9.4,10);
\draw[{Stealth[length=2mm]}-,line width=.8pt,black!75] (5.4,6) -- node[lab,below]{$\mathbf Q_k$ from $\mathcal A_k'$} (9.4,6);
\end{tikzpicture}
\caption{The alignment transformation for $K=4$ and $(\kappa,\nu)=(11,4)$, the configuration of Example~\ref{ex:reduction}. Blue marks the users' local coordinates, $\nu=4$ per user, and gray the alignment space $\nullspace(\mathbf A)$, in which each user becomes a line, $\nu'=1$. Forward (upper arrows): $\mathbf A\mathbf Q=\mathbf 0$ gives a basis $\mathbf Q$ of the $5$-dimensional alignment space, with row blocks $\mathbf Q_k=\boldsymbol\Pi_k\mathbf Q$, and $\mathcal A_k'=\nullspace(\mathbf Q_k)$. Any three users have a $1$-dimensional space of alignments among themselves. Choose a nonzero alignment from each of these four one-dimensional spaces as columns $1$ to $4$ of $\mathbf Q$. Then column $j$ involves every user except $j$, so $\mathbf Q_j$ has a zero block in column $j$ (white) and $\mathcal A_j'=\col(\boldsymbol e_j)$, as the dotted lines indicate. Reverse (lower arrows): from the lines $\mathcal A_k'$, choose $\mathbf Q_k$ with these null spaces, and $\mathbf A^\top$ from $\mathbf Q^\top\mathbf A^\top=\mathbf 0$; by~\eqref{eq:reverse-dims}, this returns $(11,4)$. In the proof, the reverse construction is applied not to this configuration directly but to each component after the base decomposition (Example~\ref{ex:reduction}).}
\label{fig:alignment}
\end{figure}

\emph{Configurations with $\kappa<(K-1)\nu$: alignments among the user subspaces.} Consider such a configuration below the balanced interval. We build a new configuration on the null space $\nullspace(\mathbf A)$ of Lemma~\ref{lem:null-balance}, with the same matrices $\boldsymbol\Pi_k$ (Figure~\ref{fig:alignment}).
\begin{itemize}
\item \emph{The alignment configuration.} In $\nullspace(\mathbf A)$, user $k$'s subspace is $\mathcal A_k'=\{\bz\in\nullspace(\mathbf A):\boldsymbol\Pi_k\bz=\boldsymbol0\}$, and the \emph{alignment configuration} is $(\nullspace(\mathbf A);\mathcal A_1',\ldots,\mathcal A_K')$. In coordinates, choose $\mathbf Q\in\Fq^{K\nu\times\kappa'}$ of full column rank with $\mathbf A\mathbf Q=\mathbf 0$, so that its columns form a basis of $\nullspace(\mathbf A)$, of dimension $\kappa'$. Its user blocks are $\mathbf Q_k=\boldsymbol\Pi_k\mathbf Q$, and in these coordinates $\mathcal A_k'$ is $\nullspace(\mathbf Q_k)\le\Fq^{\kappa'}$.
\item \emph{Its dimensions.} Generically, $\mathbf A$ has full row rank, so $\kappa'=K\nu-\kappa$, and any $K-1$ user subspaces span $\Fq^\kappa$, since $(K-1)\nu>\kappa$. Then every $\bz_k\in\Fq^\nu$ is user $k$'s entry of some element of $\nullspace(\mathbf A)$, so $\mathbf Q_k$ has full row rank $\nu$, and $\dim(\mathcal A_k')=\kappa'-\nu$. The alignment configuration is thus a configuration of $K$ subspaces of dimension $\nu'$ in a space of dimension $\kappa'$, with
\begin{equation}\label{eq:forward-dims}
\kappa'=K\nu-\kappa,\qquad \nu'=(K-1)\nu-\kappa,
\end{equation}
and $\nu'<\nu$ below the balanced interval. To see this, note that by~\eqref{eq:forward-dims}, $\nu'<\nu$ is equivalent to $\nu/\kappa<1/(K-2)$. Below the balanced interval $\nu/\kappa<\beta_\infty$, so it suffices that $\beta_\infty\le1/(K-2)$. For $K\ge4$, $1/(K-2)\le1/2$, and $K\beta(1-\beta)$, which is increasing for $\beta\le1/2$, equals $K(K-3)/(K-2)^2\ge1$ at $\beta=1/(K-2)$; hence it reaches $1$ at some $\beta_\infty\le1/(K-2)$.
\item \emph{The reverse construction, by transposition.} In Figure~\ref{fig:alignment}, the forward construction starts from $\mathbf A$ and finds $\mathbf Q$ through $\mathbf A\mathbf Q=\mathbf 0$. Transposing $\mathbf A\mathbf Q=\mathbf 0$ gives $\mathbf Q^\top\mathbf A^\top=\mathbf 0$, so the reverse construction is the same step applied to the transposed matrices: it starts from $\mathbf Q$ and finds $\mathbf A^\top$, as in the lower arrows of Figure~\ref{fig:alignment}. Concretely, let $\mathcal A_k'\le\Fq^{\kappa'}$ have dimension $\nu'$ with $\bigcap_{i\ne k}\mathcal A_i'=\{\boldsymbol0\}$ for every $k$. The dimensions are those that~\eqref{eq:forward-dims} produces, so each $\mathcal A_i'$ leaves out $\kappa'-\nu'=\nu$ dimensions of $\Fq^{\kappa'}$, and $K-1$ of them leave out $(K-1)\nu\ge K\nu-\kappa=\kappa'$ dimensions, because $\nu\le\kappa$; hence $K-1$ generic such subspaces have no overlap.

\emph{The result is a valid configuration.} Choose $\mathbf Q_k$ of full row rank with $\nullspace(\mathbf Q_k)=\mathcal A_k'$ and stack them into $\mathbf Q$, which has full column rank since $\bigcap_k\mathcal A_k'=\{\boldsymbol0\}$; then choose $\mathbf A^\top$ whose columns form a basis of $\nullspace(\mathbf Q^\top)$, so that $\nullspace(\mathbf A)=\col(\mathbf Q)$. Each $\mathbf A_k$ has full column rank: if $\mathbf A_k\bz=\boldsymbol0$, the tuple with $\bz$ in block $k$ and zeros elsewhere lies in $\col(\mathbf Q)$ with zero blocks for the other users, so it is $\mathbf Q$ times a vector of $\bigcap_{i\ne k}\mathcal A_i'=\{\boldsymbol0\}$, and $\bz=\boldsymbol0$. The rebuilt configuration also meets the conditions of the forward construction: since each $\mathbf Q_k$ has full row rank, every local vector of user $k$ is the $k$th entry of some element of $\nullspace(\mathbf A)$, so the signals of user $k$ lie in the span of the other users' signals, and these span $\Fq^\kappa$ because $\mathbf A$ has full row rank.

The dimensions follow by inverting~\eqref{eq:forward-dims}: subtracting its two equations gives $\nu=\kappa'-\nu'$, and then $\kappa=K\nu-\kappa'=(K-1)\kappa'-K\nu'$. Together with the resulting per-user fraction,
\begin{equation}\label{eq:reverse-dims}
\begin{gathered}
\kappa=(K-1)\kappa'-K\nu',\qquad \nu=\kappa'-\nu',\\
\frac\nu\kappa=\operatorname{lift}\Big(\frac{\nu'}{\kappa'}\Big),\qquad
\operatorname{lift}(\beta)=\frac{1-\beta}{K-1-K\beta},\qquad
\operatorname{lift}(\beta_i)=\beta_{i+2},
\end{gathered}
\end{equation}
The second row follows from the first: dividing, $\nu/\kappa=(\kappa'-\nu')/((K-1)\kappa'-K\nu')=(1-\beta')/(K-1-K\beta')$ with $\beta'=\nu'/\kappa'$, and the recursion $\beta_{i+1}=1/(K(1-\beta_i))$ gives $\beta_{i+2}=1/(K-1/(1-\beta_i))=(1-\beta_i)/(K-1-K\beta_i)=\operatorname{lift}(\beta_i)$. The map $\operatorname{lift}$ is increasing, and~\eqref{eq:forward-dims} is its inverse, so the pair $(\beta_i,\beta_{i+1})$ reduces to $(\beta_{i-2},\beta_{i-1})$ and lifts back. Applied to each component of a direct sum, the reverse construction lifts the sum componentwise.
\end{itemize}
The two constructions are inverse up to changes of basis (Figure~\ref{fig:alignment} shows an example), and after choosing nonzero minors (determinants of square submatrices), both use rational functions of the matrix entries. The proof uses them in sequence: it applies the forward construction until a base configuration is reached, decomposes it by~\eqref{eq:base-table}, and applies the reverse construction to each component the same number of times (Appendix~\ref{app:split}).

Balance passes through these constructions in both directions.

\Needspace{6\baselineskip}
\begin{corollary}[Alignment configurations preserve balance]\label{lem:relations-balance}
The alignment configuration of $(\Fq^\kappa;\mathcal A_1,\ldots,\mathcal A_K)$ is balanced if and only if $(\Fq^\kappa;\mathcal A_1,\ldots,\mathcal A_K)$ is balanced.
\end{corollary}
\begin{proof}
Lemma~\ref{lem:null-balance} expresses balance of $(\Fq^\kappa;\mathcal A_1,\ldots,\mathcal A_K)$ through the matrices $\boldsymbol\Pi_k$ on $\nullspace(\mathbf A)$: it holds exactly when~\eqref{eq:null-balance} does. The alignment configuration lives on the same space $\nullspace(\mathbf A)$, of dimension $\kappa'$, and its user subspaces $\mathcal A_k'$, of dimension $\nu'$, are the vectors that $\boldsymbol\Pi_k$ sends to $\boldsymbol0$. So it suffices to show that its balance condition,
\[
\sum_k\dim(\mathcal U\cap\mathcal A_k')\le\frac{K\nu'}{\kappa'}\dim(\mathcal U)\qquad\text{for every }\mathcal U\le\nullspace(\mathbf A),
\]
is~\eqref{eq:null-balance}. For such $\mathcal U$, the vectors of $\mathcal U$ that $\boldsymbol\Pi_k$ sends to $\boldsymbol0$ form $\mathcal U\cap\mathcal A_k'$, so the dimensions of $\mathcal U$ split as $\dim(\mathcal U)=\dim(\mathcal U\cap\mathcal A_k')+\dim(\boldsymbol\Pi_k\mathcal U)$. Substituting $\dim(\mathcal U\cap\mathcal A_k')=\dim(\mathcal U)-\dim(\boldsymbol\Pi_k\mathcal U)$, the condition becomes
\[
\sum_k\dim(\boldsymbol\Pi_k\mathcal U)\ge\Big(K-\frac{K\nu'}{\kappa'}\Big)\dim(\mathcal U)=\frac{K\nu}{K\nu-\kappa}\dim(\mathcal U)=\frac{K\beta}{K\beta-1}\dim(\mathcal U),
\]
using $\kappa'-\nu'=\nu$ and $\kappa'=K\nu-\kappa$ from~\eqref{eq:forward-dims} and $\beta=\nu/\kappa$. This is~\eqref{eq:null-balance}.
\end{proof}

\emph{Orthogonal complements and direct sums.} For a subspace $\mathcal U\le\Fq^\kappa$, its orthogonal complement is $\mathcal U^\perp=\{\boldsymbol x\in\Fq^\kappa:\boldsymbol x^\top\boldsymbol u=0\text{ for all }\boldsymbol u\in\mathcal U\}$, the null space of any matrix whose rows span $\mathcal U$. It has dimension $\kappa-\dim(\mathcal U)$, and $(\mathcal U^\perp)^\perp=\mathcal U$.

\begin{lemma}[Direct sums and orthogonal complements preserve balance]\label{lem:sum-perp}
\begin{enumerate}
\item A direct sum of balanced configurations with the same per-user fraction $\beta$ is balanced, with per-user fraction $\beta$.
\item If $(\Fq^\kappa;\mathcal A_1,\ldots,\mathcal A_K)$ is balanced with user dimension $\nu$, then $(\Fq^\kappa;\mathcal A_1^\perp,\ldots,\mathcal A_K^\perp)$ is balanced with user dimension $\kappa-\nu$, so the per-user fraction $\beta$ becomes $1-\beta$.
\item A configuration splits into balanced components with per-user fractions $\beta_-$ and $\beta_+$ if and only if $(\Fq^\kappa;\mathcal A_1^\perp,\ldots,\mathcal A_K^\perp)$ splits into balanced components with per-user fractions $1-\beta_+$ and $1-\beta_-$.
\end{enumerate}
\end{lemma}
\begin{proof}
For the first part, let $\mathcal W\le\mathcal E_1\oplus\mathcal E_2$, let $\mathcal W_1=\mathcal W\cap\mathcal E_1$, and let $\mathcal W_2$ be the projection of $\mathcal W$ onto $\mathcal E_2$, so $\dim(\mathcal W)=\dim(\mathcal W_1)+\dim(\mathcal W_2)$. Each user subspace splits with the components, so projecting $\mathcal W\cap\mathcal A_k$ onto $\mathcal E_2$ gives $\dim(\mathcal W\cap\mathcal A_k)\le\dim(\mathcal W_1\cap\mathcal A_k)+\dim(\mathcal W_2\cap\mathcal A_k)$. Balance of each part bounds the sum over $k$ by $K\beta(\dim(\mathcal W_1)+\dim(\mathcal W_2))=K\beta\dim(\mathcal W)$.

For the second part, $\dim(\mathcal W\cap\mathcal A_k^\perp)=\dim(\mathcal W)-\nu+\dim(\mathcal W^\perp\cap\mathcal A_k)$, so balance of the original configuration, applied to $\mathcal W^\perp$, gives $\sum_k\dim(\mathcal W\cap\mathcal A_k^\perp)\le K(\dim(\mathcal W)-\nu)+(K\nu/\kappa)(\kappa-\dim(\mathcal W))=(K(\kappa-\nu)/\kappa)\dim(\mathcal W)$.

The third part follows from the second, applied to each component after a change of basis. Choose an invertible matrix $\mathbf P$ whose columns are bases of $\mathcal E_-$ and $\mathcal E_+$. In these coordinates, $\mathcal A_k=\mathbf P(\mathcal U_{k,-}\oplus\mathcal U_{k,+})$, where the two coordinate configurations are balanced with fractions $\beta_-$ and $\beta_+$. Since $(\mathbf P\mathcal U)^\perp=\mathbf P^{-\top}\mathcal U^\perp$ for every subspace $\mathcal U$,
\[
\mathcal A_k^\perp=\mathbf P^{-\top}\bigl(\mathcal U_{k,-}^\perp\oplus\mathcal U_{k,+}^\perp\bigr),
\]
with each complement taken in its own coordinate block. By the second part, these two configurations are balanced with fractions $1-\beta_-$ and $1-\beta_+$. Invertible linear maps preserve balance and direct sums, so the complemented configuration has the required split, with the sparse and dense roles exchanged. The converse follows from $(\mathcal A_k^\perp)^\perp=\mathcal A_k$.
\end{proof}

\Needspace{8\baselineskip}
\subsection{Split between consecutive balanced fractions}\label{app:split}
Let $\nu/\kappa$ lie in $[\beta_j,\beta_{j+1}]$ for consecutive elements $\beta_j<\beta_{j+1}$ of the sequence of Theorem~\ref{thm:main}; above the balanced interval, orthogonal complements reduce to this case (Lemma~\ref{lem:sum-perp}). By~\eqref{eq:reverse-dims}, an alignment step moves the per-user fraction into $[\beta_{j-2},\beta_{j-1}]$, so after $\lfloor j/2\rfloor$ steps the configuration has $\kappa\ge(K-1)\nu$ and is a direct sum of two base configurations by~\eqref{eq:base-table}. Reversing the steps carries this split back to the original configuration, with exactly two balanced components at every step.

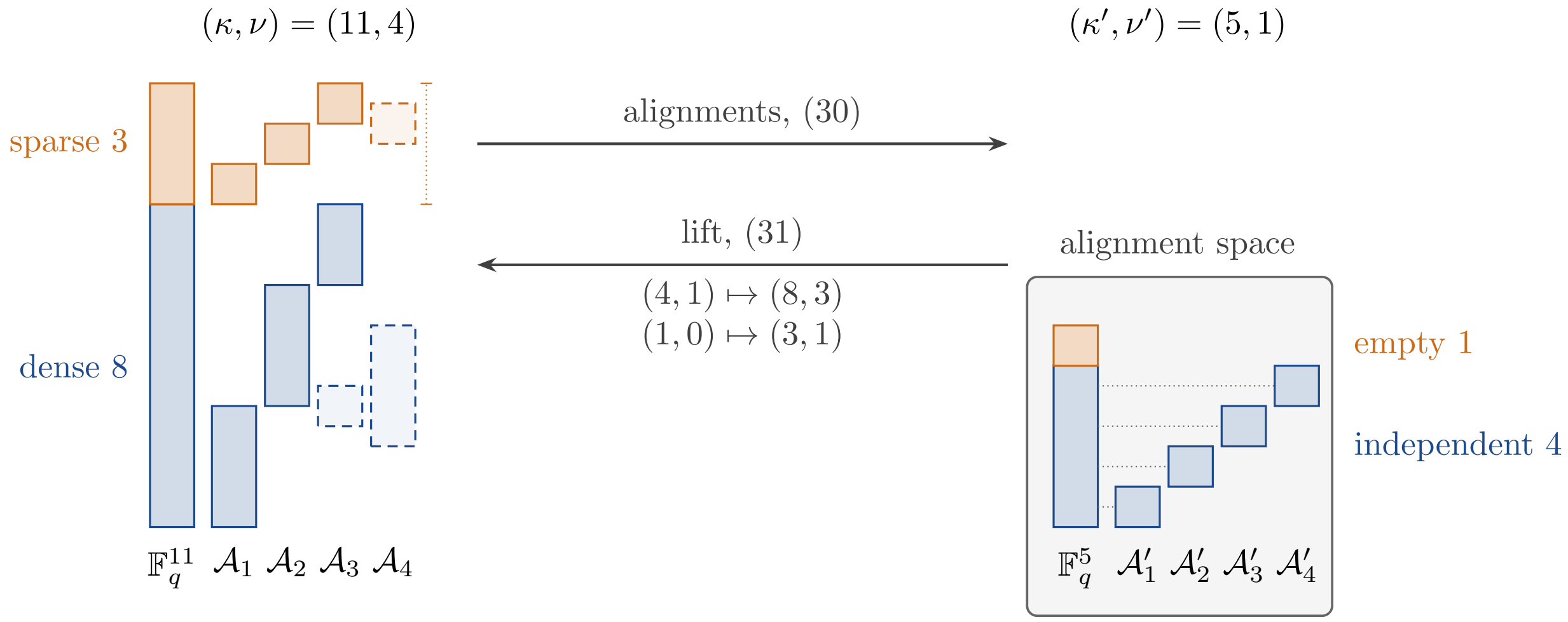
\begin{figure}[t]
\centering
\begin{tikzpicture}[x=0.46cm,y=0.42cm,>=Stealth,
  dense/.style={fill=capblue!20,draw=capblue,line width=.6pt},
  sparse/.style={fill=alignorange!25,draw=alignorange,line width=.6pt},
  densea/.style={fill=capblue!6,draw=capblue,densely dashed,line width=.6pt},
  sparsea/.style={fill=alignorange!8,draw=alignorange,densely dashed,line width=.6pt},
  lab/.style={font=\small}]
\node[lab,anchor=south] at (3.6,11.7) {$(\kappa,\nu)=(11,4)$};
\draw[dense]  (0,0) rectangle (1,8);
\draw[sparse] (0,8) rectangle (1,11);
\node[lab,anchor=north] at (0.5,-0.2) {$\Fq^{11}$};
\foreach \u/\x in {1/1.4,2/2.6,3/3.8,4/5.0}{\node[lab,anchor=north] at (\x+0.5,-0.2) {$\mathcal A_{\u}$};}
\draw[dense] (1.4,0) rectangle (2.4,3);
\draw[dense] (2.6,3) rectangle (3.6,6);
\draw[dense] (3.8,6) rectangle (4.8,8);
\draw[densea] (3.8,2.5) rectangle (4.8,3.5);
\draw[densea] (5.0,2) rectangle (6.0,5);
\draw[sparse] (1.4,8) rectangle (2.4,9);
\draw[sparse] (2.6,9) rectangle (3.6,10);
\draw[sparse] (3.8,10) rectangle (4.8,11);
\draw[sparsea] (5.0,9.5) rectangle (6.0,10.5);
\draw[alignorange,densely dotted] (6.25,8) -- (6.25,11);
\foreach \yy in {8,11}{\draw[alignorange] (6.12,\yy) -- (6.38,\yy);}
\node[lab,anchor=east,capblue] at (-0.2,4) {dense $8$};
\node[lab,anchor=east,alignorange] at (-0.2,9.5) {sparse $3$};
\begin{scope}[xshift=9.4cm]
\draw[black!60,fill=black!4,rounded corners=3pt,line width=.7pt] (-0.6,-2.2) rectangle (6.3,6.2);
\node[lab,anchor=south] at (2.8,11.7) {$(\kappa',\nu')=(5,1)$};
\node[lab,anchor=south,black!75] at (2.8,6.3) {alignment space};
\draw[dense]  (0,0) rectangle (1,4);
\draw[sparse] (0,4) rectangle (1,5);
\node[lab,anchor=north] at (0.5,-0.2) {$\Fq^{5}$};
\foreach \u in {1,...,4}{
  \pgfmathsetmacro{\x}{0.2+\u*1.2}
  \draw[black!60,densely dotted] (1,\u-0.5) -- (\x,\u-0.5);
  \draw[dense] (\x,\u-1) rectangle (\x+1,\u);
  \node[lab,anchor=north] at (\x+0.5,-0.2) {$\mathcal A_{\u}'$};
}
\node[lab,anchor=west,capblue] at (6.5,2) {independent $4$};
\node[lab,anchor=west,alignorange] at (6.5,4.5) {empty $1$};
\end{scope}
\draw[-{Stealth[length=2mm]},line width=.8pt,black!75] (7.4,9.5) -- node[lab,above]{alignments, \eqref{eq:forward-dims}} (19.4,9.5);
\draw[{Stealth[length=2mm]}-,line width=.8pt,black!75] (7.4,6.5) -- node[lab,above,align=center]{lift, \eqref{eq:reverse-dims}}
  node[lab,below,align=center]{$(4,1)\mapsto(8,3)$\\ $(1,0)\mapsto(3,1)$} (19.4,6.5);
\end{tikzpicture}
\caption{The reduction of Example~\ref{ex:reduction} for $K=4$. Bar heights are dimensions, and each user subspace is drawn beside the part of the space it occupies. Left: the relation space $\Fq^{11}$ and the four $4$-dimensional user subspaces, colored by the dense component (blue, $8$ dimensions, $3$ per user) and the sparse component (orange, $3$ dimensions, $1$ per user). Solid blocks are new dimensions, and dashed blocks lie in the span of the solid blocks they are drawn across, so each dashed dimension is an alignment: in the dense part, any three users have one alignment among themselves ($3\cdot3-8=1$), giving the dashed block of $\mathcal A_3$, and $\mathcal A_4$ lies entirely in the span of the others; in the sparse part, the line $\mathcal A_4$ is a combination of all three other lines (dotted span). Right, on the gray background: these five alignments form the alignment space $\Fq^5$, with the lines $\mathcal A_k'=\col(\boldsymbol e_k)$ as in Figure~\ref{fig:alignment}; its base decomposition~\eqref{eq:base-table} is an independent component (blue) and an empty component (orange). Lifting each component by~\eqref{eq:reverse-dims} gives the two components on the left.}
\label{fig:reduction}
\end{figure}

\begin{example}[One reduction step]\label{ex:reduction}
Let $(K,\mu,\nu)=(4,9,4)$, so $\kappa=11$ and $\nu/\kappa=4/11$. With configurations written by their dimensions $(\kappa,\nu)$ as in~\eqref{eq:base-table}, the decomposition follows the chain
\[
(11,4)
\xrightarrow[\eqref{eq:forward-dims}]{\text{alignments}}
(5,1)
\xrightarrow[\eqref{eq:base-table}]{\text{base}}
(4,1)\oplus(1,0)
\xrightarrow[\eqref{eq:reverse-dims}]{\text{lift}}
(8,3)\oplus(3,1).
\]
\begin{itemize}
\item \emph{Alignments.} By~\eqref{eq:forward-dims}, $(\kappa',\nu')=(4\cdot4-11,\ 3\cdot4-11)=(5,1)$: the alignment configuration is four lines in $\Fq^5$.
\item \emph{Base.} Since $5\ge K\cdot1=4$,~\eqref{eq:base-table} gives $(5,1)=(4,1)\oplus(1,0)$, an independent and an empty component.
\item \emph{Lift.} By~\eqref{eq:reverse-dims}, $(4,1)$ gives $(3\cdot4-4\cdot1,\ 4-1)=(8,3)$, and $(1,0)$ gives $(3\cdot1-4\cdot0,\ 1-0)=(3,1)$, both balanced.
\end{itemize}
The per-user fractions are $3/8$ and $1/3$, consecutive elements of $\mathcal S_4$ with $1/3<4/11<3/8$. So $4/11$ is a split fraction, with a dense component of dimension $\kappa_+=8$ and a sparse component of dimension $\kappa_-=3$, as in~\eqref{eq:blocks}. Figure~\ref{fig:reduction} shows the step.
\end{example}

The general case follows the same steps, with one reduction replaced by a finite sequence of reductions. The following lemma is the split part of Lemma~\ref{lem:geometry}, extended to include the endpoints $\beta_\pm$; Appendix~\ref{app:geometry-proof} deduces Lemma~\ref{lem:geometry} from it and Lemma~\ref{lem:central}.
\begin{lemma}[Split between consecutive balanced fractions]\label{lem:outer}
Let $\beta_-<\beta_+$ be consecutive elements of $\mathcal S_K$ outside the balanced interval. Generic configurations with $\beta_-\le\nu/\kappa\le\beta_+$ split into two balanced components with per-user fractions $\beta_-$ and $\beta_+$ and dimensions $\kappa_\pm$ of~\eqref{eq:blocks}; at an endpoint, one component is zero, and the configuration is balanced.
\end{lemma}
\begin{proof}
\emph{Reduction to lower pairs.} If $(\beta_-,\beta_+)$ is an upper pair of Theorem~\ref{thm:main}, lying above the balanced interval, the configuration $(\Fq^\kappa;\mathcal A_1^\perp,\ldots,\mathcal A_K^\perp)$ has fraction $1-\nu/\kappa$ between the consecutive elements $1-\beta_+<1-\beta_-$ below it, and by Lemma~\ref{lem:sum-perp} it suffices to split that configuration. So assume $(\beta_-,\beta_+)=(\beta_j,\beta_{j+1})$ for some $j\ge0$, a lower pair or, for $j=0$, the pair $(0,1/K)$.

\emph{The chain.} Let $m=\lfloor j/2\rfloor$. The decomposition follows the chain
\[
\begin{aligned}
(\kappa_0,\nu_0)
&\xrightarrow[\eqref{eq:forward-dims}]{\text{alignments}}
(\kappa_1,\nu_1)
\xrightarrow[\eqref{eq:forward-dims}]{\text{alignments}}
\cdots
\xrightarrow[\eqref{eq:forward-dims}]{\text{alignments}}
(\kappa_m,\nu_m)\\
&\xrightarrow[\eqref{eq:base-table}]{\text{base}}
\text{two components}
\xrightarrow[\eqref{eq:reverse-dims}]{\text{lift }m\text{ times}}
\text{two components},
\end{aligned}
\]
with $(\kappa_0,\nu_0)=(\kappa,\nu)$. It remains to track the parameters along the chain.
\begin{itemize}
\item \emph{Alignments.} By~\eqref{eq:forward-dims}, $(\kappa_{t+1},\nu_{t+1})=(K\nu_t-\kappa_t,\ (K-1)\nu_t-\kappa_t)$, and by~\eqref{eq:reverse-dims}, a configuration between $\beta_i$ and $\beta_{i+1}$ reduces to one between $\beta_{i-2}$ and $\beta_{i-1}$. As shown below~\eqref{eq:forward-dims}, $\nu_{t+1}<\nu_t$ below the balanced interval, so the user dimension strictly decreases, and each step stays below the balanced interval.
\item \emph{Base.} After $m$ steps the configuration lies between $\beta_0=0$ and $\beta_1=1/K$ (if $j$ is even) or between $\beta_1=1/K$ and $\beta_2=1/(K-1)$ (if $j$ is odd). It then has $\kappa\ge(K-1)\nu$, so~\eqref{eq:base-table} writes it as a direct sum of two balanced base configurations with these fractions.
\item \emph{Lift.} By~\eqref{eq:reverse-dims}, each lift acts componentwise and maps a component with fraction $\beta_i$ to one with fraction $\beta_{i+2}$; by Corollary~\ref{lem:relations-balance}, it stays balanced. After $m$ lifts, the components have fractions $\beta_j$ and $\beta_{j+1}$.
\end{itemize}
For $K=2$ and $K=3$, Theorem~\ref{thm:main} defines no lower or upper pairs; there, fractions at most $1/2$ are handled directly by~\eqref{eq:base-table}, and fractions above $1/2$ by orthogonal complements, so $m=0$.

\emph{Dimensions and changes of basis.} The chain produces $\Fq^\kappa=\mathcal E_-\oplus\mathcal E_+$ with $\mathcal A_k=(\mathcal A_k\cap\mathcal E_-)\oplus(\mathcal A_k\cap\mathcal E_+)$ and $\dim(\mathcal A_k\cap\mathcal E_\pm)=\beta_\pm\kappa_\pm$, where $\kappa_\pm=\dim(\mathcal E_\pm)$. Counting dimensions of the relation space and of one user subspace gives $\kappa_-+\kappa_+=\kappa$ and $\beta_-\kappa_-+\beta_+\kappa_+=\nu$, and solving these two equations gives the values in~\eqref{eq:blocks}. They are integers because they are dimensions of the constructed spaces. The chain uses only null spaces, complements, and matrix inverses on sets specified by nonzero minors, so the changes of basis are rational functions of the user-channel entries.

\emph{Probability, uniformly in characteristic.} The chain has a number of steps bounded in terms of $(K,\nu,\kappa)$, and each step needs finitely many matrices of full rank ($\mathbf A$, the blocks $\mathbf Q_k$, and the inverting minors). Fix $p$. Starting from base configurations with full-rank channels and lifting, we obtain, in a large enough finite field of characteristic $p$, a configuration at which all these rank conditions hold. Each lift is a valid reverse construction (Appendix~\ref{app:elementary}); its condition that no $K-1$ user subspaces share a nonzero vector holds because the component being lifted is balanced below the balanced interval, and Corollary~\ref{lem:relations-balance} keeps the lifted component balanced. Fix minors that are nonzero at this configuration. In the original entries of $\mathbf A_1,\ldots,\mathbf A_K$, after clearing denominators, their product is a polynomial $f_p$ over $\Fp$, nonzero because it does not vanish there, of degree at most a constant $\delta$ depending only on $(K,\nu,\kappa)$. The construction succeeds whenever $f_p\ne0$, so by the Schwartz--Zippel lemma it fails with probability at most $\delta/q$, uniformly in $p$.
\end{proof}

\subsection{Completing the proof of Lemma~\ref{lem:geometry}}\label{app:geometry-proof}
\emph{First part: balanced fractions.} Let $\nu/\kappa\in\mathcal S_K$. There are two cases.
\begin{itemize}
\item \emph{The balanced interval} ($K\ge4$, $\beta_\infty\le\nu/\kappa\le1-\beta_\infty$). Then $K\beta(1-\beta)\ge1$ with $\beta=\nu/\kappa$, and Lemma~\ref{lem:central} gives balance.
\item \emph{An isolated balanced fraction} ($\nu/\kappa=\beta_j$ or $1-\beta_j$). Apply Lemma~\ref{lem:outer} to a pair of consecutive elements of $\mathcal S_K$ with $\nu/\kappa$ as one of its ends. The dimension formulas~\eqref{eq:blocks} give $\kappa_+=0$ or $\kappa_-=0$, so the configuration is a single balanced component.
\end{itemize}

\emph{Second part: split fractions.} Let $\beta_-<\beta_+$ be consecutive elements of $\mathcal S_K$, which lie outside the balanced interval, and let $\beta_-<\nu/\kappa<\beta_+$. Lemma~\ref{lem:outer} gives the explicit decomposition $\Fq^\kappa=\mathcal E_-\oplus\mathcal E_+$ with $\mathcal A_k=(\mathcal A_k\cap\mathcal E_-)\oplus(\mathcal A_k\cap\mathcal E_+)$ for every user, $\dim(\mathcal E_\pm)=\kappa_\pm$ as in~\eqref{eq:blocks}, $\dim(\mathcal A_k\cap\mathcal E_\pm)=\beta_\pm\kappa_\pm$, and both components balanced. Since $\beta_-<\nu/\kappa<\beta_+$, both $\kappa_-$ and $\kappa_+$ are positive. This is the second part of Lemma~\ref{lem:geometry}.

In both parts, the events used have probability $1-O(1/q)$, uniformly in the characteristic, so they hold with probability tending to one along every sequence of prime powers. \hfill$\square$

\begin{remark}[Connections with quiver geometry]
In quiver language~\cite{DerksenWeyman}, the alignment transformation and its reverse are compositions of reflection functors, balance is King's semistability, the condition of Lemma~\ref{lem:central} says that the Tits form is nonpositive, and the split of Lemma~\ref{lem:geometry} is related to the canonical decomposition. These results are developed mainly over the complex numbers; this appendix proves what is needed here over arbitrary finite fields.
\end{remark}

\section{Proof of Corollary~\ref{cor:channel}}\label{app:channel}
\emph{Achievability.} Use the uncertainty sets $\mathcal Z_k$ of the computation scheme in Section~\ref{sec:ach} as the interferers' codebooks, with $L=1$. Receiver~$0$ fails to decode only if two inputs with different desired vectors give the same received signal, that is, $\mathbf B\by+\sum_k\mathbf A_k\bz_k=\mathbf B\by'+\sum_k\mathbf A_k\bz_k'$ with $\by\ne\by'$ and $\bz_k,\bz_k'\in\mathcal Z_k$. This gives $\mathbf B\Delta\by=-\sum_k\mathbf A_k\Delta\bz_k$ with $\Delta\bz_k\in\mathcal Z_k-\mathcal Z_k$. Since $\mathcal Z_k-\mathcal Z_k$ is unchanged under negation, replacing $\Delta\bz_k$ by $-\Delta\bz_k$ gives two solutions of the relation equations with observation differences in $\mathcal Z_k-\mathcal Z_k$ and demand difference $\Delta\by$. The decoding argument of the computation scheme rules this out unless $\Delta\by=\boldsymbol0$. Receivers $k\ge1$ observe their codewords directly. By~\eqref{eq:distinct} and Lemma~\ref{lem:alignment-rate}, the interference sum-rate $\sum_k\log_q|\mathcal T_k|$ approaches $K\nu-C$ with probability tending to one, with zero error.

\emph{Converse.} The following lemma replaces Lemma~\ref{lem:entropy-rate}. Here $\mathcal U_{\mathrm{res}}=\nullspace(\mathbf A)$ and $\boldsymbol\Pi_k$ are as in Section~\ref{sec:geo-converse}, and the interferer inputs play the role of the observations.
\begin{lemma}[An entropy bound gives an interference sum-rate upper bound]\label{lem:entropy-rate-channel}
For fixed channel matrices of full rank, suppose that for some $\zeta>1$ and constant $\gamma_\zeta$, every distribution of $\bwr$ in $\mathcal U_{\mathrm{res}}$ satisfies $\sum_kH(\boldsymbol\Pi_k\bwr)\ge\zeta H(\bwr)+\gamma_\zeta$. Then every achievable interference sum-rate satisfies
\[
\sum_kR_k\le K\nu-\frac{\zeta\mu-K\nu+\gamma_\zeta}{\zeta-1},
\]
that is, $K\nu$ minus the right side of~\eqref{eq:hidden} in Lemma~\ref{lem:entropy-rate}.
\end{lemma}
\begin{proof}
Fix a code of block length $L$ with vanishing error probabilities, and let $\boldsymbol u^L=\sum_k\mathbf A_k^{[L]}\bz_k^L$ be the combined interference, where $\mathbf A_k^{[L]}$ repeats $\mathbf A_k$ in $L$ diagonal blocks. Fano's inequality at receiver $k$ gives $H(\bz_k^L)\ge LR_k-o(L)$. The desired vector is uniform and independent of $\boldsymbol u^L$, so the entropy of receiver $0$'s observation given $\by^L$ is $H(\boldsymbol u^L)$; Fano's inequality at receiver $0$ and the bound $\kappa L$ on the entropy of its observation then give $H(\boldsymbol u^L)\le(\kappa-\nu)L+o(L)$.

For each value of $\boldsymbol u^L$, the interferer input tuples $\bz^L=(\bz_1^L,\ldots,\bz_K^L)$ with that value differ by elements of $\mathcal U_{\mathrm{res}}^L$. As in the proof of Lemma~\ref{lem:entropy-rate}, applying the entropy bound to each of the $L$ channel uses, conditioned on the earlier ones, and summing by the chain rule extends it to blocks:
\[
\sum_kH(\bz_k^L\mid\boldsymbol u^L)\ge\zeta H(\bz^L\mid\boldsymbol u^L)+L\gamma_\zeta.
\]
The inputs are independent and $\boldsymbol u^L$ is a function of them, so $H(\bz^L\mid\boldsymbol u^L)=\sum_kH(\bz_k^L)-H(\boldsymbol u^L)$, while $\sum_kH(\bz_k^L\mid\boldsymbol u^L)\le\sum_kH(\bz_k^L)$. Substituting and dividing by $\zeta-1$ gives
\[
\sum_kH(\bz_k^L)\le\frac{\zeta H(\boldsymbol u^L)-L\gamma_\zeta}{\zeta-1}.
\]
Combining this with $H(\bz_k^L)\ge LR_k-o(L)$ and $H(\boldsymbol u^L)\le(\kappa-\nu)L+o(L)$ from Fano's inequality, dividing by $L$, and letting $L\to\infty$ gives $\sum_kR_k\le(\zeta(\kappa-\nu)-\gamma_\zeta)/(\zeta-1)$, which equals the stated bound because $K\nu-\mu=\kappa-\nu$.
\end{proof}
Thus, by Lemma~\ref{lem:entropy-rate-channel}, every sum-rate lower bound $R\ge R_*$ that Section~\ref{sec:conv} obtains from Lemma~\ref{lem:entropy-rate} gives the interference bound $\sum_kR_k\le K\nu-R_*$, with the same $(\zeta,\gamma_\zeta)$, whenever the configuration is balanced or splits as stated in Lemma~\ref{lem:geometry}. The only other bound used there is the demand bound $R\ge\nu$; its counterpart is $\sum_kR_k\le K\nu-\nu$. Indeed, with $\boldsymbol z^L=(\boldsymbol z_1^L,\ldots,\boldsymbol z_K^L)$,
\begin{align*}
\sum_kH(\boldsymbol z_k^L)&=H(\boldsymbol z^L)=H(\boldsymbol u^L)+H(\boldsymbol z^L\mid\boldsymbol u^L)\\
&\le(\kappa-\nu)L+o(L)+L(K\nu-\kappa)=L(K-1)\nu+o(L).
\end{align*}
The first equality uses independence of the inputs, and the second the chain rule, since $\boldsymbol u^L$ is a function of $\boldsymbol z^L$. In the inequality, $H(\boldsymbol u^L)\le(\kappa-\nu)L+o(L)$ is the Fano bound above, and given each value of $\boldsymbol u^L$, the input tuple is determined up to an element of $\mathcal U_{\mathrm{res}}^L$, which has dimension $L(K\nu-\kappa)$, so $H(\boldsymbol z^L\mid\boldsymbol u^L)\le L(K\nu-\kappa)$. Hence every converse bound of Section~\ref{sec:conv} carries over with $R$ replaced by $K\nu-\sum_kR_k$, and $C_{\mathrm{int}}=K\nu-C$.

\bibliography{symmetric_distributed_linear_computation}
\end{document}